\documentclass[11pt,a4paper]{article}

\def\documenttitle{Modular Hamiltonian of the massive scalar field on the half line: A numerical approach}
\def\documentauthor{Christoph Minz and Erik Tonni}

\title{\Large\bf 
Modular Hamiltonian for the massive scalar field 
\\
on the half line: A numerical approach}
\author{Christoph Minz$^{a}$ and Erik Tonni$^{a,b}$}
\date{\small\em 
$^{a}$SISSA, via Bonomea 265, 34136 Trieste, Italy
\\
$^{b}$INFN, Sezione di Trieste, via Valerio 2, 34127 Trieste, Italy
}

\usepackage{paper}

\graphicspath{{figures/}}

\usepackage{csquotes}  
\usepackage[numbers,sort&compress]{natbib}
\usepackage{verbatim}
\newcommand{\citeDLMFsec}[1]{%
\cite[{\href{https://dlmf.nist.gov/#1}{Sec.~#1}}]{NIST:DLMF}}
\newcommand{\citeDLMFeq}[2]{%
\cite[{\href{https://dlmf.nist.gov/#1.E#2}{Eq.~#1.#2}}]{NIST:DLMF}}
\newcommand{\citeDLMFtab}[2]{%
\cite[{\href{https://dlmf.nist.gov/#1.T#2}{Tab.~#1.#2}}]{NIST:DLMF}}

\usepackage{soul}

\begin{document}

\maketitle

\begin{abstract} 

We study the modular Hamiltonian of an interval
for the ground state of a massive
free scalar field on the half line 
with Robin boundary conditions, 
by employing a numerical method. 
When the interval is adjacent to the boundary, 
we find numerical evidence that 
the modular Hamiltonian is non-local, 
except for the limiting cases of the massless scalar 
satisfying either Dirichlet or Neumann boundary conditions.
When the interval is separated from the boundary, 
the numerical analysis indicates that the modular Hamiltonian is 
non-local for all these boundary conditions 
and any value of the mass.

\end{abstract}

\clearpage
\setcounter{tocdepth}{2}
\tableofcontents

\section{Introduction}
\label{sec:Introduction}

The modular Hamiltonian is an important operator 
in the algebraic approach to quantum field theory (QFT), 
within the modular theory of Tomita and Takesaki \cite{Takesaki:1970, EckmannOsterwalder:1973, Haag:1996, Borchers:2000}, which provides detailed information
about entanglement in a quantum system,
connecting also to quantum information theory 
\cite{HollandsSanders:2018}.
Considering a relativistic QFT, 
it is very insightful to have an analytic expression for the modular Hamiltonian of a subsystem in terms of the fundamental fields of the model.
Few examples are known explicitly so far,
which can be collected into three classes,
depending on the degree of non-locality  
of the modular Hamiltonian. 

The first class is made by  modular Hamiltonians
that are local operators. 
The most prominent example is the modular Hamiltonian corresponding to the bipartition of a Cauchy surface in Minkowski spacetime at fixed time by a spatial hyperplane, 
when the relativistic QFT is in the ground state, due to the theorem of Bisognano and Wichmann 
\cite{BisognanoWichmann:1975, BisognanoWichmann:1976}, 
who showed that this modular Hamiltonian is proportional to the boost along the direction orthogonal to the hyperplane separating the two halves of the Cauchy surface. 
For conformal field theories (CFT), the result by Bisognano--Wichmann was adapted to other regions through various conformal transformations
\cite{Buchholz:1978, HislopLongo:1982, Hislop:1988, WongKlichZayasVaman:2013, CardyTonni:2016}.
In the case of one spatial dimension, 
analytic expressions for the modular Hamiltonians 
of an interval have been found 
also in a CFT with a boundary (BCFT),
either on the half line 
\cite{CardyTonni:2016}
or on the segment \cite{TonniRodriguezLagunaSierra:2018},
when the interval is adjacent to the boundary.
We remind that in a BCFT 
the allowed boundary conditions (b.c.\@) are determined by specific constraints 
\cite{Cardy:1984, Cardy:1986, Cardy:1989, Cardy:2004,DiFrancescoMathieuSenechal:1997}.

The other two classes contain modular Hamiltonians 
$\mathcal{K}_A$ that are non-local
and they have been obtained for QFT 
that are free and quadratic in the fundamental fields $\psi$ 
(for a bosonic scalar field, this includes both the field and its velocity); 
hence they can be written as 
\begin{equation}
\label{eq:ModularHamiltonian.FockSpace}
    \mathcal{K}_A
  = \int_{A}\int_{A}
      \psi(x)^\dagger \mult K(x, y) \mult \psi(y)
    \id{x} \id{y}
  \eqend{,}
\end{equation}
where $A$ is the spatial subsystem and 
$K(x, y)$ is a kernel such that $\mathcal{K}_A$ is self-adjoint and
whose non-local nature characterises the two other classes of known modular Hamiltonians.

The second class collects modular Hamiltonians of the form \eqref{eq:ModularHamiltonian.FockSpace}
where the  kernel $K(x, y)$ is the sum of a local term and various bilocal terms.
In a bilocal term, the integrand is also determined by a function depending only on $x$
because the second position $y=r(x)$ depends on $x$ in a non-trivial way; hence the corresponding kernel contains 
distributions that are proportional to
$\updelta\bigl( y - r(x) \bigr)$ and its derivatives.
The first example is given by  the free massless Dirac field on the line and in its ground state, when the line is bipartite 
by the union of a finite number of disjoint intervals \cite{CasiniHuerta:2009a}.
This result has been obtained by exploiting the Gaussian nature of the ground state through a formula discussed in 
\cite{Araki:1971b, Peschel:2003, EislerPeschel:2009},
which provides the modular Hamiltonian through the spectral problem associated to the two-point function restricted to the subsystem. 
This procedure has also been employed to find 
other analytic expressions of modular Hamiltonians 
for the free massless Dirac field containing also bilocal terms,
both in cases where the one-dimensional spatial slice at fixed time is 
invariant under translations \cite{KlichVamanWong:2017, KlichVamanWong:2018, BlancoPerezNadal:2019, FriesReyes:2019, CadamuroFroebPerezNadal:2024} 
and in cases where such invariance does not occur because of the presence of boundaries or defects 
\cite{MintchevTonni:2021a, MintchevTonni:2021b, TonniTrezzi:2025}.
The analytic results in the latter group
are suitable for a qualitative comparison with the 
numerical results discussed in this manuscript, 
where we consider the ground state of a scalar field on the half line.

The third class is made by the modular Hamiltonians 
of the form \eqref{eq:ModularHamiltonian.FockSpace}
whose analytic expression includes also 
a fully non-local term, where the double integration over the subsystem is not simplified by the occurrence of 
the Dirac delta or its derivatives. 
Only one example is known in the literature 
which is not based on a perturbative computation
and it corresponds to the ground state of the chiral current associated to the free massless scalar on the line, when the subsystem is the union of two disjoint intervals \cite{AriasCasiniHuertaPontello:2018}.

All the modular Hamiltonians corresponding to subsystems with a finite volume mentioned above 
have been obtained either for CFT models 
or for free massless fields.
It would be very interesting to find the explicit expression 
for a modular Hamiltonian of a compact subsystem for a massive field.
Perturbative computations in the mass have been performed 
\cite{AriasBlancoCasiniHuerta:2017, CadamuroFroebMinz:2024},
finding fully non-local expressions at first order.

While it remains difficult to obtain exact analytic expressions, numerical techniques have been developed to study the modular Hamiltonians, both to recover the existing QFT results 
in a different way and to gain insights for the cases that are not known analytically.
A numerical approach that has been largely explored 
is based on lattice models,
where the modular Hamiltonian is often called entanglement Hamiltonian. 
Numerical studies have been performed for entanglement 
Hamiltonians in the massive regime 
of some free lattice models \cite{AriasBlancoCasiniHuerta:2017, EislerDiGiulioTonniPeschel:2020, GentileRotaruTonni:2025},
based on the formulas provided in \cite{Peschel:2003, EislerPeschel:2009, CasiniHuerta:2009b}.
However, concrete predictions 
for the continuum limit of the lattice models 
in the massive regime have not been found, 
except for strong indications that the corresponding 
QFT result is fully non-local. 
Taking the continuum limit of the entanglement Hamiltonian is non-trivial, also in the massless regime
\cite{EislerPeschel:2017, EislerTonniPeschel:2019, DiGiulioTonni:2020, JaverzatTonni:2022}
for cases where the QFT result is known from CFT.

An alternative numerical approach to approximate 
the modular Hamiltonian for a scalar field 
has been introduced in \cite{BostelmannCadamuroMinz:2023},
where it has been employed to explore the ground state 
in the cases of the interval in the line 
and of the three-dimensional ball in a
Cauchy hypersurface at fixed time 
in Minkowski spacetime.
This method is not based on a lattice model,
but it directly employs the one-particle space in the continuum
of this QFT \cite{FiglioliniGuido:1989, FiglioliniGuido:1994, CiolliLongoRuzzi:2019}. 
Similarly to the lattice computation, 
this approach describes the field data on 
the initial Cauchy hypersurface
and, in contrast with the lattice computation,
the spacing of the discretization is not fixed 
throughout the space, but it can be adjusted, 
depending on the setup.
The method provides an approximation 
of the modular Hamiltonian on the full Hilbert space 
of the initial data, 
through a discretization procedure which involves both parts 
of the spatial bipartition.
The massless regime, where a zero mode occurs in the two-dimensional case, was not explored in \cite{BostelmannCadamuroMinz:2023}
and the parabolic profile found in \cite{HislopLongo:1982} 
was only used as an analytic reference for comparison.
In the regime of large mass, an interesting upper bound 
has been observed in \cite{BostelmannCadamuroMinz:2023}, 
similar to the one found in \cite{EislerDiGiulioTonniPeschel:2020} 
through a numerical analysis in free one-dimensional 
lattice models.

The QFT expressions of the modular Hamiltonian 
for the scalar and the fermion field in 
\cite{FiglioliniGuido:1989, FiglioliniGuido:1994, CiolliLongoRuzzi:2019}
are equivalent to those formulas in the continuum 
that correspond to the lattice approach
\cite{Araki:1971b,Peschel:2003, EislerPeschel:2009, CasiniHuerta:2009b, CasiniHuerta:2009a},  
as shown in \cite{CadamuroFroebMinz:2024, Froeb:2025}.

In this work we apply the numerical approach of 
\cite{BostelmannCadamuroMinz:2023} to explore the modular Hamiltonian of an interval 
for the ground state of the free scalar field 
on the half line with Robin b.c.\@  \cite{LiguoriMitchev:1998},
when the position of the interval in the half line is generic. 
The cases of Dirichlet b.c.\@ 
and Neumann b.c.\@ are limiting regimes of the Robin b.c.\@
and provide two consistent BCFT models for the massless scalar on the half line. 
The QFT model corresponding to Neumann b.c.\@ is the only one 
where a zero mode occurs, and therefore the massless regime deserves  special care. 
Our aim is to investigate the nature of 
the modular Hamiltonian of an interval for these models
in the massless and massive regimes,
as the parameter associated to the Robin b.c.\@ changes.
Since the analytic expressions for these modular Hamiltonians in terms of the fundamental field are not available in the literature, our approximations provide useful insights towards them 
or other related analytic results, like e.g.\@ upper bounds.

The outline of the paper is as follows. 
In Sec.~\ref{sec:HalfLine}, 
we describe the numerical procedure, 
after the definition of the model and 
the corresponding analytic formulas that are needed.
In Sec.~\ref{sec:Results.AdjacentInterval} and Sec.~\ref{sec:Results.SeparatedInterval}
we discuss the results of our numerical analysis, 
in the case where the interval is either adjacent to the boundary
or separated from it, respectively. 
We conclude with a summary and
possible directions for future studies 
in Sec.~\ref{sec:Conclusion}.
The appendices contain important technical details and also further results supporting the discussion in the main text.

\section{Modular Hamiltonian for the massive scalar field on the half line}
\label{sec:HalfLine}

In this section we define the model 
and describe the method employed in our numerical approach,
introducing the necessary mathematical tools. 
Further relevant details and additional results 
are discussed in 
the Appendices from \ref{appx:HilbertSpaces} to \ref{appx:Discretization}.

\subsection{The differential operator $D$ and the modular Hamiltonian}
\label{sec:HalfLine.OperatorDModularHamiltonian}

The action for the scalar field $\varphi (t, x)$ on the half line (parametrised by $x \geqslant 0$) reads 
\begin{equation}
\label{eq:Action}
    S[\varphi]
  = \frac{1}{2} \intl_{-\infty}^{\infty}
    \intl_{0}^{\infty}
    \Bigl\{
      \bigl[ \partial_t \varphi(t, x) \bigr]^2 
      - \bigl[ \partial_x \varphi(t, x) \bigr]^2 
      - m^2 \mult \varphi(t, x)^2
    \Bigr\}
    \id{x}
    \id{t}
    - \frac{\eta}{2} 
    \intl_{-\infty}^{\infty}
      \varphi(t, 0)^2
    \id{t}
  \eqend{,}
\end{equation}
where we assume $m \geqslant 0$ and $\eta \geqslant 0$ 
for the parameters corresponding to 
the mass and boundary term respectively.
The equation of motion coming from \eqref{eq:Action}
is the Klein--Gordon equation 
\begin{equation}
\label{eq:StaticKleinGordon}
    \partial_t^2 \varphi(t, x) + D \varphi(t, x)
  = 0
  \eqend{,}
  \qquad
    D
  \coloneq - \partial_x^2 + m^2
  \eqend{,}
\end{equation}
with Robin b.c.\@ (also known as mixed b.c.\@)
at the origin of the half line, namely
\begin{equation}
\label{eq:HalfLine.RobinBoundaryCondition}
    \lim_{x \to 0^+}
    \Bigl(
      \partial_x \varphi(t,x)
      - \eta \mult \varphi(t,x)
    \Bigr)
  = 0
  \eqend{.}
\end{equation}
The Neumann b.c.\@ $\lim_{x \to 0^+} \partial_x \varphi(t,x) = 0$
and the Dirichlet b.c.\@ $\varphi(t,0) = 0$ 
are the limiting cases of 
the Robin b.c.\@ \eqref{eq:HalfLine.RobinBoundaryCondition} 
given by $\eta = 0$ and $\eta \to \infty$, respectively. 
The massless regime of this model has been extensively explored in \cite{LiguoriMitchev:1998}.
We focus on the regime where $\eta \geqslant 0$ because the condition $\eta < 0$ leads to the introduction of a bound state \cite{MintchevPilo:2001, BellazziniMintchevSorba:2010}. 
Since our analysis will be carried out on the slice at $t=0$, 
we find it conveninet to lighten the notation by introducing
$\varphi(x)  \coloneq \varphi(0, x)$.

The eigenfunctions of the differential operator $D$ in \eqref{eq:StaticKleinGordon} read 
\begin{equation}
\label{eq:HalfLine.MomentumIntegral.Mode}
    \psi_p(x)
  = \e^{-\i p x}
    + \frac{p - \i \eta}{p + \i \eta} \mult \e^{\i p x}
  \eqend{,}
  \qquad
p > 0
  \eqend{,}
\end{equation}
which are the solutions found in \cite{LiguoriMitchev:1998} 
restricted to the slice $t = 0$.
Indeed, we have that 
\begin{equation}
\label{eq:Helmholtz.Spectrum}
    D \psi_p(x)
  = \widehat{D}(p) \mult \psi_p(x)
  \eqend{,}
\qquad
    \widehat{D}(p)
  \coloneq p^2 + m^2
    \eqend{,}
\qquad    
  \lim_{x \to 0^+}
    \Bigl(
      \partial_x \psi_p(x)
      - \eta \mult \psi_p(x)
    \Bigr)
  = 0
  \eqend{.}
\end{equation}
The eigenfunctions \eqref{eq:HalfLine.MomentumIntegral.Mode} satisfy  
the orthogonality and completeness conditions \cite{LiguoriMitchev:1998},
which are given respectively by 
\begin{equation}
\label{eq:HalfLine.MomentumIntegral.Mode.Properties}
    \frac{1}{2 \pi}
    \int_{0}^{\infty}
      \conj{\psi}_p(x) \mult \psi_q(x)
    \id{x}
  = \updelta(p - q)
  \eqend{,}
\qquad
    \frac{1}{2 \pi}
    \int_{0}^{\infty}
      \conj{\psi}_p(x) \mult \psi_p(y)
    \id{p}
  = \updelta(x - y)
  \eqend{.}
\end{equation}

In the following, we adapt the definition of the
one-particle Hilbert space $\mathcal{H}$ of the fields and the field velocities from \cite{BostelmannCadamuroMinz:2023} to 
the half line $\Reals_+$ at $t = 0$ 
and the one-particle structure for the modular Hamiltonian.
We start with the real Hilbert space $\mathcal{H}_{\labelReal} \coloneq \Lp2(\Reals_+, \Reals)$ of square-integrable functions on $\Reals_+$ with the inner product
\begin{equation}
\label{eq:HilbertSpace.RealInnerProduct}
    \innerProd[\labelReal]{f}{g}
  \coloneq \int_{0}^{\infty}
      f(x) \mult g(x)
    \id{x}
  \eqend{,}
\qquad
    \forall f, g \in \mathcal{H}_{\labelReal}
  \eqend{.}
\end{equation}
Since the operator $D$ is positive but unbounded, 
its domain must be carefully defined, as discussed e.g.\@ in 
\cite{BostelmannCadamuroMinz:2023,FiglioliniGuido:1989,FiglioliniGuido:1994}.
Thus, we get the two real Hilbert spaces $\mathcal{H}_{\labelReal}^{\pm1/4}$ (Sobolev spaces)
obtained by considering the closures of the operator domain 
with respect to the norms $\|\cdot\|_{\pm1/4} \coloneq \|D^{\pm1/4} \cdot\|_{\labelReal}$, namely
\begin{equation}
\label{eq:RealHilbertSpaces.FromDomains}
    \mathcal{H}_{\labelReal}^{\pm\frac{1}{4}}
  \coloneq \closure{\dom D^{\pm\frac{1}{4}}}^{\left\| \cdot \right\|_{\pm\frac{1}{4}}}
  \eqend{.}
\end{equation}
The root $D^{+1/2}$ of the differential operator $D$ naturally extends to a map $\mathcal{H}_{\labelReal}^{-1/4} \to \mathcal{H}_{\labelReal}^{+1/4}$ and it has the inverse map $D^{-1/2}$.
The complex one-particle Hilbert space $\mathcal{H}$ is defined as $\mathcal{H} \coloneq \mathcal{H}_{\labelReal}^{+1/4} \oplus \mathcal{H}_{\labelReal}^{-1/4}$.
It is constructed from the components $\mathcal{H}_{\labelReal}^{\pm 1/4}$
and  equipped with the complex structure
\begin{equation}
\label{eq:BosonHilbertSpace.ComplexStructure}
    I
  \coloneq \begin{pmatrix}
      0 & D^{-\frac{1}{2}} \\
      - D^{+\frac{1}{2}} & 0
    \end{pmatrix}
    \eqend{,}
\end{equation}
and the complex inner product 
\cite{BostelmannCadamuroMinz:2023} 
\begin{equation}
\label{eq:BosonHilbertSpace.InnerProduct}
    \innerProd[\mathcal{H}]{f}{g}
  \coloneq  
  \innerProd[\frac{1}{4} \oplus -\frac{1}{4}]{f}{U^2 g}
    - \i 
    \innerProd[\frac{1}{4} \oplus -\frac{1}{4}]{f}{\sigma g}
      \eqend{,}
  \qquad
    \forall f, g \in \mathcal{H}
      \eqend{,}
\end{equation}
defined through the real inner product $\innerProd[1/4 \oplus -1/4]{f}{U^2 g}$ (for $f, g \in \mathcal{H}$) and the operators
\begin{equation}
\label{eq:BosonHilbertSpace.Operators}
    U
  \coloneq \begin{pmatrix}
      D^{+\frac{1}{4}} & 0 \\
      0 & D^{-\frac{1}{4}}
    \end{pmatrix}
  \eqend{,}
\qquad
    \sigma
  \coloneq 
  \begin{pmatrix}
      0 & \one \\
      -\one & 0
    \end{pmatrix}
    = U^2 I
  \eqend{.}
\end{equation}

The unitary transformation $U : \mathcal{H}  \to \widetilde{\mathcal{H}} $ maps $\mathcal{H}$ into the Hilbert space $\widetilde{\mathcal{H}} \coloneq \mathcal{H}_{\labelReal} \oplus \mathcal{H}_{\labelReal}$
with the standard complex inner product
\begin{equation}
\label{eq:BosonHilbertSpace.InnerProduct.Standard}
    \innerProd[\tilde{\mathcal{H}}]{f}{g}
  \coloneq 
  \innerProd[\labelRealDouble]{f}{g}
    - \i \innerProd[\labelRealDouble]{f}{\sigma g}
  \eqend{,}
\end{equation}
where the real inner product is derived from \eqref{eq:HilbertSpace.RealInnerProduct}, so that
\begin{equation}
\label{eq:BosonHilbertSpace.InnerProduct.StandardReal}
    \innerProd[\labelRealDouble]{f}{g}
  \coloneq 
  \innerProd[\labelReal]{f_1}{g_1}
    + \innerProd[\labelReal]{f_2}{g_2}
  \eqend{,}
\end{equation}
for $f = f_1 \oplus f_2$ and $g = g_1 \oplus g_2$ as two generic elements of $\widetilde{\mathcal{H}}$.

We consider the  bipartition of the half line at $t=0$
characterised by a spatial region $A$ and its complement $B$.
This bipartition corresponds to a specific choice of the 
standard subspace $\mathcal{H}_A \subset \mathcal{H}$ 
and more details about this construction are provided in Appendix~\ref{appx:HilbertSpaces}.
In our analysis the region $A \subsetneq \mathbb{R}_+$ 
is an interval of length $\ell$
separated from the boundary by a distance $d$,
namely $(d, d + \ell)$.
We introduce also the orthogonal projector $\Theta_A$
onto the real subspace $\mathcal{H}_{\labelReal, A} \subset \mathcal{H}_{\labelReal}$, whose kernel 
depends on the endpoints of $A$ and is defined as 
\begin{equation}
\label{eq:BosonHilbertSpace.SubspaceProjectorKernel}
    \Theta_A(x, y)
  \coloneq
  \upTheta(x - d) \mult \upTheta(d + \ell - x) \mult \updelta(x - y)
  \eqend{,}
\end{equation}
in terms of the Heaviside step function $\upTheta$  
and its derivative $\updelta$ (i.e.\@ the Dirac delta).
The closed projectors $\Theta_{A, \pm1/4}$ can be defined 
in a similar way. 
In particular, their domains are 
$\mathcal{H}_{\labelReal, A}^{\pm1/4} + \mathcal{H}_{\labelReal, A}^{\perp, \pm1/4}$
(the space spanned by elements in $\mathcal{H}_{\labelReal, A}^{\pm1/4}$ and its orthogonal complement),
their images are $\mathcal{H}_{\labelReal, A}^{\pm1/4}$ 
and their kernels are 
$\mathcal{H}_{\labelReal, A}^{\perp, \pm1/4}$.
For more information on the Hilbert spaces and projectors (cutting projections), see also 
\cite{RieffelVanDaele:1977,CiolliLongoRuzzi:2019, BostelmannCadamuroDelVecchio:2022, Longo:2022}.

When the entire system is in its ground state, 
the modular Hamiltonian associated to the standard subspace $\mathcal{H}_A$ 
is an operator acting on the complex Hilbert space $\mathcal{H} \coloneq \mathcal{H}_{\labelReal}^{+1/4} \oplus \mathcal{H}_{\labelReal}^{-1/4}$,
which can be written as follows 
\cite{FiglioliniGuido:1989}
\begin{equation}
\label{eq:BosonHilbertSpace.ModularHamiltonian}
    - \log \Delta
  = - I \mult \begin{pmatrix}
      0 & M_- \\
      - M_+ & 0
    \end{pmatrix}
    \eqend{,}
\end{equation}
where the blocks $M_\pm$ are the operators 
\begin{equation}
\label{eq:BosonHilbertSpace.ModularHamiltonian.Blocks}
    M_{\pm}
  \coloneq 
  2 \mult D^{\pm\frac{1}{4}}
    \arcoth(Z)
    \mult D^{\pm\frac{1}{4}}
    \eqend{,}
\end{equation}
defined in terms of the fourth roots $D^{\pm\frac{1}{4}}$ and 
\begin{equation}
\label{eq:BosonHilbertSpace.ModularHamiltonian.ArcothArg}
    Z
  \coloneq D^{+\frac{1}{4}} \mult \Theta_{A, +\frac{1}{4}} \mult D^{-\frac{1}{4}}
    + D^{-\frac{1}{4}} \mult \Theta_{A, -\frac{1}{4}} \mult D^{+\frac{1}{4}}
    - \one
  \eqend{.}
\end{equation}
It is straightforward to observe that the operators in \eqref{eq:BosonHilbertSpace.ModularHamiltonian.Blocks} 
are related as 
\begin{equation}
\label{eq:BosonHilbertSpace.ModularHamiltonian.Blocks.Relation}
    M_+
  = 
  D^{+\frac{1}{2}} \mult M_- \mult D^{+\frac{1}{2}}
  \eqend{,}
\end{equation}
hence, we mainly focus on $M_-$ in our discussions. 
Notice that the choice of the subspace $A$ 
enters into \eqref{eq:BosonHilbertSpace.ModularHamiltonian.Blocks}  
through the projectors $\Theta_{A, \pm1/4}$.
We remind that the operator $Z$ 
is essentially self-adjoint (see Prop.~2.3 of \cite{BostelmannCadamuroMinz:2023}\footnote{In our notation, 
$D$, $I$, $Z$ and $ \Theta_{A, \pm1/4}$ 
corresponds respectively to 
$A$, $i_A$, $B$ and $\chi_{\pm1/4}$
in \cite{BostelmannCadamuroMinz:2023}. 
}).

We study the modular Hamiltonian $-\log \Delta$ 
for a free QFT in its ground state (vacuum) representation at the ``one-particle level'' with a one-particle Hilbert space $\mathcal{H}$ and a subspace $\mathcal{H}_A$ corresponding to initial data for a region $A$ of the  hypersurface at $t=0$.
In QFT, we may then consider the Weyl algebra $\mathcal{W}(\mathcal{H})$ in its Fock representation for the Hilbert space $\mathcal{H}$ and the subalgebra $\mathcal{W}(\mathcal{H}_A) \subset \mathcal{W}(\mathcal{H})$.
Since we are investigating the vacuum,
which is a pure and quasi-free state,
on the Fock space the modular operator 
is the second quantization $\Gamma(\Delta)$ 
of the one-particle modular operator $\Delta$ 
\cite{FiglioliniGuido:1989, FiglioliniGuido:1994, Longo:2022}
and also \eqref{eq:BosonHilbertSpace.ModularHamiltonian} is related to \eqref{eq:ModularHamiltonian.FockSpace} on the Fock space by second quantization.
Because of this relation, 
the Fock space is not required for our analysis 
and we focus on the one-particle structure 
with the complex Hilbert space $\mathcal{H}$.
The Hadamard property of the ground state of the model that we are considering 
has been discussed in \cite{CosteriDappiaggiJuarezAubrySingh:2025}.

\subsection{The numerical method}
\label{sec:HalfLine.NumericalMethod}

A numerical procedure has been developed 
in \cite{BostelmannCadamuroMinz:2023}
to study the modular Hamiltonian \eqref{eq:BosonHilbertSpace.ModularHamiltonian} of an interval 
for the massive scalar on the line and in its ground state. 
In the following we adapt this procedure to the case of the interval in the half line.
This numerical method is based on two sets of functions on the half line, 
denoted by $e^{(n, \Lambda)}_i(x)$ and $h_{x_k}(x)$.

\begin{figure}
  \centering
  \includegraphics{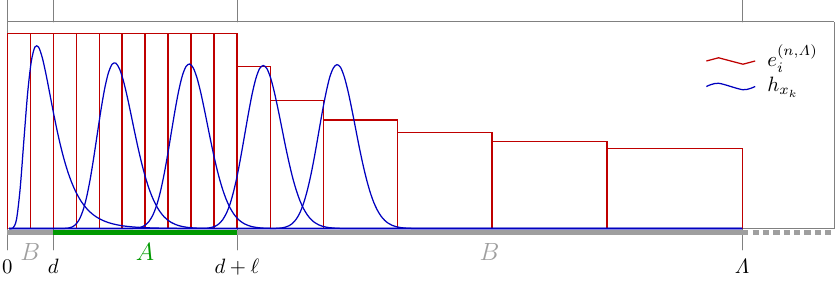}
  \caption{\label{fig:MethodFunctions} 
  Two sets of functions on the half line 
  occurring in the numerical method
  discussed in Fig.~\ref{sec:HalfLine.NumericalMethod},
  given in \eqref{eq:DiscretizationBasis} (red lines) 
  and \eqref{eq:LogGaussian} (blue lines).
  }
\end{figure}

The discretization of operators is performed through
the following $n$ piecewise constant functions
\begin{equation}
\label{eq:DiscretizationBasis}
    e^{(n, \Lambda)}_i(x)
  \coloneq \frac{\upTheta(x - a_i) \mult \upTheta(b_i - x)}{\sqrt{b_i - a_{i}}}
  \eqend{,}
\end{equation}
where $a_0 < a_1 < \dots < a_n$,
ranging from $a_0 = 0$ to $a_n = \Lambda$, 
and $b_i \coloneq a_{i + 1}$ with $0 \leqslant i \leqslant n-1$.
The number $n$ of discretization functions \eqref{eq:DiscretizationBasis},
the whole domain $[0, \Lambda]$ where they are defined
and the positions $a_i$ determine 
the degree of approximation of the corresponding numerical results. 
In the following, we refer to the discretization parameters $n$ and $\Lambda$, keeping the choice of the positions $a_i$ 
implicit. 
The parameter $\Lambda$ determines the cutoff 
for the space that we can explore in our numerical analysis. 
The functions in \eqref{eq:DiscretizationBasis}
are orthonormal with respect to the real inner product 
on the Hilbert spaces $\mathcal{H}_{\labelReal}$ of the initial data components, i.e.\@ they satisfy 
\begin{equation}
\label{eq:DiscretizationBasis.Orthonormality}
    \innerProd[\labelReal]{e^{(n, \Lambda)}_i}{e^{(n, \Lambda)}_j}
  = \updelta_{i j}
  \eqend{.}
\end{equation}
In our analysis we only consider even values for $n$, 
in order to find $a_i$'s where  
we can impose that $n/2$ functions are supported in $A$ 
and the remaining $n/2$ functions are supported in $B$.
Since we are considering the interval $A =(d, d+\ell)$, 
we choose the functions $e_i^{(n, \Lambda)}$ such that
$a_i = d$ and $a_j = d + \ell$ for some indices $i$ and $j$,
which are constrained by the condition  $j = i + \frac{n}{2}$,
in order to have the same number of discretization functions with support in $A$ and in $B$.
An explicit example of \eqref{eq:DiscretizationBasis} 
with $a_0 = 0$, $\Lambda = 4 \ell$ and $n = 16$ corresponds 
to the rectangular solid red curves is illustrated in Fig.~\ref{fig:MethodFunctions}.

The functions \eqref{eq:DiscretizationBasis} allow us to construct 
the discretized version $O^{(n, \Lambda)}$ of a generic operator $O$,
which is a $n \times n$ real matrix whose generic element is defined as follows
\begin{equation}
\label{eq:Operator.Discretization}
    O^{(n, \Lambda)}_{i j}
  \coloneq \innerProd[\labelReal]{e^{(n, \Lambda)}_i}{O \mult e^{(n, \Lambda)}_j}
  \eqend{.}
\end{equation}
For instance, the discretization of $\Theta_A(x,y)$ in \eqref{eq:BosonHilbertSpace.SubspaceProjectorKernel} 
through \eqref{eq:DiscretizationBasis}
gives $\Theta_A^{(n, \Lambda)}$,
which is the $n \times n $ diagonal matrix 
whose $i$-th element on the diagonal is equal to 
$1$ when the $i$-th interval $[a_i, b_i]$ belongs to $A$, 
while it vanishes otherwise.  
In this finite-dimensional setting of the numerical algorithm, 
the topological closure in \eqref{eq:RealHilbertSpaces.FromDomains} does not play a role, and this implies that  $\Theta_{A, \pm1/4}^{(n, \Lambda)} = \Theta_A^{(n, \Lambda)}$.

The numerical results for the modular Hamiltonian 
are smeared through some test functions.
The outcome of this smearing procedure is compared with the result obtained by integrating 
the analytic expression of the modular Hamiltonian 
with the same set of test functions.
In our analysis, this set of test functions is 
made by $\tilde{n}$ log-Gaussian functions 
defined as
\begin{equation}
\label{eq:LogGaussian}
    h_{x_k}(x)
  \coloneq 
  \frac{1}{\sqrt[4]{2 \pi \log \alpha_k}}
    \mult \frac{1}{\sqrt{x}}
    \mult \exp\Biggl(
      -\frac{ \log^2( \alpha_k\,x/x_k )}{4 \log \alpha_k}
    \Biggr)
  \eqend{,}
\qquad
    \alpha_k
  \coloneq \sqrt{1 + \frac{\sigma^2}{x_k^2}}
  \eqend{,}
\end{equation}
which are indexed by the integer $k$ such that 
$0 \leqslant k \leqslant \tilde{n} - 1$.
Since the smearing is independent from the discretization,
the parameters $\sigma$ and $x_k$ are not related 
to the parameters $a_i$, $n$ and $\Lambda$ occurring in the discretization functions \eqref{eq:DiscretizationBasis}.
The solid blue curves in Fig.~\ref{fig:MethodFunctions} provide 
an example of five log-Gaussian functions  \eqref{eq:LogGaussian}.
Since the functions \eqref{eq:LogGaussian} are employed to smear 
and probe the numerical data in order to compare them with the corresponding analytic expressions,
the positions $x_k$ can be chosen arbitrarily 
on the entire half line (and not necessarily in $[0, \Lambda]$).
Since we do not get data for $x > \Lambda$ and we are mostly interested in the results for region $A$, we consider $x_k \in [0, 2 \ell]$ and along certain curves to highlight specific features of the numerical results.
The discretization of the log-Gaussian functions \eqref{eq:LogGaussian} through \eqref{eq:DiscretizationBasis} is defined by 
\begin{equation}
\label{eq:LogGaussian.Discretization}
      h_{x_k, i}^{(n, \Lambda)}
  \coloneq 
  \innerProd[\labelReal]{e_i^{(n, \Lambda)}}{h_{x_k}}
  \eqend{,}
\end{equation}
and smearing the numerical data means to sum over these discretized log-Gaussian functions (see e.g.\@ \eqref{eq:HalfLine.ModularHamiltonian.SmearedBlock.CompactNotation} below).

Our numerical analysis has been carried out by applying 
the following sequence of steps:
\\
(a) 
Discretize $D^{-1/4}$, 
which provides the $n \times n $ matrix
$D^{-1/4,(n, \Lambda)}$ whose generic element is 
\begin{equation}
\label{eq:KernelDiscretization}
    D^{-\frac{1}{4},(n, \Lambda)}_{ij}
  \coloneq 
  \innerProd[\labelReal]{e^{(n, \Lambda)}_i}{D^{-\frac{1}{4}} \mult e^{(n, \Lambda)}_j}
  = 
  \frac{\mathcal{I}_{ij}}{\sqrt{(b_i - a_i) (b_j - a_j)}}
  \eqend{,}
\end{equation}
where 
\begin{equation}
\label{eq:KernelDiscretization.Integral}
    \mathcal{I}_{ij}
  \coloneq 
    \int_{x = a_i}^{b_i}
    \int_{y = a_j}^{b_j}
      D^{-\frac{1}{4}}(x, y)
    \id{x} \id{y}
  \eqend{.}
\end{equation}
Notice that the dependence on the boundary parameter $\eta$ 
and the mass parameter $m$ enters through the operator kernel $D^{-1/4}(x, y)$.
\\
(b) 
Invert the matrix $D^{-1/4,(n, \Lambda)}$, finding the matrix $D^{+1/4,(n, \Lambda)}$.
\\
(c) Discretize $\Theta_A$ in \eqref{eq:BosonHilbertSpace.SubspaceProjectorKernel}, getting the $n \times n $ matrix $\Theta_A^{(n, \Lambda)}$
(we remind that $\Theta_{A, \pm1/4}^{(n, \Lambda)} = \Theta_A^{(n, \Lambda)}$).
\\
(d) 
From \eqref{eq:BosonHilbertSpace.ModularHamiltonian.ArcothArg},
compute the matrix 
\begin{equation}
\label{eq:BosonHilbertSpace.ModularHamiltonian.ArcothArg.Matrix}
    Z^{(n, \Lambda)} 
  \coloneq D^{+\frac{1}{4},(n, \Lambda)}
    \mult \Theta_A^{(n, \Lambda)}
    D^{-\frac{1}{4},(n, \Lambda)}
    + D^{-\frac{1}{4},(n, \Lambda)}
    \mult \Theta_A^{(n, \Lambda)}
    D^{+\frac{1}{4},(n, \Lambda)}
    - \one
    \eqend{.}
\end{equation}
\\
(e) 
Determine the spectral decomposition of 
$Z^{(n, \Lambda)}$ 
and compute $\arcoth(Z^{(n, \Lambda)})$.
\\
(f) 
From \eqref{eq:BosonHilbertSpace.ModularHamiltonian.Blocks},
compute the matrices 
\begin{equation}
\label{eq:BosonHilbertSpace.ModularHamiltonian.Blocks.Matrix}
    M_{\pm}^{(n, \Lambda)} 
  \coloneq 2 D^{\pm \frac{1}{4},(n, \Lambda)}
    \arcoth\left( Z^{(n, \Lambda)} \right) 
    D^{\pm \frac{1}{4},(n, \Lambda)}
    \eqend{.}
\end{equation}
(g) 
Smear the numerical data for $M_{\pm}^{(n, \Lambda)}$ through 
\eqref{eq:LogGaussian.Discretization};
namely, given the set of $\tilde{n}$ positions $x_0 < x_1 < \dots < x_{\tilde{n}-1}$ in the half line,
compute the $\tilde{n} \times \tilde{n}$ matrices 
whose generic elements are
\begin{equation}
\label{eq:HalfLine.ModularHamiltonian.SmearedBlock.CompactNotation}
    \left( h_{x_k}, M_{\pm} h_{x_l} \right)^{(n, \Lambda)}
  \coloneq \sum_{i, j = 0}^{n - 1}
      h_{x_k, i}^{(n, \Lambda)}
      M_{\pm, i j}^{(n, \Lambda)}
      h_{x_l, j}^{(n, \Lambda)}
  \eqend{.}
\end{equation}

\begin{figure}[t!]
  \centering
  \includegraphics{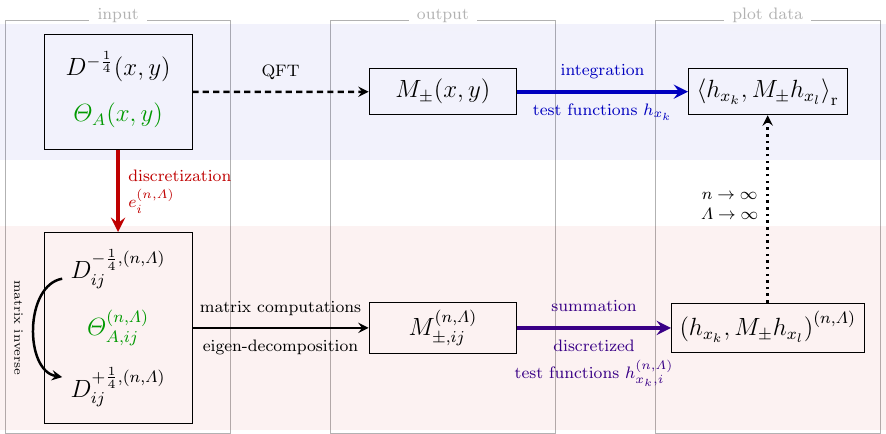}
  \caption{\label{fig:MethodScheme} 
  Scheme for the construction of 
  the kernels through the analytic approach of QFT (top) 
  and the numerical approach employed in this work (bottom).}
\end{figure}

The $\tilde{n} \times \tilde{n}$ matrices defined by \eqref{eq:HalfLine.ModularHamiltonian.SmearedBlock.CompactNotation}
are the result of our numerical procedure
and they can be compared with 
the analytic expressions coming from the corresponding computation 
in QFT of the free massive scalar. 
We stress that changing the number $\tilde{n}$ of the log-Gaussian functions does not improve the numerical result
and the quality of the numerical approximation only depends on the
discretization parameters $n$ and $\Lambda$, combined with the choice of the positions $a_i$.
In  some cases, we also smear along a real curve $\gamma(x)$, which yields $(h_{x_k}, M_{\pm} h_{\gamma(x_k)})^{(n, \Lambda)}$
(see e.g.\@ the curve provided by the red crosses in Fig.~\ref{fig:DoubleCone.Mm.Massless.SmearedMatrices}).

The scheme in Fig.~\ref{fig:MethodScheme} illustrates   
the main steps of the procedure. 
The upper part of the scheme (light blue background)
and the lower part of the scheme (light red background) 
correspond to the QFT
and to the numerical computations 
respectively.
The numerical computation, in the 
bottom part of the scheme in Fig.~\ref{fig:MethodScheme},
begins with the discretization of the analytic operator kernels (input),
which is based on the discretization functions \eqref{eq:DiscretizationBasis} in our analysis. 
The step (b)--(f) in the algorithm outlined before 
lead to construct the matrices 
$M_{\pm}^{(n, \Lambda)}$ in \eqref{eq:BosonHilbertSpace.ModularHamiltonian.Blocks.Matrix}, 
which are the output of the procedure. 
In order to compare these numerical results 
with the analytic results coming from QFT computations, 
the output matrices must be integrated against some test functions, 
which are the log-Gaussian functions \eqref{eq:LogGaussian} 
discretized as in \eqref{eq:LogGaussian.Discretization}
in our analysis (see the step (g) in the sequence).
This numerical procedure provides insightful results whenever
convergence of the numerical results \eqref{eq:HalfLine.ModularHamiltonian.SmearedBlock.CompactNotation} 
is observed as the values of the parameters $n$ and $\Lambda$ 
become very large (dotted arrow).
As for the analytic counterpart of the procedure in the 
top part of the scheme in Fig.~\ref{fig:MethodScheme},
starting from the kernels $D^{-1/4}(x,y)$ and $\Theta_A(x, y)$ (input), the kernels $M_{\pm}(x,y)$ 
are obtained through QFT computations (dashed arrow).
This step is usually very difficult.
Indeed, the outcome of this analytic computation 
is unknown for the massive scalar, even on the line.
An analytic result has been found only
for some cases where the underlying model is a CFT \cite{HislopLongo:1982, CasiniHuertaMyers:2011, WongKlichZayasVaman:2013, CardyTonni:2016}, like e.g.\@ for the massless scalar,
by heavily relying on the symmetry of the model.

The numerical procedure is implemented by 
taking an even number $n$ of discretization functions \eqref{eq:DiscretizationBasis}, with half of them 
supported in the interval $A$.
In the complement region, which extends to infinity, 
we introduce a cutoff $\Lambda$ where 
the numerical discretization terminates.
The quality of our approximation is explored 
by increasing the discretization parameters $n$ and $\Lambda$.
In particular, we consider $n \in \{64, 128, 256\}$ and 
$\Lambda \in \{4 \ell, 8 \ell, 16 \ell, 32 \ell\}$ in our analyses,
while the best results have been obtained with $n = 256$ and $\Lambda = 16 \ell$, as discussed in Sec.~\ref{sec:Results.AdjacentInterval}.
The grid points $a_i$ for the discretization are chosen such that they are equidistant in the interval $[0, d + \ell]$, while 
the grid spacing increases in a linear way in $[d + \ell, \Lambda]$
(see Fig.~\ref{fig:MethodFunctions} for an explicit example).
We use the interval width $\ell$ as the length scale and consider different values for the separation distances from the boundary
$d \in \{0, \frac{\ell}{16}, \frac{\ell}{8}, \frac{\ell}{4}, \frac{\ell}{2}\}$, where $d=0$ corresponds to $A$ adjacent to the boundary (see Sec.~\ref{sec:Results.AdjacentInterval}), 
while the other values of $d$ have been chosen 
to study the case where $A$ is separated from the boundary (see Sec.~\ref{sec:Results.SeparatedInterval}).
As for the smearing functions \eqref{eq:LogGaussian}, 
that are employed to probe both 
the analytic expressions and the numerical data
(blue and red shaded sections in the scheme, Fig.~\ref{fig:MethodScheme}, respectively),
we consider $\tilde{n} = 47$ of these functions,
such that the positions $x_k$ of the corresponding peak
are equidistantly distributed over $[0, 2 \ell]$
($0< x_0 < x_1 < \dots < x_{46} < 2 \ell$)
and they have the same width $\sigma \approx 0.096\ell$ 
throughout the analysis.

\subsection{Fractional powers of the differential operator $D$
}
\label{sec:HalfLine.FractionalD}

In the following we discuss the fractional powers $\nu$ of the differential operator $D$ introduced in \eqref{eq:StaticKleinGordon}.
The special cases relevant in our analysis 
are given by $\nu=-1$, corresponding to the Green's function, 
by the roots $\nu = \pm\frac{1}{2}$, which provide 
the complex structure \eqref{eq:ComplexStructure.Blocks}
and by the fourth roots $\nu = \pm\frac{1}{4}$,
which occur in the operators $M_\pm$ defined in \eqref{eq:BosonHilbertSpace.ModularHamiltonian.Blocks}
entering in the modular Hamiltonian 
\eqref{eq:BosonHilbertSpace.ModularHamiltonian}.
We recall that $D$ depends on the parameters $m$ and $\eta$ but not on any bipartition of the system.

The integral kernels $D^{\nu}(x, y)$ for the fractional powers $\nu$ are computed as tempered distributions.
By employing the complete set of orthogonal eigenfunctions \eqref{eq:HalfLine.MomentumIntegral.Mode},
that satisfy \eqref{eq:Helmholtz.Spectrum},
one realises that $D$ becomes a multiplicative operator in this momentum space representation,
where its generic power is given by 
$\widehat{D}^{\nu}(p) \mult g(p) = (p^2 + m^2)^\nu g(p)$.
Thus, in the configurations space, the kernel of a generic power of $D$ 
is the distribution
\begin{equation}
\label{eq:FractionalModifiedHelmholtz.HalfLine.MomentumIntegral.Definition}
    D^{\nu}(x, y)
  = \frac{1}{2 \pi}
    \int_{0}^{\infty}
      \conj{\psi}_p(x) \mult \widehat{D}^{\nu}(p) \mult \psi_p(y)
    \id{p}
  \eqend{,}
\end{equation}
which is equivalent to
\begin{eqnarray}
\label{eq:FractionalModifiedHelmholtz.HalfLine.MomentumIntegral}
    D^{\nu}(x, y)
  &=& 
  \frac{1}{2 \pi}
    \int_{0}^{\infty}
      \widehat{D}^{\nu}(p)
      \left(
        \e^{\i p (x - y)}
        + \frac{p + \i \eta}{p - \i \eta} \mult \e^{-\i p (x + y)}
      \right)
    \id{p}
    \nonumber
\\
& & {}
    + \frac{1}{2 \pi}
    \int_{-\infty}^{0}
      \widehat{D}^{\nu}(-p)
      \left(
        \e^{\i p (x - y)}
        + \frac{p + \i \eta}{p - \i \eta} \mult \e^{-\i p (x + y)}
      \right)
    \id{p}
  \eqend{.}
\end{eqnarray}
These are tempered distributions in the sense of \cite{GelfandShilov:1964}.
Since $\widehat{D}^{\nu}(-p)=\widehat{D}^{\nu}(p)$, 
the expression in \eqref{eq:FractionalModifiedHelmholtz.HalfLine.MomentumIntegral}
can be written as the following sum 
\begin{equation}
\label{eq:FractionalModifiedHelmholtz.HalfLine.MomentumIntegral.Split}
    D^{\nu}(x, y)
  = D^{\nu}_{\labelMink}(x, y)
    + D^{\nu}_{\textlabel{bdy}}(x, y)
  \eqend{,}
\end{equation}
where the two terms in the r.h.s.\@ are defined as 
\begin{equation}
\label{eq:FractionalModifiedHelmholtz.HalfLine.MinkowskianPart}
    D^{\nu}_{\labelMink}(x, y)
  \coloneq
    \frac{1}{2 \pi}
    \int_{-\infty}^{\infty}
      \widehat{D}^{\nu}(p)
      \mult \e^{\i p (x - y)}
    \id{p}
  \eqend{,}
\end{equation}
and 
\begin{equation}
\label{eq:FractionalModifiedHelmholtz.HalfLine.BoundaryPart}
    D^{\nu}_{\textlabel{bdy}}(x, y)
  \coloneq
    \frac{1}{2 \pi}
    \int_{-\infty}^{\infty}
      \widehat{D}^{\nu}(p)
      \mult \frac{p + \i \eta}{p - \i \eta} \mult \e^{-\i p (x + y)}
    \id{p}
  \eqend{.}
\end{equation}
Notice that \eqref{eq:FractionalModifiedHelmholtz.HalfLine.MinkowskianPart} is the definition of the same operator on the line, which is independent on the 
parameter $\eta$ associated to the boundary conditions;
while \eqref{eq:FractionalModifiedHelmholtz.HalfLine.BoundaryPart} encodes the whole dependence 
of $D^{\nu}(x, y)$ on $\eta$.
The integrand of \eqref{eq:FractionalModifiedHelmholtz.HalfLine.BoundaryPart} contains the unitary phase factor $(p + \i \eta) / (p - \i \eta)$, 
which is a one-dimensional scattering matrix
\cite{BellazziniMintchevSorba:2010}, hence one may give the Fourier factor $\e^{\i p y}$ the physical interpretation of an outgoing wave that picked up a phase factor when reflected at the boundary.

The case $\nu=-1$, i.e.\@ $\widehat{D}^{-1}(p) = (p^2 + m^2)^{-1}$,
is well known because it 
corresponds to the Green function.
In Appendix~\ref{appx:GreensFunctions}
the kernels of the Green functions $D^{-1}(x, y)$ for the massless and massive cases with different boundary parameters $\eta$ are discussed.

The operators $D^{\pm1/2}$ 
are related to the two-point function of the scalar fields,
as discussed in general e.g.\@ in \cite{Froeb:2025}.
The specific case that we are exploring is discussed 
in Appendix~\ref{appx:ComplexStructure.TwoPointFunctions},
where we include a comparison with the standard form of the 
two-point functions \cite{LiguoriMitchev:1998,MintchevPilo:2001}.
Our numerical approach is based on the modular Hamiltonian given by \eqref{eq:BosonHilbertSpace.ModularHamiltonian} and  \eqref{eq:BosonHilbertSpace.ModularHamiltonian.Blocks}.
Although $D^{\pm 1/4}(x, y)$ occur in \eqref{eq:BosonHilbertSpace.ModularHamiltonian.Blocks},
the numerical procedure requires only $D^{-1/4}(x, y)$
because the contribution from $D^{+1/4}(x, y)$ is evaluated numerically  as a matrix inverse after discretization
(step (b) in the sequence described in Sec.~\ref{sec:HalfLine.NumericalMethod}).
In the following we report the explicit expressions of the kernel 
$D^{-1/4}(x, y)$ that are needed, referring to Appendix~\ref{appx:InverseFourthRoots} for their derivations.

For the massless scalar with Robin b.c.\@, we find 
\begin{equation}
\label{eq:InverseFourthRootLaplace}
    D^{-\frac{1}{4}}(x, y)
  = \frac{1}{\sqrt{2 \pi \abs{x - y}}}
    + \frac{1}{\sqrt{2 \pi (x + y)}}
    - \sqrt{2 \eta} \mult \e^{\eta (x + y)} \mult
    \erfc\sqrt{\eta (x + y)}
  \eqend{,}
\end{equation}
where $\erfc(x)$ is the complementary error function.
In the cases of Neumann b.c.\@ and Dirichlet b.c.\@, 
corresponding to $\eta = 0$ and $\eta \to \infty$
respectively, 
the kernel \eqref{eq:InverseFourthRootLaplace} 
becomes 
\begin{equation}
\label{eq:InverseFourthRootLaplace.SpecialBoundary}
    D^{-\frac{1}{4}}(x, y)
  = \frac{1}{\sqrt{2 \pi \abs{x - y}}}
    \pm 
    \frac{1}{\sqrt{2 \pi (x + y)}}
  \eqend{,}
\end{equation}
where the sign $+$ and $-$ is associated to the Neumann b.c.\@ and Dirichlet b.c.\@, respectively.

The expansions of the kernel 
\eqref{eq:InverseFourthRootLaplace} as $\eta \to 0^+$
and $\eta \to \infty$ are given respectively by 
\begin{equation}
\label{eq:InverseFourthRootLaplace.RobinBoundary.NeumannExpansion}
    D^{-\frac{1}{4}}(x, y)
  = \frac{1}{\sqrt{2 \pi \abs{x - y}}}
    + \frac{1}{\sqrt{2 \pi (x + y)}}
    - \sqrt{2 \eta}
    + 2 \eta \mult \sqrt{\frac{2 (x + y)}{\pi}}
    + \Ord\Bigl( \eta^{\frac{3}{2}} \Bigr)
    \eqend{,}
\end{equation}
and 
\begin{equation}
\label{eq:InverseFourthRootLaplace.RobinBoundary.DirichletExpansion}
    D^{-\frac{1}{4}}(x, y)
  = \frac{1}{\sqrt{2 \pi \abs{x - y}}}
    - \frac{1}{\sqrt{2 \pi (x + y)}}
    + \frac{1}{\sqrt{2\pi} \mult \eta \mult (x+y)^{3/2}}
      + \Ord\Bigl( \eta^{-2} \Bigr)
  \eqend{,}
\end{equation}
which has been obtained by using \citeDLMFeq{7.6}{2}.
An important feature of \eqref{eq:InverseFourthRootLaplace.RobinBoundary.NeumannExpansion},
which does not occur in \eqref{eq:InverseFourthRootLaplace.RobinBoundary.DirichletExpansion}, 
is that the first correction at order $\mathcal{O}(\sqrt{\eta})$
is constant, i.e.\@ independent of the arguments of the kernel $x$ and $y$, and its derivative w.r.t.\@ $\eta$ diverges as $\eta \to 0^+$.
This corresponds to the fact that the theory with Neumann b.c.\@ has a zero mode, which is not present in the case of generic Robin b.c.\@, hence we also expect a high sensitivity of the modular Hamiltonian for very small values of $\eta$, especially in the limit $\eta \to 0^+$.

For the massive regime and 
when Robin b.c.\@ are imposed, we compute the kernel in Appendix~\ref{appx:InverseFourthRoots.Massive} and find
\begin{equation}
\label{eq:InverseFourthRootHelmholtz.RobinBoundary}
    D^{-\frac{1}{4}}(x, y)
  = \frac{\sqrt[4]{2 m} }{\sqrt{\pi} \mult \upGamma(1/4)}
    \left(
      \frac{
        \BesselK{\frac{1}{4}}\Bigl( m \abs{x - y} \Bigr)
      }{ \sqrt[4]{\abs{x - y}}}
      + \frac{
        \BesselK{\frac{1}{4}}\Bigl( m (x + y) \Bigr)
        - 2 B_{-\frac{1}{4}}(\eta, m; x + y)
      }{ \sqrt[4]{x + y} }
    \right)
  \eqend{,}
\end{equation}
where $B_\nu (\eta, m; z)$ is a generalization of Basset's integral discussed in Appendix~\ref{appx:GeneralizedBassetIntegral}.
An important consistency check for 
\eqref{eq:InverseFourthRootHelmholtz.RobinBoundary} is that 
the kernel \eqref{eq:InverseFourthRootLaplace} is recovered 
in the massless limit.
In the limit of either $\eta \to 0^+$ or $\eta \to \infty$,
which correspond to the Neumann and Dirichlet b.c.\@ respectively,
the kernel \eqref{eq:InverseFourthRootHelmholtz.RobinBoundary} simplifies to
\begin{equation}
\label{eq:InverseFourthRootHelmholtz.SpecialBoundary}
    D^{-\frac{1}{4}}(x, y)
  = \frac{\sqrt[4]{2 m}}{\sqrt{\pi} \mult \upGamma(1/4)}
    \left(
      \frac{
        \BesselK{\frac{1}{4}}\Bigl( m \abs{x - y} \Bigr)
      }{ \sqrt[4]{\abs{x - y}}}
      \pm \frac{
        \BesselK{\frac{1}{4}}\Bigl( m (x + y) \Bigr)
      }{ \sqrt[4]{x + y} }
    \right)
  \eqend{,}
\end{equation}
where the signs $+$ and $-$ stand for
the Neumann and Dirichlet b.c.\@ respectively.
In Appendix~\ref{appx:InverseFourthRoots.Massive} 
we show that \eqref{eq:InverseFourthRootHelmholtz.SpecialBoundary}
also follows from a direct computation that does not employ \eqref{eq:InverseFourthRootHelmholtz.RobinBoundary},
providing another consistency check for \eqref{eq:InverseFourthRootHelmholtz.SpecialBoundary}.

For a generic finite and non-vanishing value $\eta \neq m$, 
we only have an integral representation for 
$B_{-\frac{1}{4}}(\eta, m, x + y)$ and 
we would have to evaluate this integral numerically. 
Since such a numeric evaluation is too slow, considering the high floating point precision required for our numerical algorithm, 
we study only the case $\eta = m$.
Indeed, when $\eta = m$ the function $B_{-1/4}$ occurring in \eqref{eq:InverseFourthRootHelmholtz.RobinBoundary} 
can be written as follows
\begin{equation}
\label{eq:InverseFourthRootHelmholtz.RobinMassBoundary.BoundaryInteral}
    B_{-\frac{1}{4}}(m, m, x + y)
  = 2 m \mult (x + y)
    \mult \Bigl[
      \BesselK{\frac{3}{4}}\bigl( m (x + y) \bigr)
      - \BesselK{-\frac{1}{4}}\bigl( m (x + y) \bigr)
    \Bigr]
  \eqend{,}
\end{equation}
which is \eqref{eq:BoundaryIntegral.ArgumentCoincidence} 
for $\nu = -1/4$ and that allows us to write \eqref{eq:InverseFourthRootHelmholtz.RobinBoundary} 
through Bessel functions, making this special case manageable 
in our numerical procedure.

\subsection{Discretization of $D^{-1/4}$}
\label{sec:HalfLine.Discretization}

In the sequence of operations determining our numerical approach and discussed in Sec.~\ref{sec:HalfLine.NumericalMethod}
(see Fig.~\ref{sec:HalfLine.NumericalMethod}),
the first step consists in discretizing the operator $D^{-1/4}(x,y)$, 
which leads to the matrix whose elements are given by 
\eqref{eq:KernelDiscretization} and \eqref{eq:KernelDiscretization.Integral}.
Since $D^{-1/4}(x, y)$ 
is symmetric under the exchange of its arguments $x$ and $y$, 
the corresponding matrix element \eqref{eq:KernelDiscretization} 
is symmetric under the exchange $i \leftrightarrow j$;
hence it is enough to compute \eqref{eq:KernelDiscretization}
for $i \leqslant j$.
In the following we provide the  expressions for the implementation of this discretization in the numerical procedure.

In Appendix~\ref{appx:Discretization}, the double integral \eqref{eq:KernelDiscretization.Integral} is split into
\begin{equation}
\label{eq:KernelDiscretization.Integral.Split}
    \mathcal{I}_{ij}^{-}
  \coloneq \int_{a_i}^{b_i} \int_{a_j}^{b_j}
      D^{-\frac{1}{4}}_{\labelMink}(x, y)
    \id{x} \id{y}
  \eqend{,}
  \qquad
    \mathcal{I}_{ij}^{+}
  \coloneq \int_{a_i}^{b_i} \int_{a_j}^{b_j}
      D^{-\frac{1}{4}}_{\textlabel{bdy}}(x, y)
    \id{x} \id{y}
  \eqend{,}
\end{equation}
which depends on $x-y$ and $x+y$ respectively, as remarked above. 
This dependence leads us to employ $\pm$ for the $\mathcal{I}_{ij}^{\pm}$ in \eqref{eq:KernelDiscretization.Integral.Split}; 
hence we introduce $s_{\textlabel{bdy}}$ in this subsection to denote 
the Neumann b.c.\@ with $s_{\textlabel{bdy}} = 1$
and the Dirichlet b.c.\@ with $s_{\textlabel{bdy}} = -1$. 
In Appendix~\ref{appx:Discretization}, we find that  \eqref{eq:KernelDiscretization.Integral.Split} can be written as
\begin{equation}
\label{eq:KernelDiscretization.ElementIntegrals}
    \mathcal{I}^{\mp}_{i j}
  = \pm 
    \big[ F^{\mp}(a_j \mp b_i)
    - F^{\mp}(a_j \mp a_i)
    - F^{\mp}(b_j \mp b_i)
    + F^{\mp}(b_j \mp a_i)
    \big]
  \eqend{,}
\end{equation}
for all $i \leqslant j$ in the case of $\mathcal{I}^{+}_{i j}$,
while for the diagonal of $\mathcal{I}^{-}_{i j}$ we have 
\begin{equation}
\label{eq:KernelDiscretization.ElementIntegrals.ConvolutionDiagonal}
    \mathcal{I}^-_{i i}
  = 2 \big[ F^-(b_i - a_i) - F^-(0) \big]
  \eqend{,}
\end{equation}
in terms of 
\begin{equation}
\label{eq:KernelDiscretization.Antiderivatives}
        F^{\mp}(x)
  \coloneq
    x \mult F^{\mp}_0(x) - F^{\mp}_1(x)
    \eqend{,}
  \qquad
    F^{\mp}_k(x)
  \coloneq \int_{0}^{x} t^k f^{\mp}(\mp t) \id{t}
    \eqend{,}
    \qquad
    k \in \big\{0,1\big\}
  \eqend{,}
\end{equation}
where $f^{-}(x - y)$ stands for the convolution kernel $D^{-1/4}_{\labelMink}(x, y)$ and $f^{+}(x + y)$ for $D^{-1/4}_{\textlabel{bdy}}(x, y)$.

In the massless regime, 
the discretization of the term $D^{-1/4}_{\labelMink}(x, y)$ of the integral kernel \eqref{eq:InverseFourthRootLaplace} 
with \eqref{eq:KernelDiscretization.Integral.Split} 
is given by
\begin{equation}
\label{eq:KernelDiscretization.Antiderivatives.Massless.Minkowskian}
    f^-(-t)
  = \frac{1}{\sqrt{2 \pi t}}
  \eqend{,}
  \qquad
    F^-_k(x)
  = \sqrt{\frac{2}{\pi}}
    \mult \frac{x^{k + \frac{1}{2}}}{2 k + 1}
  \eqend{.}
\end{equation}
In the case of the  Neumann b.c.\@ and Dirichlet b.c.\@
(corresponding to $s_{\textlabel{bdy}} = 1$ and $s_{\textlabel{bdy}} = -1$ respectively), 
for the discretization of the second term of the kernel \eqref{eq:InverseFourthRootLaplace.SpecialBoundary}
we find 
\begin{equation}
\label{eq:KernelDiscretization.Antiderivatives.Massless.SpecialBoundary}
    f^+(t)
  = \frac{s_{\textlabel{bdy}}}{\sqrt{2 \pi t}}
  \eqend{,}
  \qquad
    F^+_k(x)
  = s_{\textlabel{bdy}} \sqrt{\frac{2}{\pi}}
    \mult \frac{x^{k + \frac{1}{2}}}{2 k + 1}
  \eqend{.}
\end{equation}
Similarly, for the generic case of Robin b.c.\@, we have that the last two terms in the r.h.s.\@ of \eqref{eq:InverseFourthRootLaplace} are implemented with
\begin{eqnarray}
    f^+(t)
  &=& \frac{1}{\sqrt{2 \pi t}}
    - \sqrt{2 \eta} \mult \e^{\eta t} 
    \erfc\bigl( \sqrt{\eta t}\, \bigr)
  \eqend{,}
\\
    F^+_0(x)
  &=& - \sqrt{\frac{2 x}{\pi}}
    - \sqrt{\frac{2}{\eta}} \mult \e^{\eta x} 
    \erfc\bigl( \sqrt{\eta t}\, \bigr)
    + \sqrt{\frac{2}{\eta}}
  \eqend{,}
\\
\label{eq:KernelDiscretization.Antiderivatives.Massless.RobinBoundary}
    F^+_1(x)
  &=& \sqrt{\frac{2 x}{\pi}}
    \left( \frac{2}{\eta} - \frac{x}{3} \right)
    - \sqrt{\frac{2}{\eta}}
    \left( x - \frac{1}{\eta} \right) \e^{\eta x} 
    \erfc\bigl( \sqrt{\eta t}\, \bigr)
    - \sqrt{\frac{2}{\eta^3}}
  \eqend{.}
\end{eqnarray}

In the massive regime, the operator kernel $D^{-1/4}_{\labelMink}(x, y)$ 
as the first term in the r.h.s.\@ of \eqref{eq:InverseFourthRootHelmholtz.RobinBoundary} 
has an implementation given by
\begin{eqnarray}
    f^-(-t)
  &=& \frac{1}{\sqrt{\pi} \mult \upGamma(1/4)}
    \mult \sqrt[4]{\frac{2 m}{t}}
    \mult \BesselK{-\frac{1}{4}}(m t)
  \eqend{,}
\\
\label{eq:KernelDiscretization.Antiderivatives.Massive.Minkowskian}
      F^-_k(x)
  &=& x^k
    \Biggl(
      \frac{\sqrt{2x}}{\sqrt{\pi} \mult (2 k + 1)} \mult F^{1,3}_k(x)
      - \frac{\sqrt{m} \mult \upGamma(3/4)}{\sqrt{\pi} \mult \upGamma(1/4)}
      \mult \frac{2 x }{k + 1} \mult F^{2,5}_k(x)
    \Biggr)
  \eqend{.}
\end{eqnarray}
Here we use 
\begin{equation}
\label{eq:HypergeometricFunction.ShortNotation}
    F^{a,b}_k(x)
  \coloneq \hypF12\left(
      \frac{2 k + a}{4}; \frac{b}{4}, \frac{2 k + a + 4}{4}; \frac{m^2 x^2}{4}
    \right)
  \eqend{,}
\end{equation}
as a short notation for the generalized hypergeometric function $\hypF12$ (see e.g.\@ \citeDLMFsec{16.2}),
which should not be confused with the standard hypergeometric function $\hypF21$ (see e.g.\@ \citeDLMFsec{15.2}).
When either Neumann b.c.\@ ($s_{\textlabel{bdy}} = 1$) or Dirichlet b.c.\@ ($s_{\textlabel{bdy}} = -1$) hold, 
for the contribution given by $D^{-1/4}_{\textlabel{bdy}}$ we implement
\begin{eqnarray}
    f^+(t)
  &=& \frac{s_{\textlabel{bdy}}}{\sqrt{\pi} \mult \upGamma(1/4)}
    \mult \sqrt[4]{\frac{2 m}{t}}
    \mult \BesselK{-\frac{1}{4}}(m t)
  \eqend{,}
\\
\label{eq:KernelDiscretization.Antiderivatives.Massive.SpecialBoundary}
      F^+_k(x)
  &=& s_{\textlabel{bdy}} \mult x^k
    \Biggl(
      \frac{\sqrt{2x}}{\sqrt{\pi}\,(2 k + 1)} \mult F^{1,3}_k(x)
      - \frac{\sqrt{m} \mult \upGamma(3/4)}{\sqrt{\pi} \mult \upGamma(1/4)}
      \mult \frac{2 x }{k + 1} \mult F^{2,5}_k(x)
    \Biggr)
  \eqend{.}
\end{eqnarray}
For Robin b.c.\@, 
we consider only the special case given by $\eta = m$. 
In particular, the second term in the r.h.s.\@ 
of \eqref{eq:InverseFourthRootHelmholtz.RobinBoundary} 
is implemented with
\begin{eqnarray}
\label{eq:KernelDiscretization.Antiderivatives.Massive.RobinMassBoundary}
    f^+(t)
  &=& \frac{\sqrt[4]{2} \mult \sqrt{m}}{\sqrt{\pi} \mult \upGamma(1/4)}
    \mult \frac{
      (1 + 4 m t) \mult \BesselK{\frac{1}{4}}(m t)
      - 4 m t \mult \BesselK{\frac{3}{4}}(m t)
    }{\sqrt[4]{m t}}
  \eqend{,}
\\
    F^{\pm}_k(x)
  &=& x^k \bigggl(
      \sqrt{\frac{2 x}{\pi}}
      \mult \Biggl[
        \frac{1}{2 k + 1} \mult F^{1,3}_k(x)
        + \frac{4 m x }{2 k + 3} \mult F^{3,3}_k(x)
        + \frac{8 m^2 x^2}{6 k + 15} \mult F^{5,7}_k(x)
      \Biggr]
  \nonumber\\
  && \qquad{}
      - \frac{\sqrt{m} \mult \upGamma(3/4)}{\sqrt{\pi} \mult \upGamma(1/4)}
      \mult 2 x
      \mult \Biggl[
        \frac{1}{k + 1} \mult F^{2,5}_k(x)
        + \frac{2 }{k + 1} \mult F^{2,1}_k(x)
        + \frac{4 m x }{k + 2} \mult F^{4,5}_k(x)
      \Biggr]
    \bigggr)
  \eqend{.}
  \qquad
\end{eqnarray}
These functions discretize $D^{-1/4}$, 
allowing us to compute the numerical approximation for $M_-$, 
as described in the scheme shown in Fig.~\ref{fig:MethodScheme}.

\section{Interval adjacent to the boundary}
\label{sec:Results.AdjacentInterval}

In this section we discuss the case where $A$ is adjacent to the boundary, i.e.\@ when $d=0$.
The massless and massive regimes are considered in
Sec.~\ref{sec:Results.AdjacentInterval.Massless} and Sec.~\ref{sec:Results.AdjacentInterval.Massive}, respectively.

\subsection{Massless regime}
\label{sec:Results.AdjacentInterval.Massless}

The massless scalar field on the half line is a BCFT
when either Dirichlet or Neumann b.c.\@ are imposed. 
In these two cases, the modular Hamiltonian of an interval adjacent to the boundary (i.e.\@ for $d=0$)
can be found from the well known 
result of Bisognano and Wichmann 
\cite{BisognanoWichmann:1975,BisognanoWichmann:1976},
through a suitable conformal mapping \cite{CardyTonni:2016} and it 
corresponds to the following kernel 
\begin{equation}
\label{eq:LeftWedge.BCFTReference}
    M_-^{\textlabel{BCFT}}(x, y)
  \coloneq 2 \pi \mult \frac{\ell^2 - x^2}{2\ell} \mult \updelta(x - y)
  \eqend{.}
\end{equation}
This analytic result provides an important benchmark for our numerical procedure.
Even though we do not show the numerical results for the component 
$M_+$ in the modular Hamiltonian \eqref{eq:BosonHilbertSpace.ModularHamiltonian}, 
in Appendix~\ref{appx:ModularHamiltonian.BCFT.Mp}
we compute $M_+^{\textlabel{BCFT}}$ 
corresponding to the reference \eqref{eq:LeftWedge.BCFTReference}.
The BCFT reference $M_{\mp}^{\textlabel{BCFT}}$ is formally identical to the corresponding component in the modular Hamiltonian of the double cone 
for the massless scalar in two-dimensional Minkowski space \cite{HislopLongo:1982},
which becomes obvious in the formulation of
\cite{LongoMorsella:2023}\footnote{In our notation, $M_-$ and $M_+$ 
correspond respectively to $M$ and $-L$ in \cite{LongoMorsella:2023}.}.
Instead, when Robin b.c.\@ hold 
and $\eta$ takes a non-zero finite value, 
the model is not a BCFT and therefore 
the corresponding methods cannot be applied. 
The analytic expression for the modular Hamiltonian of an interval adjacent to the boundary in the case of Robin b.c.\@ 
is not available in the literature, to our best knowledge.

Another important reference in our analysis is 
given by the linear kernel 
\begin{equation}
\label{eq:LeftWedge.LinearReference}
    M_-^{\textlabel{LW}}(x, y)
  = 2 \pi (\ell - x) \mult \updelta(x - y)
  \eqend{,}
\end{equation}
suggested by the case of Bisognano and Wichmann,
which corresponds to the modular Hamiltonian of the left wedge (LW) in Minkowski space having its vertex at $x = \ell$.
Indeed, \eqref{eq:LeftWedge.LinearReference} 
is an insightful approximation of the kernels for the interval $A$ adjacent to the boundary
in the neighbourhood of the entangling point.

In our analysis we compare the results obtained 
from the BCFT kernel \eqref{eq:LeftWedge.BCFTReference}
and from \eqref{eq:LeftWedge.LinearReference}
against the ones corresponding to the kernels that cannot be studied through the BCFT methods.
The QFT expression is expected to correspond to the numerical approximation in the limit $n \to \infty$ and $\Lambda \to \infty$ only in the weak sense, i.e.\@ when smeared against test functions, which are 
the log-Gaussians \eqref{eq:LogGaussian} in our analysis
(see the dotted arrow in Fig.~\ref{fig:MethodScheme}).
Hence, these two reference kernels are integrated
against the log-Gaussian  functions \eqref{eq:LogGaussian}.
We choose $\tilde{n} = 47$ distinct positions $x_k$ and $\sigma \approx 0.096\ell$; hence we have $\tilde{n} \times \tilde{n}$ matrices with elements $\innerProd[\labelReal]{h_{x_k}}{M_-^{\textlabel{BCFT}} h_{x_l}}$ and $\innerProd[\labelReal]{h_{x_k}}{M_-^{\textlabel{LW}} h_{x_l}}$
(see Appendix~\ref{appx:Discretization} for a detailed discussion).
The elements of these matrices along the diagonal  and the anti-diagonal (where $l = k$ and $l = 23 - k$, respectively)
from both reference kernels are plotted against 
the corresponding numerical data points 
of the modular Hamiltonian.
Along the diagonal, for the smeared kernels obtained from \eqref{eq:LeftWedge.BCFTReference} and \eqref{eq:LeftWedge.LinearReference}, we find respectively
\begin{equation}
\label{eq:LeftWedge.References.Smearing.Diagonal}
        \innerProd{h_{x_i}}{M_-^{\textlabel{BCFT}} h_{x_i}}
  = 
  2\pi\,
  \frac{\ell^2 - x_i^2}{2\ell} 
  \eqend{,}
  \qquad
      \innerProd{h_{x_i}}{M_-^{\textlabel{LW}} h_{x_i}}
  = 
  2 \pi (\ell - x_i) + \Ord(\sigma^2)
  \eqend{.}
\end{equation}
Thus, the smearing procedure 
of local BCFT kernel \eqref{eq:LeftWedge.BCFTReference}
along the diagonal provides exactly 
the function multiplying the Dirac delta,
and for the left wedge kernel \eqref{eq:LeftWedge.LinearReference}
the result is approximately linear 
(since $\sigma^2 \approx 0.009 \ell^2$).
As for the off-diagonal element, we refer to 
\eqref{eq:LeftWedge.BCFTReference.Smearing} and \eqref{eq:LeftWedge.LinearReference.Smearing}, respectively.
We remark again that the positions $x_k$ 
and the number $\tilde{n}$ of log-Gaussian functions 
determine the data presentation and comparison, 
but do not influence the numerical approximation, 
which improves instead with increasing discretization parameters $n$ and $\Lambda$ 
(see the dotted arrow in Fig.~\ref{fig:MethodScheme}).

\begin{figure}[t!]
  \centering
  \includegraphics{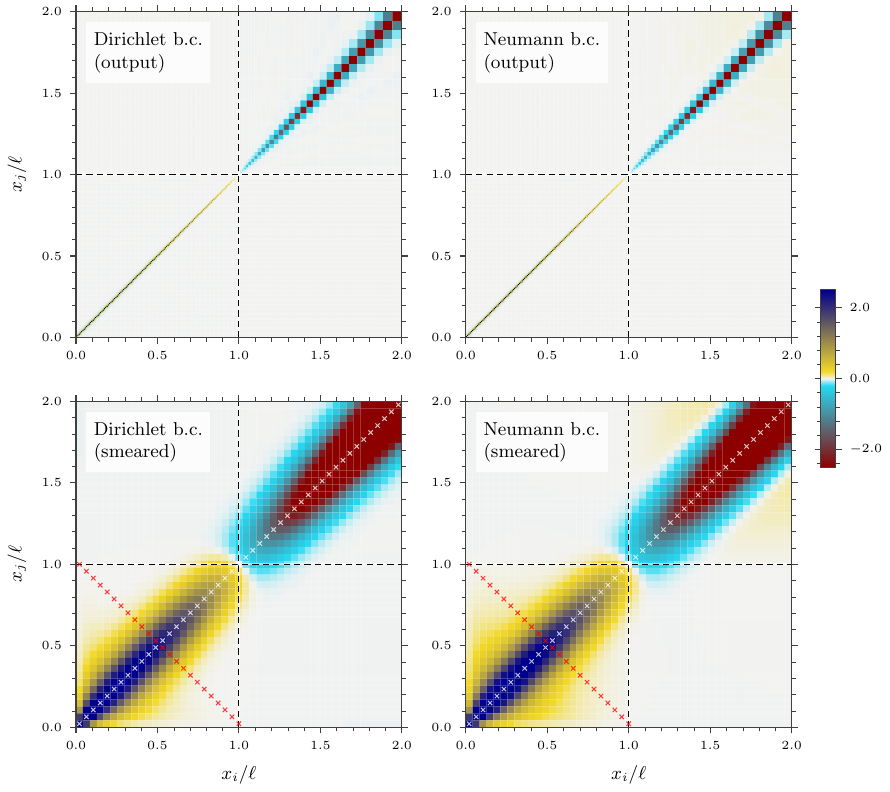}
  \caption{
  \label{fig:LeftWedge.Mm.Discretization.Matrices} 
  Interval adjacent to the boundary, massless regime 
  for either Dirichlet b.c.\@ (left) or Neumann b.c.\@ (right). 
  Output of the numeric algorithm obtained with discretization parameters 
  $n = 256$ and $\Lambda = 16 \ell$.
  The top (bottom) panels report the data before (after) the smearing procedure 
  (see \eqref{eq:HalfLine.ModularHamiltonian.SmearedBlock.CompactNotation}) through the log-Gaussian test functions \eqref{eq:LogGaussian.Discretization}.
  }
\end{figure}

In the top panels of Fig.~\ref{fig:LeftWedge.Mm.Discretization.Matrices},
we show two examples of the numerical 
$n \times n$ output matrices $M_-^{(256, 16 \ell)} / \ell$,
either for the Dirichlet b.c.\@ (left panel) 
or the Neumann b.c.\@ (right panel).
Then, such matrices are smeared by summing with the discretized log-Gaussian functions \eqref{eq:LogGaussian.Discretization}, which yields the smeared matrices with elements $(h_{x_k}, M_- h_{x_l})^{(256, 16 \ell)}$ as defined in \eqref{eq:HalfLine.ModularHamiltonian.SmearedBlock.CompactNotation}.
The smeared results are displayed 
in the corresponding bottom panels. 
In all the panels of Fig.~\ref{fig:LeftWedge.Mm.Discretization.Matrices},
the data scales have been cut in order to highlight the non-vanishing elements that do not belong to the main diagonal.
Furthermore, we remark that also the results obtained in the complementary domain $B$ have been reported, where negative matrix elements occur. 
The main visible difference between the two cases considered in this figure is provided by the small non-vanishing positive  
data away from the diagonal in the bottom right panel, 
which do not occur in the bottom left panel.
However, only in this case we consider the non-diagonal terms to be an artefact of the discretization, as discussed in the following.

\begin{figure}[t!]
  \centering
\includegraphics{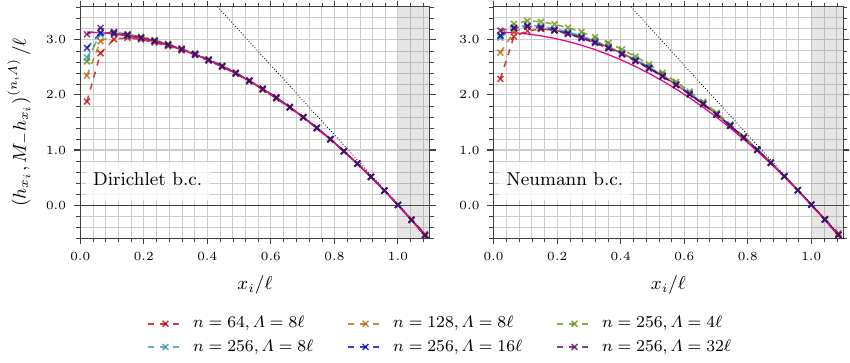}
  \caption{
  \label{fig:LeftWedge.Mm.Discretization.Diagonal} 
    Interval adjacent to the boundary, massless regime 
  for either Dirichlet b.c.\@ (left) or Neumann b.c.\@ (right). 
  Smeared expression 
  from \eqref{eq:HalfLine.ModularHamiltonian.SmearedBlock.CompactNotation} along the diagonal (white crosses in Fig.~\ref{fig:LeftWedge.Mm.Discretization.Matrices}), 
  for different values of discretization parameters $n$ and $\Lambda$.}
\end{figure}

In Fig.~\ref{fig:LeftWedge.Mm.Discretization.Diagonal},
we report the numerical results for 
\eqref{eq:HalfLine.ModularHamiltonian.SmearedBlock.CompactNotation}
along the diagonal,
corresponding to the white crosses in 
Fig.~\ref{fig:LeftWedge.Mm.Discretization.Matrices},
for either Dirichlet b.c.\@ or Neumann b.c.\@ in the left and right panel, respectively. 
Different colours are associated to different choices for the pair $(n, \Lambda)$ 
characterising the degree of approximation 
of the QFT result (magenta curve), 
which corresponds to the limit 
where $n \to \infty$ and $\Lambda \to \infty$.
Instead, increasing the number $\tilde{n}$ of log-Gaussian functions for the smearing does not improve further the approximation,
but only increases the number of data points in the plots.
The solid magenta curve and the dotted black straight line,
which correspond to the BCFT kernel 
\eqref{eq:LeftWedge.BCFTReference}
and to the linear kernel \eqref{eq:LeftWedge.LinearReference}
respectively, 
are also obtained through the white crosses in 
Fig.~\ref{fig:LeftWedge.Mm.Discretization.Matrices}
and computed in 
\eqref{eq:LeftWedge.References.Smearing.Diagonal}.
To highlight the numeric approximations in all plots, 
we use cross marks only for the numeric data and plot the analytic references without marks.
In both panels of Fig.~\ref{fig:LeftWedge.Mm.Discretization.Diagonal} 
the agreement with the curve 
coming from the BCFT kernel \eqref{eq:LeftWedge.BCFTReference}
close to the boundary improves as $n$ and $\Lambda$ increase.
The numerical algorithm requires high floating point 
precision (implemented with \cite{mpmath}) that increases 
with $n$ in a non-linear way
and our highest resolution corresponds to $n = 256$.
Since we have observed additional numerical artifacts
going from $\Lambda = 16 \ell$ to $\Lambda = 32 \ell$ 
with fixed resolution $n = 256$
in the case of Dirichlet b.c.\@, 
we mainly consider
the values $n = 256$ and $\Lambda = 16 \ell$ 
in our numerical analyses.

In the left panel of Fig.~\ref{fig:LeftWedge.Mm.Discretization.Diagonal},
we observe a remarkable agreement between the numerical results
and the solid magenta curve for the BCFT kernel \eqref{eq:LeftWedge.BCFTReference}
in its smeared form as given in the first expression in \eqref{eq:LeftWedge.References.Smearing.Diagonal}.
A discrepancy occurs close to the boundary, 
but its amplitude becomes smaller 
when the approximation improves, namely 
when both $n$ and $\Lambda$ increase, 
as expected.
This provides an important consistency check 
for the validity of our numerical approach. 
In the right panel of Fig.~\ref{fig:LeftWedge.Mm.Discretization.Diagonal},
where the data points corresponding to the Neumann b.c.\@
are displayed, beside the discrepancy 
with the solid magenta curve close to the boundary 
already highlighted in the left panel, 
a slight deviation 
between our approximations and the BCFT data points 
occurs over a wider domain adjacent to the boundary 
with respect to the Dirichlet b.c.\@ case,
while a much better match is observed in the domain 
close to the entangling point at $x = \ell$.
We think that this discrepancy is due to the fact 
that our best values for the parameters $n$ and $\Lambda$ 
are not high enough to reach the convergence to the BCFT curve in the first formula in \eqref{eq:LeftWedge.References.Smearing.Diagonal},
since the data curves still improve very slowly 
as $n$ and $\Lambda$ increase.

\begin{figure}[t!]
  \centering
  \includegraphics{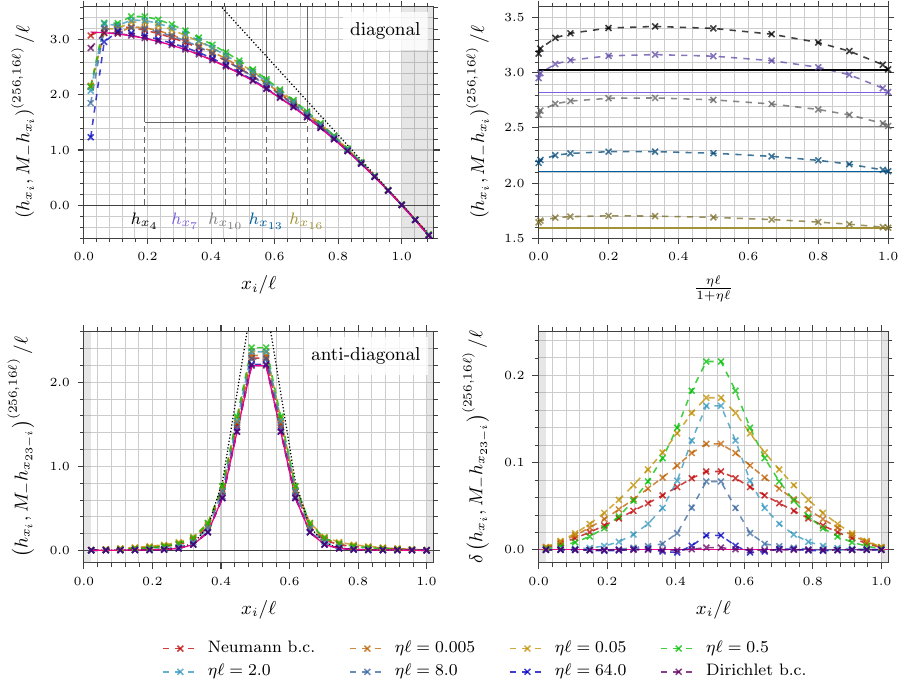}
  \caption{\label{fig:LeftWedge.Mm.Boundaries.Diagonals} 
  Interval adjacent to the boundary: 
  Massless regime when Robin b.c.\@ are imposed.
  Numerical results for 
  \eqref{eq:HalfLine.ModularHamiltonian.SmearedBlock.CompactNotation}
  along the diagonal (top) and anti-diagonal (bottom), 
  which correspond to the white and red crosses in Fig.~\ref{fig:LeftWedge.Mm.Discretization.Matrices}, 
  respectively.
  }
\end{figure}

\begin{figure}[t!]
  \centering
  \includegraphics{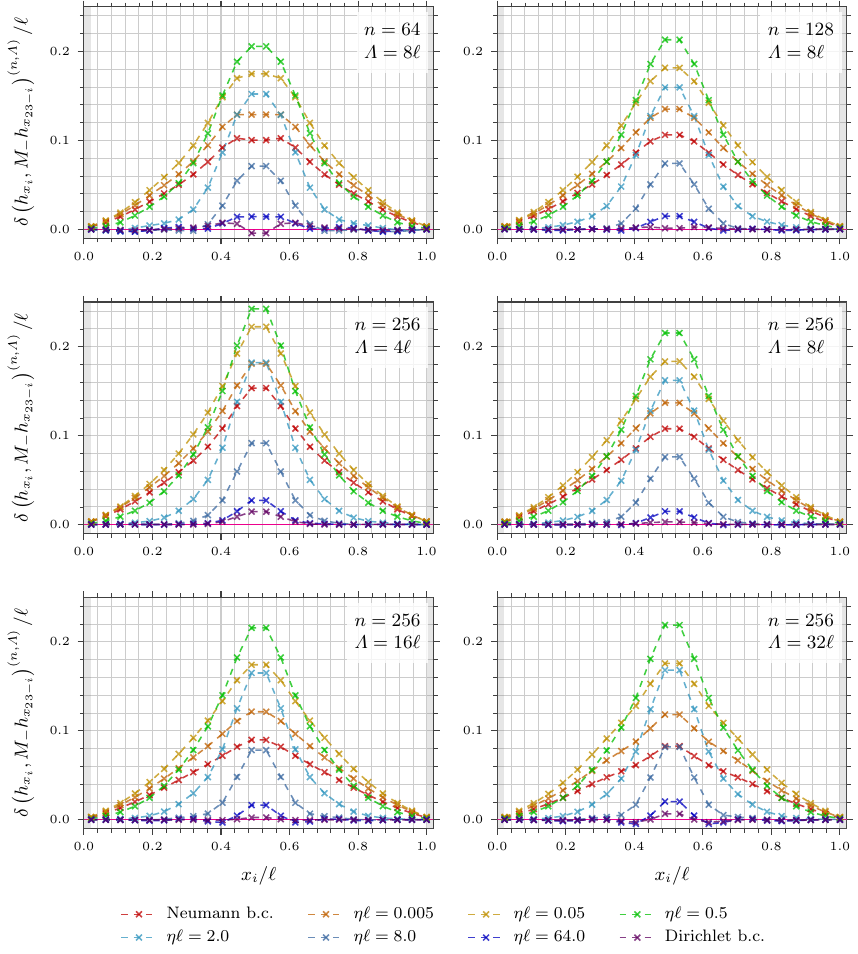}
  \caption{\label{fig:LeftWedge.errMm.Boundaries.Antidiagonal} 
  Interval adjacent to the boundary: 
  Massless regime when Robin b.c.\@ are imposed.
  Difference \eqref{eq:LeftWedge.Mm.Error} along 
  the anti-diagonal of the smeared matrix in $A$,
  for increasing values of the discretization parameters 
  $n$ and $\Lambda$.}
\end{figure}

When the Robin b.c.\@ \eqref{eq:HalfLine.RobinBoundaryCondition} are imposed in the massless regime, 
the dependence on the parameter $\eta$ of the modular Hamiltonian for the interval adjacent to the boundary
is explored through the numerical data points shown in Fig.~\ref{fig:LeftWedge.Mm.Boundaries.Diagonals}
and Fig.~\ref{fig:LeftWedge.errMm.Boundaries.Antidiagonal},
by employing the dimensionless quantity $\eta \ell$.
The extreme values of $\eta =0$ and $\eta \to \infty$
correspond to the  Neumann b.c.\@ and the Dirichlet b.c.\@, respectively. 
In Fig.~\ref{fig:LeftWedge.Mm.Boundaries.Diagonals},
the data points have been obtained 
for $n=256$ and $\Lambda = 16\ell$, 
while in Fig.~\ref{fig:LeftWedge.errMm.Boundaries.Antidiagonal}
different values for these discretization parameters are considered,
in order to study the convergence to stable results.
Recall that, as in all plots, we show smeared results \eqref{eq:HalfLine.ModularHamiltonian.SmearedBlock.CompactNotation}.
In particular, we explore the elements along either the diagonal 
or the anti-diagonal, 
which are taken respectively at the white and red crosses 
in the bottom panels of Fig.~\ref{fig:LeftWedge.Mm.Discretization.Matrices}.

In the top left panel of Fig.~\ref{fig:LeftWedge.Mm.Boundaries.Diagonals},
the diagonal elements display a non-trivial dependence on the b.c.\@ dimensionless parameter $\eta \ell$.
Indeed, considering the half parabola 
provided by the BCFT kernel \eqref{eq:LeftWedge.BCFTReference},
namely the solid magenta curve,
the most evident deviation from it 
corresponds to $\eta \ell = 0.5$ (see the green marks).
Such deviation for the 
extreme cases of Neumann b.c.\@ and Dirichlet b.c.\@ 
has been already described above, in the discussion of  Fig.~\ref{fig:LeftWedge.Mm.Discretization.Diagonal}.
We cannot exclude that, 
like for the Neumann b.c.\@ case discussed above, 
also for the Robin b.c.\@ cases that we have studied
our best approximation is not enough to catch the QFT limit
and therefore higher values of $n$ and $\Lambda$ should be considered.
Nonetheless, the deviation for the Robin b.c.\@ is more pronounced 
and appears to be much more than a numerical artefact.

In order to highlight this dependence on $\eta \ell$ along the diagonal, 
in the right panel of Fig.~\ref{fig:LeftWedge.Mm.Boundaries.Diagonals}
we report a zoom into the data of the left panel
for five different positions in the interval $A$ 
corresponding to the peaks of the log-Gaussians 
$\big\{h_{x_4}, h_{x_7}, h_{x_{10}}, h_{x_{13}}, h_{x_{16}}\big\}$, which are indicated through vertical lines in the top left panel.
In the top right panel, these numerical values are shown in terms of 
$\eta \ell / (1 + \eta \ell)$, which compactifies monotonically the domain 
$\eta \geqslant 0$ to $\eta \ell / (1 + \eta \ell) \in [0,1)$;
indeed, the Neumann and Dirichlet b.c.\@ correspond to the
endpoints, i.e.\@ 0 and 1 respectively. 
For each one of these five positions, 
we find an arc with a single maximum located around $\eta \ell / (1 + \eta \ell) = 0.33$, which corresponds to the above mentioned value $\eta\ell = 0.5$.
In the right panel of Fig.~\ref{fig:LeftWedge.Mm.Boundaries.Diagonals}
some observations made for Fig.~\ref{fig:LeftWedge.Mm.Discretization.Diagonal} 
become more evident. 
Indeed, 
considering the BCFT value for each position,
obtained from the first expression in \eqref{eq:LeftWedge.References.Smearing.Diagonal},
denoted by the horizontal solid line having the corresponding colour, we observe a perfect match for Dirichlet b.c.\@
and a slight deviation occurs for Neumann b.c.\@,
which decreases as the position gets closer to the entangling point at $x=\ell$.
The latter observation can be applied to the whole arc;
indeed, considering the various arcs for 
$0< x_4 < x_7 < x_{10} < x_{13} < x_{16} < \ell$, 
the distance between each of them and the corresponding horizontal solid line from BCFT decreases in a monotonic way.

To explore the possible non-local nature of the modular Hamiltonian when Robin b.c.\@ with finite $\eta>0$ are imposed, 
we show the smeared results for 
\eqref{eq:HalfLine.ModularHamiltonian.SmearedBlock.CompactNotation}
along the anti-diagonal (red crosses in Fig.~\ref{fig:LeftWedge.Mm.Discretization.Matrices})
in the bottom panels of Fig.~\ref{fig:LeftWedge.Mm.Boundaries.Diagonals}.
In the bottom left panel the numerical results 
for the matrix elements are reported, 
while in the bottom right panel we consider their difference 
with respect to the corresponding matrix elements obtained from the BCFT kernel \eqref{eq:LeftWedge.BCFTReference}, namely
\begin{equation}
\label{eq:LeftWedge.Mm.Error}
    \delta\left( h_{x_i}, M_- h_{x_{23 - i}} \right)^{(n, \Lambda)}
  \coloneq \left( h_{x_i}, M_- h_{x_{23 - i}} \right)^{(n, \Lambda)}
    - \innerProd[\labelReal]{h_{x_i}}{M_-^{\textlabel{BCFT}} h_{x_{23 - i}}}
  \eqend{,}
\end{equation}
where, in the r.h.s.,  
the first and second term come from \eqref{eq:HalfLine.ModularHamiltonian.SmearedBlock.CompactNotation} 
and \eqref{eq:LeftWedge.BCFTReference.Smearing}, respectively. 
The difference \eqref{eq:LeftWedge.Mm.Error} is also explored in 
Fig.~\ref{fig:LeftWedge.errMm.Boundaries.Antidiagonal}
for various values of the discretization parameters $n$ and $\Lambda$
(notice that the bottom right panel of Fig.~\ref{fig:LeftWedge.Mm.Boundaries.Diagonals} is also the bottom left panel of Fig.~\ref{fig:LeftWedge.errMm.Boundaries.Antidiagonal}).
We highlight the remarkable agreement between the numerical data for the Dirichlet b.c.\@ and the BCFT kernel \eqref{eq:LeftWedge.BCFTReference};
indeed, the corresponding line for the difference \eqref{eq:LeftWedge.Mm.Error} is very close to zero 
in  Fig.~\ref{fig:LeftWedge.Mm.Boundaries.Diagonals}
and Fig.~\ref{fig:LeftWedge.errMm.Boundaries.Antidiagonal}.
Instead, for Robin b.c.\@ with finite $\eta > 0$
the numerical data points for \eqref{eq:LeftWedge.Mm.Error} 
do not resemble a Gaussian profile (of width $\sigma \approx 0.096 \ell$) coming from the smeared local term, 
where the diagonal intersects the anti-diagonal and the local term dominates over non-local contributions.
This effect is more pronounced for small values of $\eta > 0$.
Indeed, note the rapid change of the numerical results when moving away from the Neumann b.c.\@ with $\eta = 0$ to small values $\eta > 0$, which is to be expected, because the derivative in $\eta$ of the expansion of $D^{-1/4}$ as $\eta \to 0^+$
(see \eqref{eq:InverseFourthRootLaplace.RobinBoundary.NeumannExpansion}) diverges like $\eta^{-1/2}$ at $\eta = 0$.
Since these results for $\eta > 0$ are mostly stable 
when increasing the discretization parameters (see Fig.~\ref{fig:LeftWedge.errMm.Boundaries.Antidiagonal}), 
we expect that 
the modular Hamiltonian of the interval adjacent to the half line 
for the massless scalar with Robin b.c.\@ is  non-local.

We report a comparison of the difference \eqref{eq:LeftWedge.Mm.Error}
for different discretization parameters in Fig.~\ref{fig:LeftWedge.errMm.Boundaries.Antidiagonal}.
Comparing the top two panels with the middle right panel, 
we find that increasing $n$ provides a slight decrease in the width of the curves, which remain non-vanishing 
for finite values $\eta \geqslant 0$.
Interestingly, from the middle and bottom panels, 
where the cutoff parameter changes,
$\Lambda \in \big\{ 4 \ell, 8 \ell, 16 \ell, 32 \ell \big\}$
with $n = 256$ kept fixed, 
we observe that the data points for large values of $\eta$ 
rapidly stabilize, while the results corresponding to small $\eta$ 
reach stable values more slowly.
In particular, for the Neumann b.c.\@ at $\eta = 0$, the change in the data points from panel to panel indicates that these results have not converged yet;
hence we conclude again that higher values of the discretization parameters are needed to capture the QFT limit in this case, as already highlighted in the discussion of the right panel of Fig.~\ref{fig:LeftWedge.Mm.Discretization.Diagonal}.
It would require much larger values of $\Lambda$ 
to get a significant improvement.
However, at $\Lambda = 32 \ell$ 
we already observe a slight increase of the curve 
for Dirichlet b.c.\@ and $\eta\ell =64.0$, which is due to the fact that, 
when $\Lambda$ becomes large with $n$ fixed,
discretization functions $e_i^{(n, \Lambda)}$ occur
near the value of $\Lambda$ with a large width,
which also influences the results within the region $A$.
In order to get more reliable numerical results, 
also $n$ must increase, together with $\Lambda$,
but this requires a higher floating point precision 
and a much higher run-time.
In our implementation of the algorithm, 
we compromise on $n = 256$ and $\Lambda = 16 \ell$ for the remaining part of our analysis. 
This guarantees that for Dirichlet b.c.\@ the QFT limit is sufficiently well approximated and, presumably, also the results for Robin b.c.\@ with large values of $\eta$ are sufficiently precise,
while the results for very small values of $\eta$ and for 
Neumann b.c.\@ are less accurate (in particular in the massless regime).

\subsection{Massive regime}
\label{sec:Results.AdjacentInterval.Massive}

\begin{figure}[t!]
  \centering
  \includegraphics{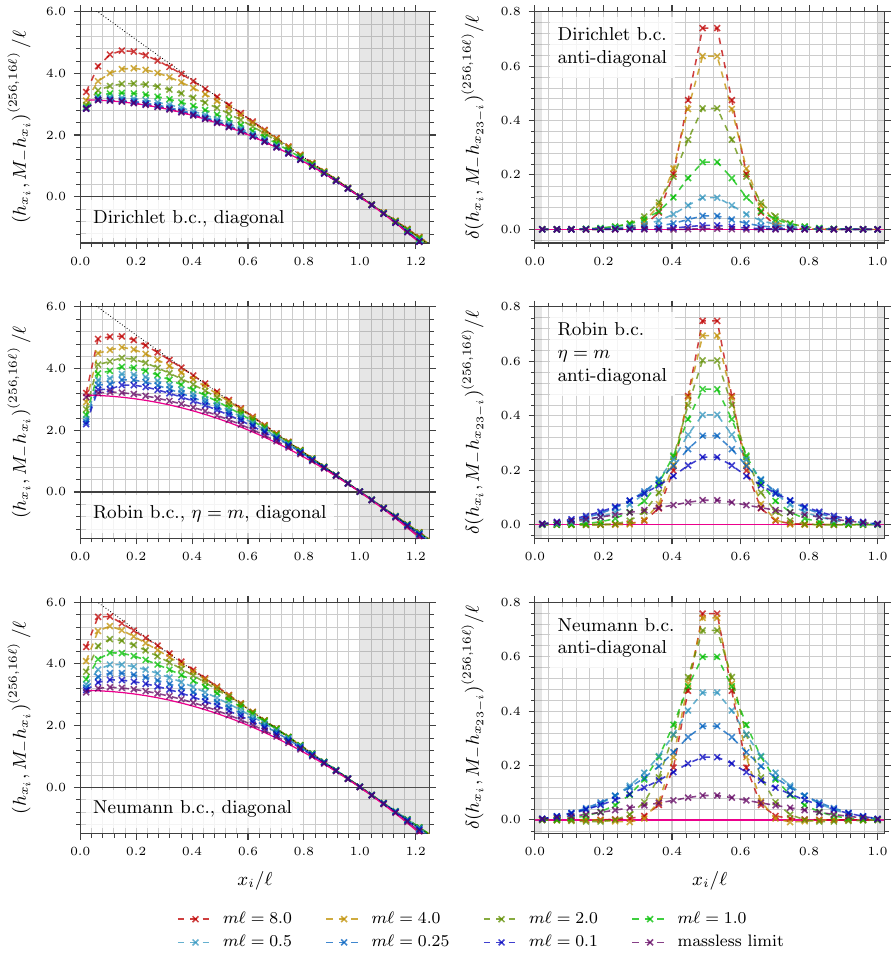}
  \caption{\label{fig:LeftWedge.Mm.Massive.Diagonals} 
  Interval adjacent to the boundary: Massive regime
  for various values of $m \ell$,
  when either Dirichlet b.c.\@ (top) 
  or Robin b.c.\@ with $\eta = m$ (middle) 
  or Neumann (bottom) are imposed. 
  Numerical results along the diagonal (left) 
  and along the anti-diagonal (right),
  corresponding to the white and red crosses in Fig.~\ref{fig:LeftWedge.Mm.Discretization.Matrices}, respectively.
  }
\end{figure}

The most interesting insights that we have obtained through our numerical approach regard the modular Hamiltonian in the massive regime. 
In Fig.~\ref{fig:LeftWedge.Mm.Massive.Diagonals} we report our numerical results for Dirichlet b.c.\@ (upper panels),
Robin b.c.\@ when $\eta=m$ (central panels)
and  Neumann b.c.\@ (lower panels),
which have been found by employing 
the expression for $D^{-1/4}(x, y)$
given by 
\eqref{eq:InverseFourthRootHelmholtz.SpecialBoundary} with the minus sign,
\eqref{eq:InverseFourthRootHelmholtz.RobinBoundary}--\eqref{eq:InverseFourthRootHelmholtz.RobinMassBoundary.BoundaryInteral}
and \eqref{eq:InverseFourthRootHelmholtz.SpecialBoundary} with the plus sign, respectively.
In the left and right panels of Fig.~\ref{fig:LeftWedge.Mm.Massive.Diagonals},
we report the results along the main diagonal 
(see the white crosses in Fig.~\ref{fig:LeftWedge.Mm.Discretization.Matrices}) 
and the difference \eqref{eq:LeftWedge.Mm.Error} 
on the anti-diagonal (see the red crosses in Fig.~\ref{fig:LeftWedge.Mm.Discretization.Matrices}), 
as also done in Sec.~\ref{sec:Results.AdjacentInterval.Massless}
for the massless regime (see Fig.~\ref{fig:LeftWedge.Mm.Boundaries.Diagonals}).
As in the massless regime, we obtain the 
numerical data points by choosing 
$n = 256$ and $\Lambda = 16 \ell$ for the discretization parameters
and performing a smearing against the log-Gaussian functions.
Notice that the numerical data becomes inaccurate 
again near the boundary, 
at least for the first few data points 
and this deviation becomes more pronounced as $m$ increases.
However, we expect that it vanishes in limit  
$n \to \infty$ and $\Lambda \to \infty$.

An important observation regarding the data points 
along the main diagonal,
shown in the left panels of 
Fig.~\ref{fig:LeftWedge.Mm.Massive.Diagonals},
 is that the straight black dotted line 
 given by the second expression in \eqref{eq:LeftWedge.References.Smearing.Diagonal}
plays the role of an upper bound that appears to become 
saturated in the large mass limit.
The saturation occurs 
through a rate that decreases as $\eta$ increases. 
A similar result has been first observed on the lattice 
for a block of consecutive sites in a one-dimensional chain \cite{EislerDiGiulioTonniPeschel:2020}, 
and later confirmed also for the interval on the line 
\cite{BostelmannCadamuroMinz:2023} through the numerical procedure that we are also employing for our analyses. 
The case of two disjoint blocks in the infinite harmonic chain 
in the large mass regime has been recently explored \cite{GentileRotaruTonni:2025}, 
finding interesting sharp transitions.

The modular Hamiltonian of an interval for the massive scalar in the ground state is non-local. 
However, our numerical data points seem to suggest that  
the modular Hamiltonian becomes local in the large mass regime 
$m \to \infty$.
Indeed, considering the red curves in the right panels of 
Fig.~\ref{fig:LeftWedge.Mm.Massive.Diagonals}, 
which correspond to the largest value of $m \ell$ 
that we have explored, 
they have the smallest width 
and it is comparable to a Gaussian profile that has a width given by the smearing parameter $\sigma$, i.e.\@ a local contribution.
Moreover, these curves are almost identical, 
indicating that in the large mass limit this quantity seems independent of the boundary conditions.

As for the small mass regime for the three values of $\eta$ 
considered in Fig.~\ref{fig:LeftWedge.Mm.Massive.Diagonals},
we remind that, while for Dirichlet b.c.\@ the local BCFT result 
(solid magenta curve) is matched in the massless limit, 
this is not the case for Neumann b.c.\@ because higher values for the discretization parameters are required, as already remarked in Sec.~\ref{sec:Results.AdjacentInterval.Massless}.
This explains the qualitative difference between 
the top right panel and the other two right panels as $m \ell$
vanishes.
Moreover, notice that the curves spread over the entire anti-diagonal 
more than the smearing size $\sigma$, at least up to $m \ell = 1$,
and this numerical observation further supports the non-local nature of the modular Hamiltonian in the massive regime.
The non-local contributions are strongest for a Robin b.c.\@ with small $\eta > 0$ and for small mass, since we have to expect a sharp transition away from the Neumann b.c.\@ in the massless regime ($\eta = m = 0$), in particular when increasing $\eta$, 
as already discussed in Sec.~\ref{sec:Results.AdjacentInterval.Massless}.

Finally, we remark that the numerical approach 
that we have employed 
provides results also in the complementary domain $B$ 
in the half line, 
which is a half line itself with origin at $x=\ell$.
The corresponding results are mostly not reported in the figures 
of this section and shaded in gray.
Only Fig.~\ref{fig:LeftWedge.Mm.Discretization.Matrices} 
includes a larger part of $B$ of length $\ell$.
These numerical results in $B$ suggest similar features to the ones described above for the interval $A$. However, since $B$ has infinite volume and the cutoff $\Lambda$ must be introduced, our numerical procedure becomes inaccurate away from the entangling point at $x=\ell$, 
as indicated by the increasing pixel size towards the upper right corners in the top panels of Fig.~\ref{fig:LeftWedge.Mm.Discretization.Matrices},
and reasonable results are obtained only for $\ell < x \ll \Lambda$.

\section{Interval separated from the boundary}
\label{sec:Results.SeparatedInterval}

In this section we study the modular Hamiltonian of
the interval $A = (d, d + \ell)$ of length $\ell$ 
at a certain finite distance $d > 0$ 
from the origin of the half line.
The massless and massive regimes are discussed in
Sec.~\ref{sec:Results.SeparatedInterval.Massless} and Sec.~\ref{sec:Results.SeparatedInterval.Massive}, respectively.
Additional results for the massive regime are included in Appendix~\ref{appx:AdditionalNumericalResults}.

\subsection{Massless regime}
\label{sec:Results.SeparatedInterval.Massless}

In the case of the massless scalar on the half line 
and in its ground state, 
the analytic expression for the modular Hamiltonian of an interval 
at a certain distance from the origin is unknown. 
We expect that this operator is fully non-local.
Indeed, for instance, in the case of the chiral current 
associated to the massless scalar on the half line and in its ground state, 
the modular Hamiltonian of an interval could be found by applying the folding procedure \cite{DiFrancescoMathieuSenechal:1997} to the corresponding 
modular Hamiltonian of two equal and disjoint intervals in the line,
which is fully non-local \cite{AriasCasiniHuertaPontello:2018}.

To have some analytic reference as comparison in
our numerical analysis, 
we consider the analytic expression for the modular Hamiltonian 
of an interval in a generic position in the half line 
for the massless Dirac field in the ground state 
\cite{MintchevTonni:2021a},
which has been derived through the modular Hamiltonian 
of two disjoint equal intervals 
for the same field in the infinite line 
and in its ground state \cite{CasiniHuerta:2009a}.
This expression is the sum of a local 
and a bilocal operator,
where the former one is the integral over $A$ 
of the energy density of the massless Dirac field, 
multiplied by the proper weight function,
and the latter one is the integral over $A$ of
a bilocal operator
multiplied by a different weight function.
The bilocal term is a quadratic operator in the field where 
one fermionic field is evaluated at the position $x$
and the other one is supported at its conjugate position 
\begin{equation}
\label{eq:ConjugateCurve}
    x_{\textlabel{c}}(x)
  \coloneq \frac{d (d + \ell)}{x}
  \eqend{.}
\end{equation}
The self-conjugate point $x_{\textlabel{sc}} \in A$ is such that $x_{\textlabel{c}}(x_{\textlabel{sc}}) = x_{\textlabel{sc}}$, i.e.\@ 
\begin{equation}
\label{eq:SelfConjugatePoint}
    x_{\textlabel{sc}}
  = \sqrt{d (d + \ell)}
  \eqend{.}
\end{equation}

As a reference for our analysis,
the weight function of the local term in the fermionic modular Hamiltonian found in \cite{MintchevTonni:2021a}
corresponds to the kernel
\begin{equation}
\label{eq:DoubleCone.LocalReference}
    M_{-}^{\textlabel{F,loc}}(x, y)
  \coloneq \pi \mult \frac{
      \bigl[ x^2 - d^2 \bigr]
      \bigl[ (d + \ell)^2 - x^2 \bigr]
    }{ \ell \big[ d (d + \ell) + x^2 \big]}
    \mult \updelta(x - y)
  \eqend{.}
\end{equation}
In the limit of large $d$, this kernel 
approaches the known result for a double cone over an 
interval of length $\ell$ 
on the initial time Cauchy surface (the real line) 
in Minkowski space \cite{HislopLongo:1982}.
In Appendix~\ref{appx:Discretization.TestFunctions}
we derive the analytic expression for the smearing of \eqref{eq:DoubleCone.LocalReference}
through a generic pair of log-Gaussian test functions \eqref{eq:LogGaussian},
finding \eqref{eq:DoubleCone.LocalReference.Smearing}.
Along the diagonal, where $i=j$,
and for small values of the width $\sigma$,
it becomes the weight function multiplying the Dirac delta in \eqref{eq:DoubleCone.LocalReference}, namely
\begin{equation}
\label{eq:DoubleCone.LocalReference.Smeared.Diagonal}
    \innerProd[\labelReal]{h_{x_i}}{M_{-}^{\textlabel{F,loc}} \mult h_{x_i}}
  = \pi \mult \frac{
      \bigl[ x_i^2 - d^2 \bigr]
      \bigl[ (d + \ell)^2 - x_i^2 \bigr]
    }{ \ell \bigl[ d (d + \ell) + x_i^2 \bigr]}
    + \Ord\bigl( \sigma^2 \bigr)
  \eqend{.}
\end{equation}
We do not report a comparison with the bilocal term of the fermionic case 
\cite{MintchevTonni:2021a} because its dimension is not compatible 
with a bilocal expression in terms of the massless scalar field.

Further analytic references for our comparison with the numerical results are obtained by considering the linear approximations in the neighbourhoods of the endpoints at $x=d$ and $x=d+\ell$, given respectively by 
\begin{eqnarray}
\label{eq:DoubleCone.InnerLinearReference}
    M_-^{\textlabel{RW}}(x, y)
  &=& 2 \pi (x - d) \mult \updelta(x - y)
  \eqend{,}
\\
\label{eq:DoubleCone.OuterLinearReference}
    M_-^{\textlabel{LW}}(x, y)
  &=& 2 \pi (d + \ell - x) \mult \updelta(x - y)
  \eqend{.}
\end{eqnarray}
The expressions for these reference kernels integrated against the log-Gaussian functions are given in \eqref{eq:DoubleCone.OuterLinearReference.Smearing} and \eqref{eq:DoubleCone.InnerLinearReference.Smearing}, respectively.
Along the diagonal, 
we compute the smearing as done in \eqref{eq:LeftWedge.References.Smearing.Diagonal}, obtaining 
\begin{eqnarray}
\label{eq:DoubleCone.InnerLinearReference.Smeared.Diagonal}
    \innerProd[\labelReal]{h_{x_i}}{M_-^{\textlabel{RW}} h_{x_i}}
  &=& 2 \pi (x_i - d) + \Ord\bigl( \sigma^2 \bigr)
  \eqend{,}
\\
\label{eq:DoubleCone.OuterLinearReference.Smeared.Diagonal}
    \innerProd[\labelReal]{h_{x_i}}{M_-^{\textlabel{LW}} h_{x_i}}
  &=& 2 \pi (d + \ell - x_i) + \Ord\bigl( \sigma^2 \bigr)
  \eqend{,}
\end{eqnarray}
for \eqref{eq:DoubleCone.InnerLinearReference}
and \eqref{eq:DoubleCone.OuterLinearReference} respectively.

\begin{figure}
  \centering
  \includegraphics{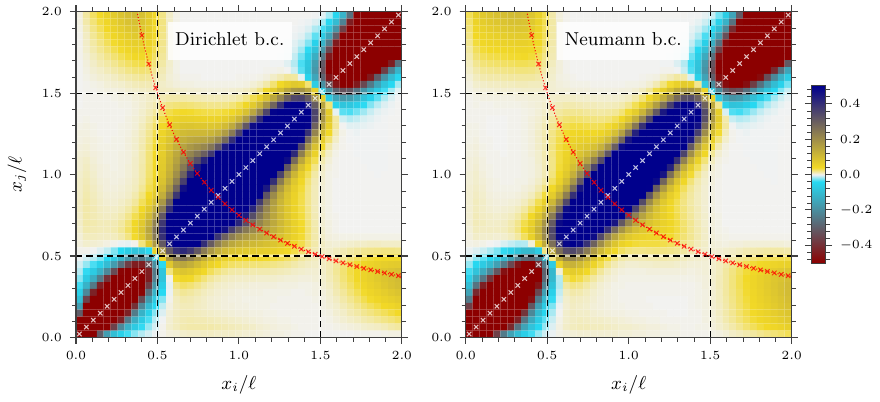}
  \caption{\label{fig:DoubleCone.Mm.Massless.SmearedMatrices} 
    Interval separated from the boundary by a distance 
    $d = \ell /2$, massless regime: 
  Output of the numeric algorithm 
  after the smearing procedure 
  (see \eqref{eq:HalfLine.ModularHamiltonian.SmearedBlock.CompactNotation}),
  obtained with discretization parameters 
  $n = 256$ and $\Lambda = 16 \ell$, 
  for either Dirichlet b.c.\@ (left) or Neumann b.c.\@ (right).
  The corresponding results before the smearing are shown in 
  the top left panel of 
  Fig.~\ref{fig:DoubleCone.Mm.Dirichlet.Matrices}
  and Fig.~\ref{fig:DoubleCone.Mm.Neumann.Matrices},
  for Dirichlet b.c.\@ and Neumann b.c.\@ respectively. 
  }
\end{figure}

In Fig.~\ref{fig:DoubleCone.Mm.Massless.SmearedMatrices} 
we show the output \eqref{eq:HalfLine.ModularHamiltonian.SmearedBlock.CompactNotation}
of the numerical algorithm 
after the smearing procedure through the log-Gaussian functions 
for a bipartition of the half line 
determined by the interval $A = (d, d + \ell)$ with $d =\ell/2$,
when either Dirichlet b.c.\@ (left panel) 
or Neumann b.c.\@ (right panel) are imposed. 
The two axes of each panel represent the positions $x_i$ and $x_j$
of the peaks of the log-Gaussian functions \eqref{eq:LogGaussian.Discretization} and $A$ is outlined by the black straight dashed lines.
The white and red crosses in Fig.~\ref{fig:DoubleCone.Mm.Massless.SmearedMatrices} 
correspond to $x_j = x_i$ (i.e.\@ the diagonal) 
and $x_j = x_{\textlabel{c}}(x_i)$ as given in \eqref{eq:ConjugateCurve}, respectively.
The corresponding results before the smearing procedure 
are reported in the first panel of Fig.~\ref{fig:DoubleCone.Mm.Dirichlet.Matrices}
and Fig.~\ref{fig:DoubleCone.Mm.Neumann.Matrices}, respectively.
The numerical results extend into the complement region $B$, where they take negative values along the diagonal and positive values close to the curve $x_{\textlabel{c}}$.

\begin{figure}[t!]
  \centering
  \includegraphics{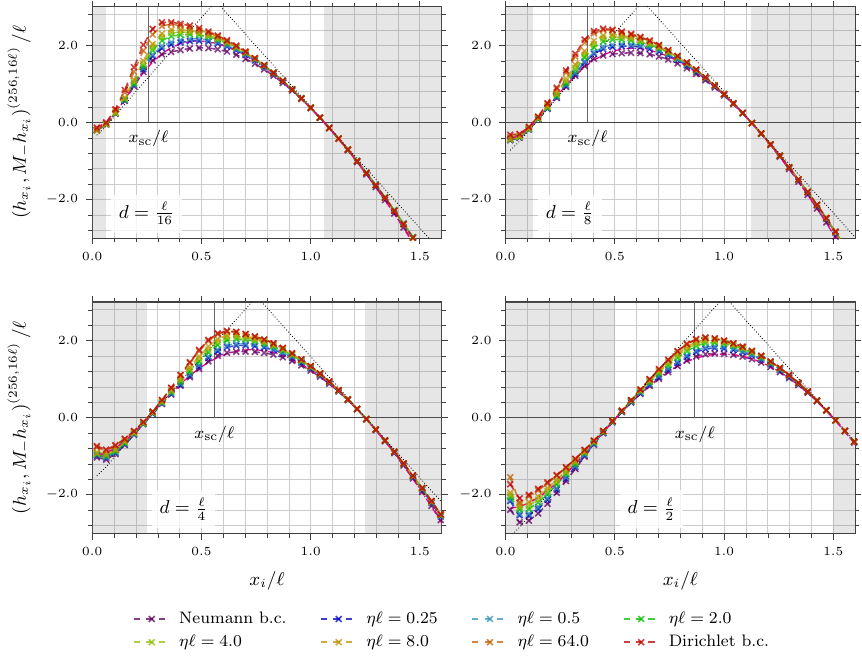}
  \caption{\label{fig:DoubleCone.Mm.Boundaries.Diagonal} 
      Interval separated from the boundary for increasing values of $d/\ell$: Massless regime when Robin b.c.\@ are imposed. 
      Modular Hamiltonian along the diagonal 
  (see the white crosses in 
    Fig.~\ref{fig:DoubleCone.Mm.Massless.SmearedMatrices}) 
    when $n = 256$ and $\Lambda = 16 \ell$,
  for different values of $\eta \ell$.
  }
\end{figure}

In the four panels of Fig.~\ref{fig:DoubleCone.Mm.Boundaries.Diagonal}
we report our numerical results for  $(h_{x_i}, M_- h_{x_i})^{(256, 16\ell)}$, i.e.\@ along the diagonal (see the white crosses in Fig.~\ref{fig:DoubleCone.Mm.Massless.SmearedMatrices}),
for four different bipartitions, where $d \in \{\ell/16, \ell/8, \ell/4, \ell/2 \}$.
For each of them, the same set of different values of $\eta \ell$ 
is considered.  
The grey shaded region corresponds to $B$.
The straight dotted lines stand for the references \eqref{eq:DoubleCone.InnerLinearReference.Smeared.Diagonal} and \eqref{eq:DoubleCone.OuterLinearReference.Smeared.Diagonal},
while the position of the self-conjugate point 
\eqref{eq:SelfConjugatePoint} is highlighted by a vertical gray line. 
In each panel, the numerical results change monotonically  
from the Neumann b.c.\@ (violet curve) 
to the Dirichlet b.c.\@ (red curve).
We recall that a universal result exists in a BCFT on the half line 
for the modular Hamiltonian of an interval adjacent to the boundary \cite{CardyTonni:2016}, while this is not true anymore when the interval is separated from the boundary.
However, in the latter case, one might expect that at least the weight function of the local term in the modular Hamiltonian for the BCFT model given by the massless Dirac field \cite{MintchevTonni:2021a} 
(see the dash-dotted magenta curve in Fig.~\ref{fig:DoubleCone.Mm.Boundaries.Diagonal})
agrees with the corresponding weight function 
for the massless scalar field, which is a BCFT as well.
The numerical results in 
Fig.~\ref{fig:DoubleCone.Mm.Boundaries.Diagonal} show that 
only the curves for small $\eta$ are close to the dashed magenta curve corresponding to \eqref{eq:DoubleCone.LocalReference.Smeared.Diagonal},
while for larger values of $\eta$, 
the results are above the local fermionic reference
(especially around the self-conjugate point $x_{\textlabel{sc}}$)
and for the Dirichlet b.c.\@ this difference is the strongest.
At this point, we cannot establish whether this behaviour 
is due to a different local term w.r.t. 
the one occurring for the massless Dirac field 
or to a non-local term depending on $\eta$ 
that also adds to the diagonal plotted in Fig.~\ref{fig:DoubleCone.Mm.Boundaries.Diagonal}.
As $d$ increases, the dependence on $\eta \ell$ becomes less evident in $A$ 
but it remains relevant in $B$, close to the boundary
(see the bottom right panel of Fig.~\ref{fig:DoubleCone.Mm.Boundaries.Diagonal}).
Note that the (first three) data points very close to the boundary 
in each panel have similar numerical artefacts as in the $d = 0$ case 
(see Fig.~\ref{fig:LeftWedge.Mm.Discretization.Diagonal}
and the top left panel of Fig.~\ref{fig:LeftWedge.Mm.Boundaries.Diagonals})
because of the limited resolution.

\begin{figure}
  \centering
  \includegraphics{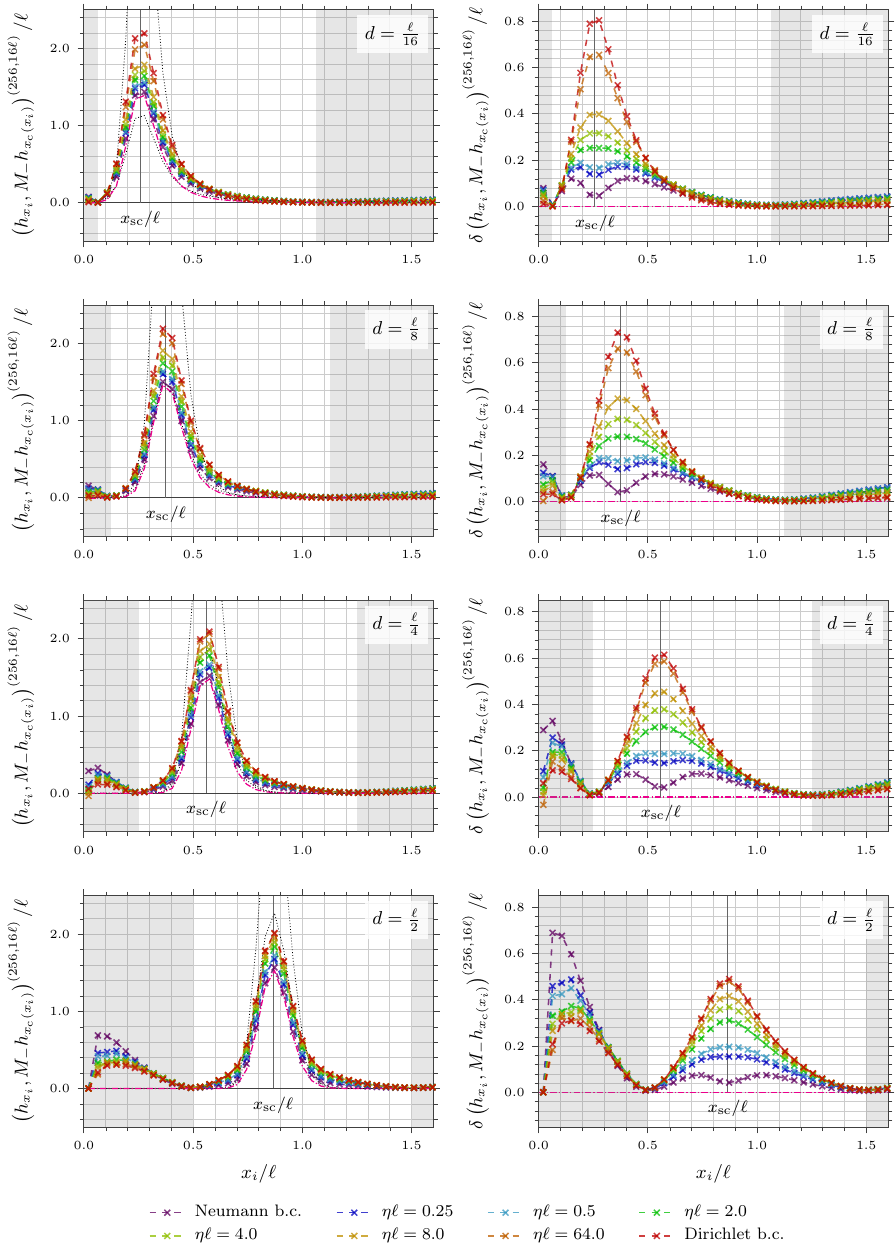}
  \caption{\label{fig:DoubleCone.Mm.Boundaries.Conjugate} 
  Interval separated from the boundary, massless regime: 
  Numerical results along the red crosses in Fig.~\ref{fig:DoubleCone.Mm.Massless.SmearedMatrices}, 
  for the same choices of $\eta \ell$ and $d/\ell$ of 
  Fig.~\ref{fig:DoubleCone.Mm.Boundaries.Diagonal}.}
\end{figure}

A major difficulty consists in separating the contributions 
of the local term from the one of the non-local term in the modular Hamiltonian. 
Since the analytic expression for the modular Hamiltonian of the interval for the massless Dirac field \cite{MintchevTonni:2021a} has a bilocal term along the conjugate curve \eqref{eq:ConjugateCurve}, 
we have also sampled our results for the scalar field 
along this curve (see the red crosses in Fig.~\ref{fig:DoubleCone.Mm.Massless.SmearedMatrices}).
The numerical results of this analysis are reported 
in the four left panels of Fig.~\ref{fig:DoubleCone.Mm.Boundaries.Conjugate},
for the same bipartitions and b.c.\@ considered in Fig.~\ref{fig:DoubleCone.Mm.Boundaries.Diagonal}.
The position of the self-conjugate point $x_{\textlabel{sc}}$
as defined in \eqref{eq:SelfConjugatePoint}, 
which corresponds to the intersection between the curves identified by the red and white crosses in Fig.~\ref{fig:DoubleCone.Mm.Massless.SmearedMatrices},
is highlighted again by the vertical gray line.
The dash-dotted magenta curve corresponds to 
$\langle h_{x_i},\, M_{-}^{\textlabel{F,loc}} h_{x_{\textlabel{c}}(x_i)} \rangle_{\labelReal}$, 
while the dotted black curves are obtained from 
$\langle h_{x_i},\, M_{-}^{\textlabel{RW}} h_{x_{\textlabel{c}}(x_i)} \rangle_{\labelReal}$ 
(upper dotted curve) and 
$\langle h_{x_i},\, M_{-}^{\textlabel{LW}} h_{x_{\textlabel{c}}(x_i)} \rangle_{\labelReal}$ 
(lower dotted curve).

The numerical data points in $A$ have a local maximum at the self-conjugate point caused by the local contributions, 
but the curves spread out more than the local reference \eqref{eq:DoubleCone.LocalReference} (dash-dotted magenta curve) and they do not vanish in the region $B$.
Thus, the plots strongly indicate the occurrence 
of a non-local term.
At this point, we cannot safely establish whether the modular Hamiltonian contains a bilocal term or whether it is fully non-local. However, 
we expect full non-locality for any finite value of $\eta > 0$,
as suggested by non-vanishing results far away from the diagonal and from the curve \eqref{eq:ConjugateCurve} in the unsmeared data, shown in Fig.~\ref{fig:DoubleCone.Mm.Massless.Matrices} below.
Focusing on the results within the region $A$, 
the curves decrease as $d$ increases,
indicating that the non-local term might vanish at large distance 
from the boundary. 
Moreover, they depend less on the boundary parameter $\eta$ 
for increasing values of the separating distance $d$.
Notice that the non-local contributions within the complementary region $B$ 
increase towards the origin and infinity. 
However, the points close to the boundary are not reliable because the curve \eqref{eq:ConjugateCurve} diverges, taking values larger than $\Lambda$ 
(where we do not have any numerical data)
close to the boundary; hence the numerical approximation 
becomes less precise and the curve drop to zero 
in the bottom panels, where $d = \ell / 2$.

\begin{figure}[t!]
  \centering
  \includegraphics{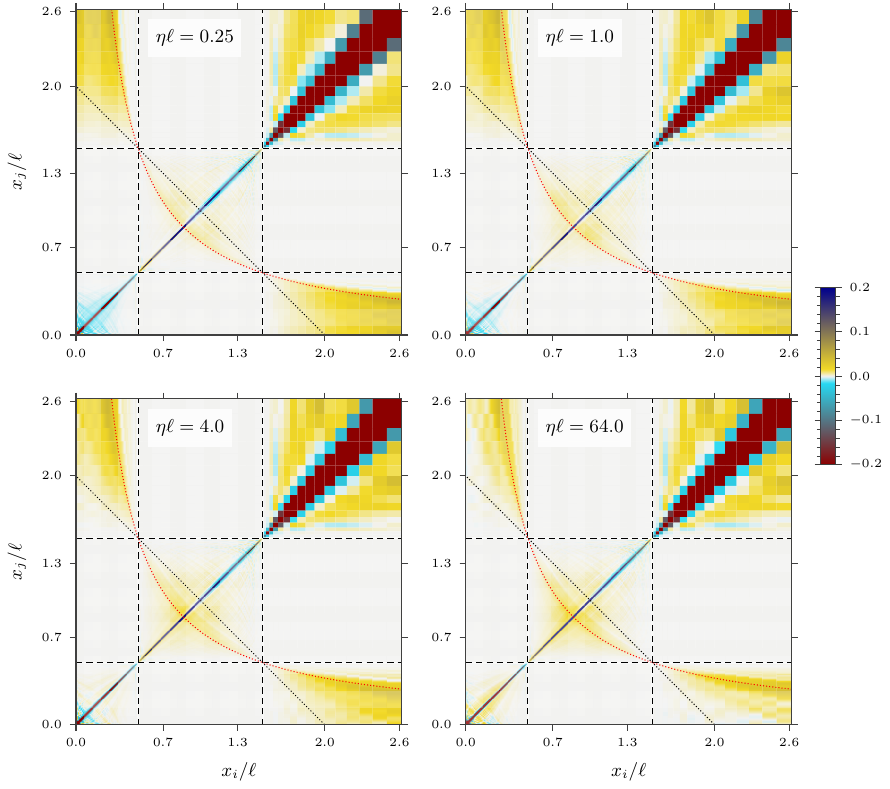}
\caption{\label{fig:DoubleCone.Mm.Massless.Matrices} 
Interval at distance $d=\ell/2$ from the boundary: 
Massless regime when Robin b.c.\@ are imposed.
Output of the numerical algorithm for 
$M_-^{(256, 16\ell)} / \ell$ before the smearing procedure 
  (see \eqref{eq:HalfLine.ModularHamiltonian.SmearedBlock.CompactNotation}) through the log-Gaussian test functions \eqref{eq:LogGaussian.Discretization}.
  The results for Dirichlet and Neumann b.c.\@ are reported in the top left panels of Fig.~\ref{fig:DoubleCone.Mm.Dirichlet.Matrices} and 
  Fig.~\ref{fig:DoubleCone.Mm.Neumann.Matrices} respectively.}
\end{figure}

For the interval adjacent to the boundary (see Sec.~\ref{sec:Results.AdjacentInterval}), 
the difference \eqref{eq:LeftWedge.Mm.Error} 
has been introduced
to explore further the non-local nature
of the modular Hamiltonian (see Fig.~\ref{fig:LeftWedge.errMm.Boundaries.Antidiagonal}). 
Similarly, for $d > 0$, we use the local fermionic reference \eqref{eq:DoubleCone.LocalReference} to define
\begin{equation}
\label{eq:DoubleCone.Mm.Error}
    \delta\left( h_{x_i}, M_- h_{x_{\textlabel{c}}(x_i)} \right)^{(n, \Lambda)}
  \coloneq \left( h_{x_i}, M_- h_{x_{\textlabel{c}}(x_i)} \right)^{(n, \Lambda)}
    - \innerProd[\labelReal]{h_{x_i}}{M_-^{\textlabel{F,loc}} \mult h_{x_{\textlabel{c}}(x_i)}}
  \eqend{.}
\end{equation}
The numerical results for this difference 
are shown in the four left panels of 
Fig.~\ref{fig:DoubleCone.Mm.Boundaries.Conjugate},
with the same configurations and the choices of b.c.\@ 
as in the corresponding left panels.
We observe that all results are positive for \eqref{eq:DoubleCone.Mm.Error}
and that the derivatives of the numeric results appear to vanish at $x = d$ and $x = d + \ell$; hence the local fermionic reference $M_-^{\textlabel{F,loc}}$ is a lower bound along the curve \eqref{eq:ConjugateCurve}.
In $A$, the difference \eqref{eq:DoubleCone.Mm.Error} 
monotonically increases from Neumann b.c.\@ to Dirichlet b.c.\@,
while in $B$ we find the opposite. 
For large values of $\eta$,
the difference \eqref{eq:DoubleCone.Mm.Error} 
has a local maximum in $A$ at the self-conjugate point $x_{\textlabel{sc}}$, which turns into a local minimum for small values of $\eta$.
Assuming that we should always get a local maximum from the local contribution, this suggests that the local fermionic reference \eqref{eq:DoubleCone.LocalReference} lies above the actual local contribution for small $\eta$, while for Dirichlet b.c.\@
it could well describe the local contribution.

We report some numerical results (before the smearing)
of $M_-^{(256, 16\ell)} / \ell$ 
for Robin b.c.\@  in 
Fig.~\ref{fig:DoubleCone.Mm.Massless.Matrices},
in the case of $d = \ell / 2$ 
and for different finite values of $\eta \ell > 0$.
The corresponding results for Dirichlet b.c.\@ and 
Neumann b.c.\@ are shown in the top left panel of Fig.~\ref{fig:DoubleCone.Mm.Dirichlet.Matrices}
and in the top left panel of Fig.~\ref{fig:DoubleCone.Mm.Neumann.Matrices},
respectively.
The range of values $[-0.2, 0.2]$ has been considered to 
highlight the small values and identify a possible structure provided
by  non-vanishing elements as a footprint of non-local terms. 
Indeed, a front described by \eqref{eq:ConjugateCurve} (red dotted curve)
can be observed all over the half line, not only in $A$, 
and for all the values of $\eta \ell$ that we have considered.
Although this front plays an interesting role, 
we cannot separate it into individual terms 
to establish whether the non-local contribution 
in the modular Hamiltonian for the massless scalar 
also contains a bilocal term.
We have also highlighted the anti-diagonal $y = \ell + 2 d - x$
(black dotted line)
because it will become relevant in the regime of large mass (see the bottom right panel of Fig.~\ref{fig:DoubleCone.Mm.Dirichlet.Matrices}).
Comparing the four panels of Fig.~\ref{fig:DoubleCone.Mm.Massless.Matrices}, corresponding to different values of $\eta \ell$, we notice that 
the differences are more significant in the region close to the boundary (see the diagonal and off-diagonal blocks corresponding to the domain $(0, d)$).

\subsection{Massive regime}
\label{sec:Results.SeparatedInterval.Massive}

In the following we discuss some numerical results for the modular Hamiltonian of an interval separated from the boundary, in the massive regime and when different b.c.\@ are imposed at the origin of the half line.

\begin{figure}
  \centering
  \includegraphics{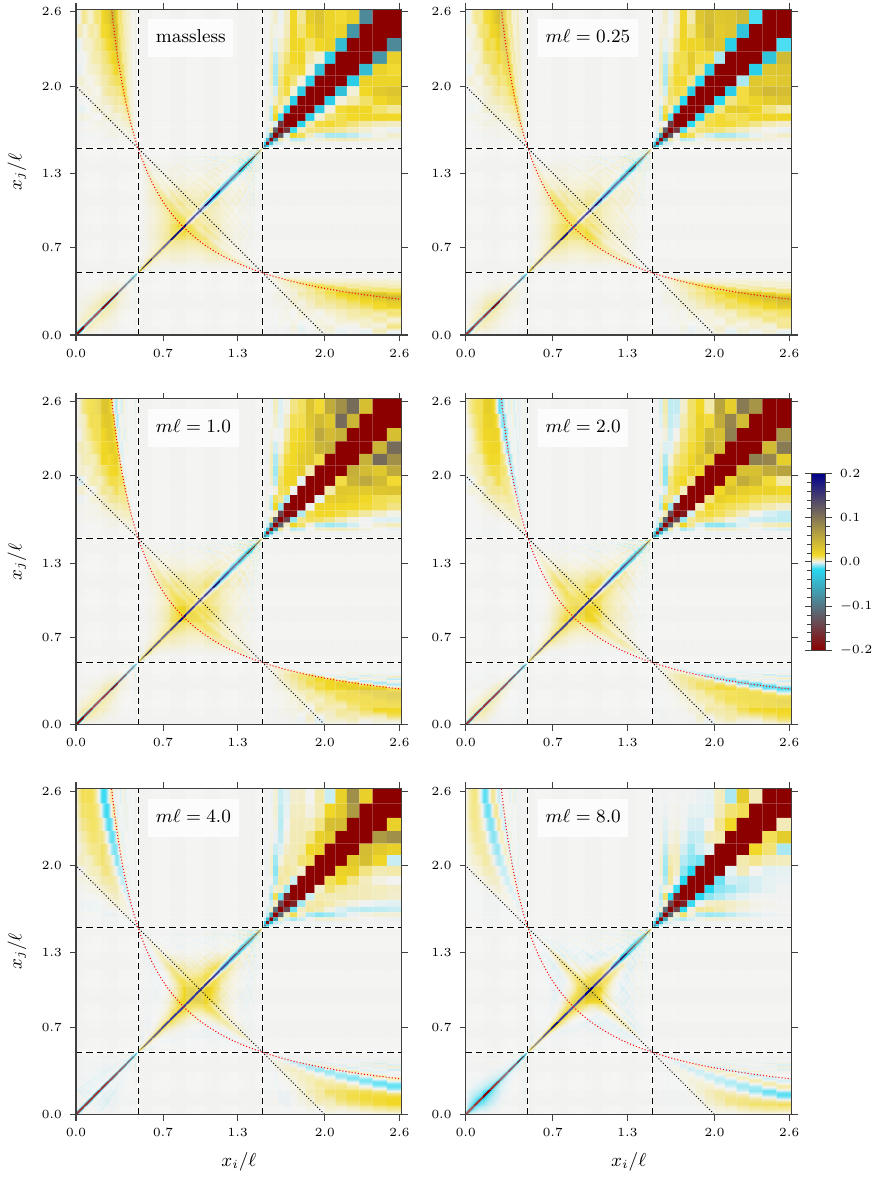}
  \caption{\label{fig:DoubleCone.Mm.Dirichlet.Matrices} 
Interval at distance $d=\ell/2$ from the boundary, where Dirichlet b.c.\@ are imposed:
Massive regime.
Output of the numerical algorithm for 
$M_-^{(256, 16\ell)} / \ell$ before the smearing \eqref{eq:HalfLine.ModularHamiltonian.SmearedBlock.CompactNotation} 
through the log-Gaussian test functions \eqref{eq:LogGaussian.Discretization},
for various values of $m\ell$.}
\end{figure}

In Fig.~\ref{fig:DoubleCone.Mm.Dirichlet.Matrices} 
we show $M_-^{(256, 16\ell)} / \ell$ for Dirichlet b.c.\@ 
before the smearing procedure in the massive regime, 
in the case of $d=\ell/2$ and for various values of $m\ell$.
In particular, the top left panel corresponds to the massless case;
hence it could be compared with the various panels of 
Fig.~\ref{fig:DoubleCone.Mm.Massless.Matrices} to gain insights in the 
dependence on the b.c.\@ of the modular Hamiltonian for the massless scalar. 
Considering the central block in each panel that corresponds to the interval $A$,
we observe that in the massless regime the non-vanishing elements outside the main diagonal form the ``front'' that follow along 
\eqref{eq:ConjugateCurve} (red dotted curve).
As $m \ell$ increases and takes finite values, the front moves away from the curve \eqref{eq:ConjugateCurve} and towards the anti-diagonal $y = \ell + 2 d - x$ (black dotted line), 
reaching it for very large values of the mass.
This behaviour indicates that the modular Hamiltonian of the interval 
for a generic finite value of $m \ell > 0$ is fully non-local.
However, we expect that it becomes local in the limit $m \to \infty$.
This expectation is also supported by numerical studies 
for the modular Hamiltonian of a block 
in the one-dimensional harmonic chain 
when the mass parameter takes large values 
\cite{EislerDiGiulioTonniPeschel:2020,Baranov:2024,GentileRotaruTonni:2025}.
As for the part of the complement region $B$ with $x > d + \ell$, 
we get a rapidly increasing pixel size as the support of the discretization functions \eqref{eq:DiscretizationBasis} increases towards the cutoff at $\Lambda = 16\ell$, so that the results beyond the outer edge of the interval region quickly become less precise.
To get a more detailed comparison for the modular Hamiltonian in the massive regime, including also the region $(0, d) \subsetneq B$, 
below we consider line plots along the diagonal and the curve \eqref{eq:ConjugateCurve}.

\begin{figure}
  \centering
  \includegraphics{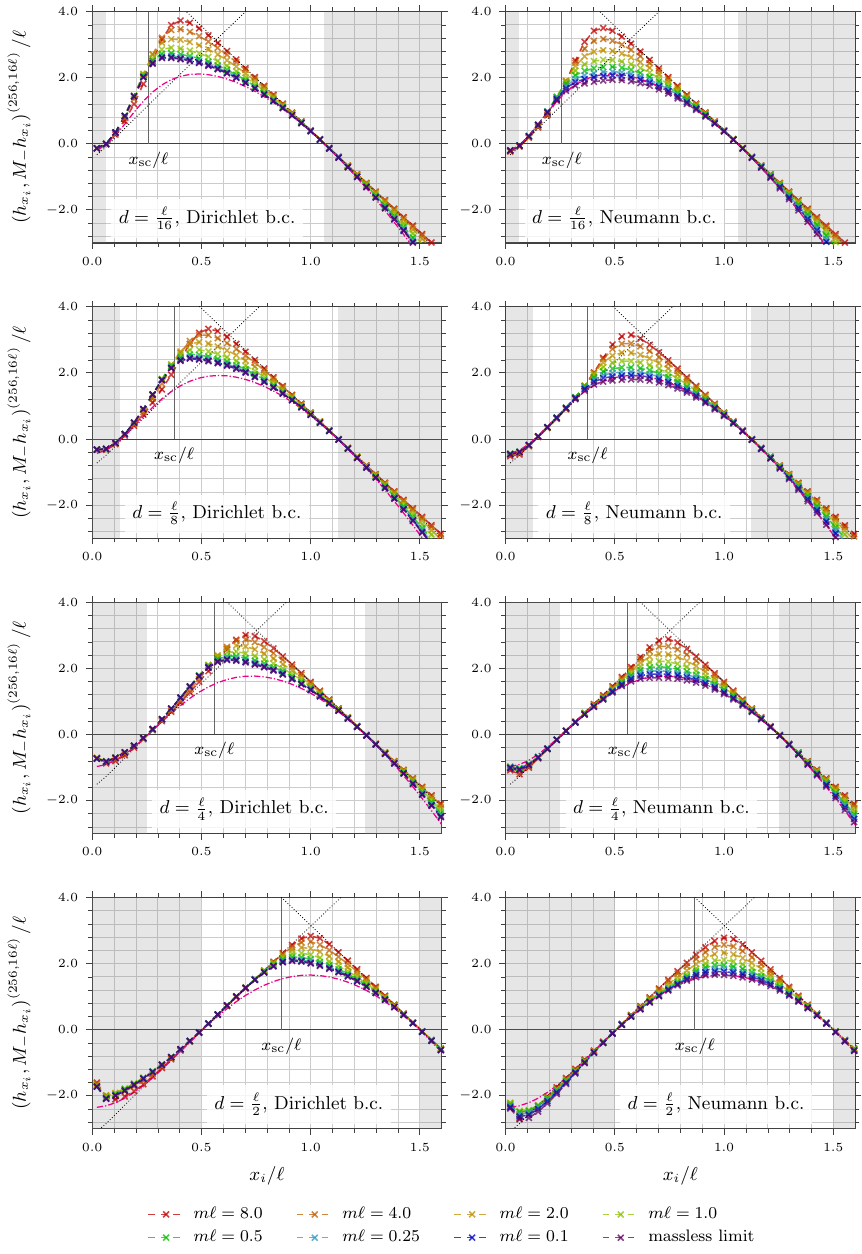}
  \caption{\label{fig:DoubleCone.Mm.Massive.Diagonal} 
    Interval separated from the boundary for various $d/\ell$: 
    Massive regime for either Dirichlet b.c.\@ (left) or Neumann b.c.\@ (right). 
    Diagonal (white crosses in 
    Fig.~\ref{fig:DoubleCone.Mm.Massless.SmearedMatrices}) 
    of the modular Hamiltonian for different values of $m \ell$.}
\end{figure}

In Fig.~\ref{fig:DoubleCone.Mm.Massive.Diagonal}, 
we explore the dependence on the mass of the 
modular Hamiltonian along the diagonal, 
similar to Fig.~\ref{fig:DoubleCone.Mm.Boundaries.Diagonal} for the same bipartitions 
and focussing on the Dirichlet b.c.\@ (left panels) 
and on the Neumann b.c.\@ (right panels). 
The dotted black lines are obtained from  \eqref{eq:DoubleCone.InnerLinearReference.Smeared.Diagonal} and \eqref{eq:DoubleCone.OuterLinearReference.Smeared.Diagonal},
where the latter appears to provide an upper bound for the numerical data,
which is reached in the limit of very large mass.
A similar behaviour for large masses 
was also observed in lattice computations \cite{GentileRotaruTonni:2025}
for the entanglement Hamiltonian of two disjoint blocks 
in the harmonic chain on the line.
The results within the region $A$ in the bottom panels suggest that, 
when $d$ becomes very large (see the bottom panels),
the effect of the boundary becomes less important 
and both dotted lines become bounds of the modular Hamiltonian, 
in agreement with the observation made in 
\cite{BostelmannCadamuroMinz:2023} for the interval in the line. 
Notice that the numerical results for the modular Hamiltonian 
in the interval $(0, d) \subsetneq B$ are 
qualitatively comparable (up to a minus sign) 
to the ones for the adjacent interval discussed in Sec.~\ref{sec:Results.AdjacentInterval}. 
Indeed, in the limit $\ell \to \infty$, 
the region $B$ becomes $(0, d)$
(see also \cite{MintchevTonni:2021a}, where this limit has been studied analytically in the case of the massless Dirac field).
In the case of Robin b.c.\@ with $m = \eta$, 
the results analogous to the ones shown in 
Fig.~\ref{fig:DoubleCone.Mm.Massive.Diagonal} 
are reported in Fig.~\ref{fig:DoubleCone.Mm.Massive.Robin.Diagonal} and discussed in Appendix~\ref{appx:AdditionalNumericalResults}.

\begin{figure}
  \centering
  \includegraphics{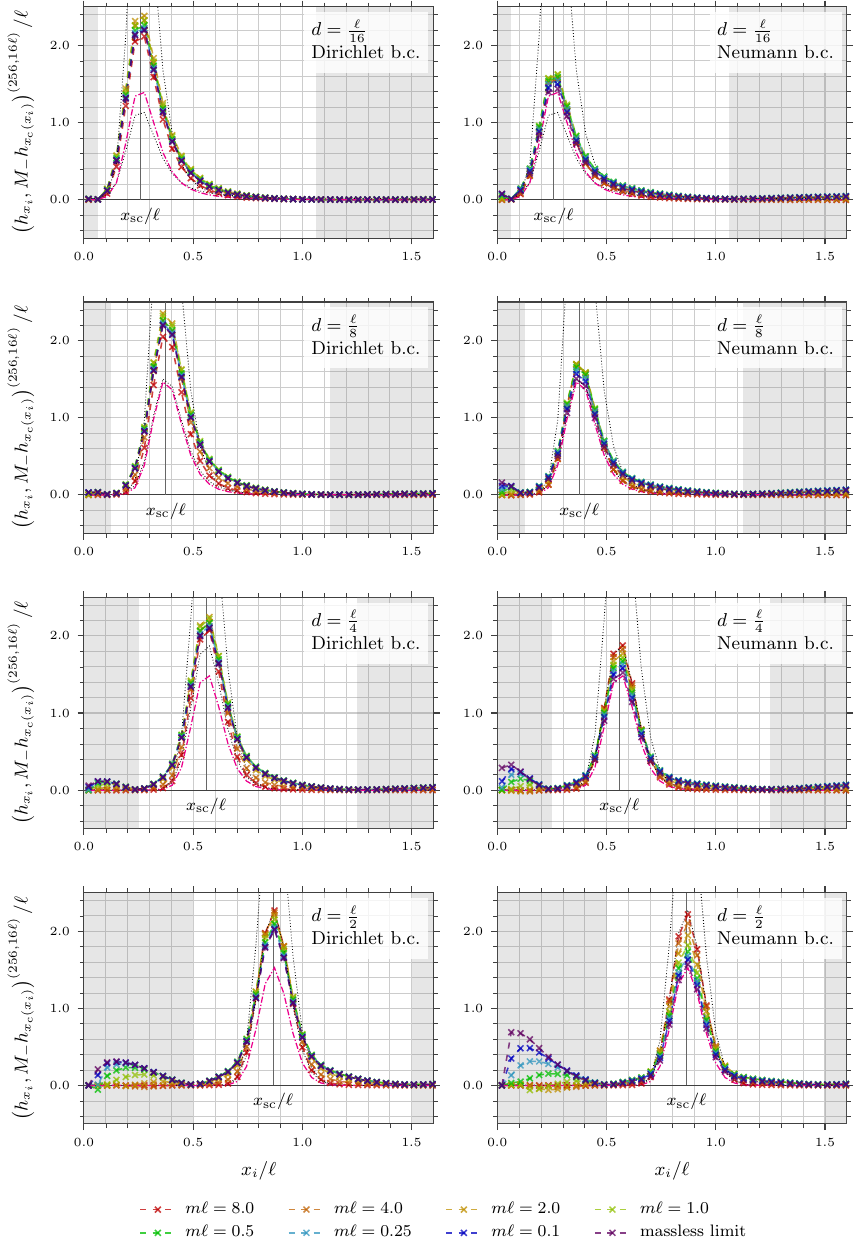}
  \caption{\label{fig:DoubleCone.Mm.Massive.Conjugate} 
    Interval separated from the boundary and massive regime.
    Numerical results along the red crosses in Fig.~\ref{fig:DoubleCone.Mm.Massless.SmearedMatrices}, 
    in the same setups of Fig.~\ref{fig:DoubleCone.Mm.Massive.Diagonal}.}
\end{figure}

In Fig.~\ref{fig:DoubleCone.Mm.Massive.Conjugate}, 
we consider the numerical results for the modular Hamiltonian
along the red crosses in 
Fig.~\ref{fig:DoubleCone.Mm.Massless.SmearedMatrices},
determined by \eqref{eq:ConjugateCurve},
for the same setups exploited in 
Fig.~\ref{fig:DoubleCone.Mm.Massive.Diagonal} 
and in analogy to the left panels of  Fig.~\ref{fig:DoubleCone.Mm.Boundaries.Conjugate} in the massless regime.
Since the local contribution is the strongest part, 
the peak of the data points occurs at $x_{\textlabel{sc}}$, 
which is the intersection of the curve given by \eqref{eq:ConjugateCurve}
with the main diagonal. 
The dash-dotted magenta curve corresponds to 
$\langle h_{x_i},\, M_{-}^{\textlabel{F,loc}} h_{x_{\textlabel{c}}(x_i)} \rangle_{\labelReal}$, 
while the dotted black curves are obtained from 
$\langle h_{x_i},\, M_{-}^{\textlabel{RW}} h_{x_{\textlabel{c}}(x_i)} \rangle_{\labelReal}$ (uppermost curve) and 
$\langle h_{x_i},\, M_{-}^{\textlabel{LW}} h_{x_{\textlabel{c}}(x_i)} \rangle_{\labelReal}$ (lowermost curve).
In the regime of small mass,
the dash-dotted magenta curve captures 
the data points obtained for Neumann b.c.\@ (right panels) better than the ones found for Dirichlet b.c.\@ (left panels),
as already discussed in Sec.~\ref{sec:Results.SeparatedInterval.Massless},
but all results for the massive regime lie above this reference within the region $A$.
As additional results, we also include the difference \eqref{eq:DoubleCone.Mm.Error} 
for the setups considered in Fig.~\ref{fig:DoubleCone.Mm.Massive.Conjugate},
shown in 
Fig.~\ref{fig:DoubleCone.errMm.Massive.Conjugate}
and  discussed in the Appendix~\ref{appx:AdditionalNumericalResults}.
Comparing the panels in the left column with the corresponding ones in the right column, it is straightforward to realise that the difference between Dirichlet b.c.\@ and Neumann b.c.\@ is stronger for small separation distance from the boundary and for small masses, as expected. 
Our numerical procedure gives access to results also in the complementary domain $B$ (see the gray shaded region in Fig.~\ref{fig:DoubleCone.Mm.Massive.Conjugate}) 
and we also highglight the behaviour of the data in the finite interval separating $A$ from the origin in the bottom panels as the mass changes.
Similarly to the massless case, these results are positive everywhere and monotonically increase towards the boundary, 
up to the three data points close to the boundary.
For these points the curve \eqref{eq:ConjugateCurve} goes beyond the cutoff $\Lambda$, 
where we do not have numerical data anymore, 
causing the plots to drop down.
We also investigated the case of Robin b.c.\@ with $\eta = m$. Since these results are qualitatively similar to the Neumann and Dirichlet b.c.\@, we report them in Fig.~\ref{fig:DoubleCone.Mm.Massive.Robin.Conjugate},
which is discussed in Appendix~\ref{appx:AdditionalNumericalResults}.

\section{Conclusions}
\label{sec:Conclusion}

In this paper we studied numerically 
the modular Hamiltonian of an interval $A$ of length $\ell$ 
for the massive scalar on the half line and in its ground state,
when Robin b.c.\@ are imposed at the origin of the half line, 
which include the Neumann b.c.\@ and the Dirichlet b.c.\@ 
as limiting cases.
Our work extends the numerical approach developed in \cite{BostelmannCadamuroMinz:2023} to the case where a boundary occurs and therefore the spacetime is not invariant under spatial translations. 
Denoting by $d$ the distance of $A$ from the boundary, 
we have considered both the cases given by $d=0$ and finite $d>0$
(see Sec.~\ref{sec:Results.AdjacentInterval} 
and Sec.~\ref{sec:Results.SeparatedInterval} respectively), 
which are qualitatively very different.  
The numerical method, 
which is based on the analytic expressions
\eqref{eq:BosonHilbertSpace.ModularHamiltonian}--\eqref{eq:BosonHilbertSpace.ModularHamiltonian.Blocks.Relation}
discussed in  
\cite{RieffelVanDaele:1977, FiglioliniGuido:1989, FiglioliniGuido:1994, CiolliLongoRuzzi:2019, BostelmannCadamuroMinz:2023, Longo:2022},
has been described in Sec.~\ref{sec:HalfLine.NumericalMethod},
where we highlighted the key steps in Fig.~\ref{fig:MethodScheme}.

For the massless scalar field with conformally invariant b.c.\@ 
(namely either Dirichlet or Neumann b.c.\@)
and the interval adjacent to the boundary,
the modular Hamiltonian is a local operator with kernel components \eqref{eq:LeftWedge.BCFTReference}
and \eqref{eq:LeftWedge.ModularHamiltonian.Mp} 
(see Appendix~\ref{appx:ModularHamiltonian.BCFT.Mp}),
which has been  obtained through a BCFT analysis \cite{CardyTonni:2016} 
and it has been employed 
as a benchmark of the numerical procedure. 
While a very good agreement has been found for the Dirichlet b.c.\@,
for the Neumann b.c.\@ we expect that higher values of the discretization parameter 
are needed to converge on the BCFT result,
as shown in Fig.~\ref{fig:LeftWedge.Mm.Discretization.Diagonal}.
We remind that, in the massless regime, 
there is a zero mode for Neumann b.c.\@,
while it is not present when either Dirichlet b.c.\@ 
or Robin b.c.\@ are imposed. 
Instead, in the massive regime, the zero mode does not occur 
in general. 
It would be interesting to explore quantitatively the role played by 
the zero mode in determining the rate of convergence to the BCFT result. 
Our analytic expressions for the Neumann b.c.\@ 
agree with the ones for Robin b.c.\@
in the limit $\eta \to 0^+$ and $m \to 0^+$.
For the massless scalar field with Robin b.c.\@ 
and the interval adjacent to the boundary,
our numerical results in 
Fig.~\ref{fig:LeftWedge.Mm.Boundaries.Diagonals} and 
Fig.~\ref{fig:LeftWedge.errMm.Boundaries.Antidiagonal} 
indicate that the modular Hamiltonian is non-local. 
It would be very insightful to support further 
this prediction through other methods. 
In the massive regime, from the numerical results reported 
in Fig.~\ref{fig:LeftWedge.Mm.Massive.Diagonals}, 
we infer that, independently of the b.c.\@,
the modular Hamiltonian of the interval adjacent to the boundary
is non-local.
The same figure suggests also that,
independently of the b.c.\@,
this modular Hamiltonian becomes local
for large mass with a kernel given by \eqref{eq:LeftWedge.LinearReference} and that, 
interestingly, its weight function appears to provide 
an upper bound for 
the quantity displayed in 
Fig.~\ref{fig:LeftWedge.Mm.Massive.Diagonals},  
which depends on the mass in a non-trivial way.

We also investigated the
modular Hamiltonian of an interval at finite distance from the boundary.
Our numerical results for this operator 
have been obtained by considering the white and red crosses 
in Fig.~\ref{fig:DoubleCone.Mm.Massless.SmearedMatrices}.
For the massless scalar,
they are shown in Fig.~\ref{fig:DoubleCone.Mm.Boundaries.Diagonal}
and Fig.~\ref{fig:DoubleCone.Mm.Boundaries.Conjugate}
respectively,
while for the massive regime 
they are reported in 
Fig.~\ref{fig:DoubleCone.Mm.Massive.Diagonal}
and Fig.~\ref{fig:DoubleCone.Mm.Massive.Conjugate} respectively.
Our analysis strongly indicates that 
the modular Hamiltonian of an interval at finite distance from the boundary 
for the scalar is non-local, independently of the 
finite value of the mass 
and of the b.c.\@ imposed at the origin of the half line.

The numerical procedure allows to explore also 
the spatial domain $B$ complementary to the interval in the half line. 
When the separation distance $d$ is large, 
the role of the b.c.\@ becomes evident 
in the part of $B$ between the boundary and $A$,
as one can observe in the bottom right panel
of Fig.~\ref{fig:DoubleCone.Mm.Boundaries.Diagonal}
and the bottom panels of 
Fig.~\ref{fig:DoubleCone.Mm.Boundaries.Conjugate},
Fig.~\ref{fig:DoubleCone.Mm.Massive.Diagonal}
and Fig.~\ref{fig:DoubleCone.Mm.Massive.Conjugate}.
However, within the interval $A$, the effects of the boundary 
become less relevant as $d$ increases.

The results presented in this work can be extended in various directions. 
It would be insightful to adapt the numerical procedure employed in our analysis 
to evaluate other interesting quantities that are heavily influenced by the choice of the b.c.\@ and the occurrence of a non-vanishing mass,
like the entanglement spectra 
\cite{Laeuchli:2013, CardyTonni:2016, AlbaCalabreseTonni:2018, SuraceTagliacozzoTonni:2020, RoyLukyanovSaleur:2025}
and the entanglement entropies
\cite{CalabreseCardy:2004, BerthiereSolodukhin:2016, EstienneIkhlefRotaruTonni:2024}. 
It is worth studying numerically the modular Hamiltonians 
of the union of two or more disjoint intervals 
for simple models like the massless compact boson and the Ising model,
where analytic expressions for the entanglement entropies are available \cite{CalabreseCardyTonni:2009, CalabreseCardyTonni:2011, CoserTagliacozzoTonni:2014, GravaKelsTonni:2021}, 
but their replica limit providing the entanglement entropy 
is difficult to perform analytically
(for which also another numerical approach based on rational extrapolations has been developed \cite{AgonHeadrickJafferisSkyler:2014, DeNobiliCoserTonni:2015}).
It would be interesting also to explore natural 
generalisations of our analysis 
where the scalar field is either on the segment
\cite{TonniRodriguezLagunaSierra:2018,TonniTrezzi:2025}, 
in a higher dimensional space \cite{Cardy:2013,Shiba:2012},
in curved spacetimes \cite{TonniRodriguezLagunaSierra:2018, TonniTrezzi:2025, Froeb:2025}
or in an out-of-equilibrium setup
\cite{CardyTonni:2016, DiGiulioAriasTonni:2019, RottoliRylandsCalabrese:2025, BonsignoriEisler:2025}.
It is worth mentioning the possibility to extend our numerical approach to explore the bipartite entanglement of mixed state, 
e.g.\@ by considering the logarithmic negativity 
\cite{Peres:1996, VidalWerner:2002, Plenio:2005, CalabreseCardyTonni:2012, CalabreseCardyTonni:2013, CalabreseCardyTonni:2015, CoserTonniCalabrese:2015, CoserTonniCalabrese:2016, MurcianoVitaleDalmonteCalabrese:2022, RottoliMurcianoTonniCalabrese:2023}.
Finally, interesting insights about the modular Hamiltonians 
could come from the gauge/gravity correspondence, 
where the entanglement entropy has been largely explored 
\cite{RyuTakayanagi:2006a, RyuTakayanagi:2006b} 
and connections have been found 
\cite{MintchevTonni:2022, CaggioliGentileSeminaraTonni:2024}
between the geodesic bit threads \cite{FreedmanHeadrick:2017,AgonDeBoerPedraza:2019} 
and the modular conjugation \cite{Haag:1996,HislopLongo:1982}.

\vskip 20pt
\centerline{\bf Acknowledgments} 
\vskip 5pt

We are very grateful to Mihail Mintchev for many important discussions,
suggestions, constant encouragement and useful comments on the draft.
ET acknowledges the Isaac Newton Institute (Cambridge),
within the program {\it Quantum field theory with boundaries, impurities, and defects},
and the Yukawa Institute for Theoretical Physics (Kyoto),
within the workshop {\it Extreme Universe 2025} (YITP-T-25-01)
and the long-term workshop {\it Progress of Theoretical Bootstrap},
for hospitality and financial support 
during the last part of this work.

This work was funded by the European Union -- NextGenerationEU, Mission 4, Component 2, Inv.1.3, in the framework of the PNRR Project National Quantum Science and Technology Institute (NQSTI) PE00023; CUP: G93C22001090006.

\vskip 20pt

\appendix
\addtocontents{toc}{\protect\setcounter{tocdepth}{1}}

\section{Details on the Hilbert spaces and the standard subspaces}
\label{appx:HilbertSpaces}

In this appendix we provide further technical details 
about the construction discussed in Sec.~\ref{sec:HalfLine.OperatorDModularHamiltonian}.
In particular, a complex Hilbert space $\mathcal{H}$ is defined from the real Hilbert spaces for the two pieces of initial data on the initial time hypersurface (the half line). 

Consider first the real Hilbert space $\mathcal{H}_{\labelReal} = \Lp2\bigl( [0, \infty), \Reals \bigr)$ with standard inner product, 
that describes fields on the hypersurface at $t = 0$.
We consider the closed subspace 
$\mathcal{H}_{\labelReal, A} = \Lp2\bigl( A, \Reals \bigr) \subset \mathcal{H}_{\labelReal}$ 
of functions with support on a region $A \subset [0, \infty)$, and its orthogonal complement given by
\begin{equation}
\label{eq:SubspaceComplement}
    \mathcal{H}_{\labelReal, A}^{\perp}
  \coloneq \bigl\{
      h \in \mathcal{L}
    \bigm\vert
      \forall f \in \mathcal{H}_{\labelReal}:
      \innerProd[\labelReal]{f}{h} = 0
    \bigr\}
  \eqend{.}
\end{equation}
Note that $D$ is a positive and unbounded operator on $\mathcal{H}_{\labelReal}$, and we define the Hilbert spaces $\mathcal{H}_{\labelReal}^{\pm1/4}$ as in \eqref{eq:RealHilbertSpaces.FromDomains}.
Then the subspace and its orthogonal complement become
\begin{equation}
\label{eq:RealHilbertSpaces.Subspaces.FromDomains}
    \mathcal{H}_{\labelReal, A}^{\pm\frac{1}{4}}
  \coloneq \closure{\mathcal{H}_{\labelReal, A} \cap \dom D^{\pm\frac{1}{4}}}^{\left\| \cdot \right\|_{\pm\frac{1}{4}}}
  \subset \mathcal{H}_{\labelReal}^{\pm\frac{1}{4}}
  \eqend{,}
\qquad
    \mathcal{H}_{\labelReal, A}^{\perp,\pm\frac{1}{4}}
  \coloneq \closure{\mathcal{H}_{\labelReal, A}^{\perp} \cap \dom D^{\pm\frac{1}{4}}}^{\left\| \cdot \right\|_{\pm\frac{1}{4}}}
  \subset \mathcal{H}_{\labelReal}^{\pm\frac{1}{4}}
  \eqend{.}
\end{equation}
The three conditions in Def.~2.1 of \cite{BostelmannCadamuroMinz:2023} 
are fulfilled so that $\mathcal{H}_A = \mathcal{H}_{\labelReal, A}^{+1/4} \oplus \mathcal{H}_{\labelReal, A}^{-1/4}$ (with the symplectic complement $\mathcal{H}_A' = \mathcal{H}_{\labelReal, A}^{\perp,+1/4} \oplus \mathcal{H}_{\labelReal, A}^{\perp,-1/4}$) is a standard and factorial subspace (see Lemma~2.2 in \cite{BostelmannCadamuroMinz:2023}).
The proof of this property follows the same argument as in the four-dimensional case without a boundary \cite{FiglioliniGuido:1989};
indeed, the expressions for $D^{-1/2}$ splits into the two parts given in \eqref{eq:FractionalModifiedHelmholtz.HalfLine.MomentumIntegral.Split}, where the first contribution \eqref{eq:FractionalModifiedHelmholtz.HalfLine.MinkowskianPart} is the same non-local operator as in the case where the initial time hypersurface is the entire real line, 
while the second term \eqref{eq:FractionalModifiedHelmholtz.HalfLine.BoundaryPart}, 
which depends on the boundary condition, is regular.

\section{Green's function on the half line with Robin b.c.}
\label{appx:GreensFunctions}

In this appendix, we discuss the computation 
of the Green functions of the scalar field
satisfying Robin b.c.\@ \eqref{eq:HalfLine.RobinBoundaryCondition},
by considering \eqref{eq:FractionalModifiedHelmholtz.HalfLine.MomentumIntegral.Definition}
in the special case of $\nu =-1$.
These results can be found also through other 
standard techniques \cite{MelnikovMelnikov:2012, Duffy:2015}.

\begin{figure}
  \centering
  \begin{subfigure}[t]{0.45\textwidth}
    \centering
    \includegraphics{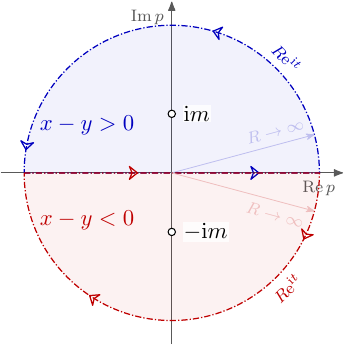}
    \caption{\label{fig:InverseHelmholtz.Minkowskian.Contour} 
    Semi-circle contour in the upper and lower half plane 
    when $x - y > 0$ and $x - y < 0$ respectively,
    occurring in $D^{-1}_{\labelMink}(x, y)$, see \eqref{eq:InverseHelmholtz.Minkowskian.Computation}.}
  \end{subfigure}
  \hfill
  \begin{subfigure}[t]{0.45\textwidth}
    \centering
    \includegraphics{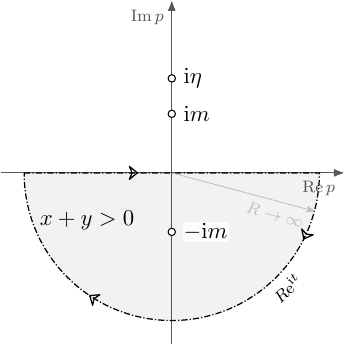}
    \caption{\label{fig:InverseHelmholtz.Boundary.Contour} 
    Semi-circle contour in the lower half plane, 
    since $x + y > 0$, in $D^{-1}_{\textlabel{bdy}}$, see \eqref{eq:InverseHelmholtz.Boundary.Robin}.
    For Neumann or Dirichlet b.c.\@, 
    the pole at $\i \eta$ does not occur.}
  \end{subfigure}
  \caption{\label{fig:InverseHelmholtz.Contour} Contours and poles in the complex momentum plane occurring in the  Fourier-type integral of $D^{-1}$ (inverse of the modified Helmholtz operator) for the massive regime.}
\end{figure}

In the massless regime, 
the term $D^{-1}_{\labelMink}$ as defined in \eqref{eq:FractionalModifiedHelmholtz.HalfLine.MinkowskianPart} has a double pole at $p = 0$, which is a tempered distribution with Fourier transform (see Ch.~II, Sec.~2.3, Eq.~(20) of \cite{GelfandShilov:1964}),
\begin{eqnarray}
\label{eq:InverseLaplace.Minkowskian}
    D^{-1}_{\labelMink}(x, y)
  &=& \frac{1}{2 \pi}
    \int_{-\infty}^{\infty}
      \frac{\e^{\i p (x - y)}}{p^2}
    \id{p}
  \,=\, - \frac{\abs{x - y}}{2}
  \eqend{.}
\end{eqnarray}
As for the term $D^{-1}_{\textlabel{bdy}}$ defined in \eqref{eq:FractionalModifiedHelmholtz.HalfLine.BoundaryPart} in the massless regime, we first integrate by parts, finding
\begin{eqnarray}
    D^{-1}_{\textlabel{bdy}}(x, y)
  &=& \frac{1}{2 \pi} \mult
    \int_{-\infty}^{\infty}
      \frac{1}{p^2}
      \mult \frac{p + \i \eta}{p - \i \eta}
      \mult \e^{-\i p (x + y)}
    \id{p}
\nonumber\\
  &=& \frac{1}{2 \pi}
    \left(
      - \frac{1}{p}
      \mult \frac{p + \i \eta}{p - \i \eta}
      \mult \e^{-\i p (x + y)}
    \right)_{-\infty}^{\infty}
    + \frac{1}{2 \pi}
    \int_{-\infty}^{\infty}
      \frac{1}{p} \mult
      \partial_p \biggl(
        \frac{p + \i \eta}{p - \i \eta}
        \mult \e^{-\i p (x + y)}
      \biggr)
    \id{p}
\nonumber\eqend{.}
\end{eqnarray}
Then, taking the derivative and performing 
a partial fraction decomposition, we get
\begin{eqnarray}
\label{eq:InverseLaplace.Boundary.Robin}
    D^{-1}_{\textlabel{bdy}}(x, y)
  &=& - \frac{\i (x + y)}{2 \pi}
    \int_{-\infty}^{\infty}
      \left(
        \frac{\i}{\eta}
        - \frac{1}{p}
        - \frac{\i}{\eta}
        \mult \frac{p + \i \eta}{p - \i \eta}
      \right)
      \e^{-\i p (x + y)}
    \id{p}
  \nonumber
\\
  & & {} + \frac{\i}{\pi \eta}
    \int_{-\infty}^{\infty}
      \left(
        \frac{1}{p}
        - \frac{1}{p - \i \eta}
        + \frac{\i \eta}{(p - \i \eta)^2}
      \right)
      \e^{-\i p (x + y)}
    \id{p}
  \nonumber
\\
  &=& \left( \frac{x + y}{2} + \frac{1}{\eta} \right)
    \frac{2}{\pi} \int_{0}^{\infty}
      \frac{\sin\bigl( (x + y) p \bigr)}{p}
    \id{p}
  \,=\, \frac{x + y}{2}
    + \frac{1}{\eta}
  \eqend{.}
\end{eqnarray}
Note that all distributional integrals without the $p^{-1}$ factor vanish and we take the remaining integral as a contour integral along a semi-circle in the lower half plane that does not cover the pole at $p = \i \eta$, which is identical to the contour shown in Fig.~\ref{fig:InverseHelmholtz.Boundary.Contour} but without the poles at $\pm \i m$.
Combining \eqref{eq:InverseLaplace.Minkowskian} and 
\eqref{eq:InverseLaplace.Boundary.Robin},
one obtains the same result found
through standard techniques for Green's functions 
(see e.g.\@ Ex.~1.3 in \cite{MelnikovMelnikov:2012}).
In the case of a Neumann ($s_{\textlabel{bdy}} = 1$) or a Dirichlet boundary ($s_{\textlabel{bdy}} = -1$), the integral for $D^{-1}_{\textlabel{bdy}}(x, y)$ reduces to one that is completely analogous to \eqref{eq:InverseLaplace.Minkowskian},
namely
\begin{equation}
\label{eq:InverseLaplace.Boundary.Special}
    D^{-1}_{\textlabel{bdy}}(x, y)
  = \frac{s_{\textlabel{bdy}}}{2 \pi}
    \int_{-\infty}^{\infty}
      \frac{\e^{-\i p (x + y)}}{p^2}
    \id{p}
  = -s_{\textlabel{bdy}} \mult \frac{x + y}{2}
  \eqend{.}
\end{equation}
The Dirichlet case agrees with \eqref{eq:InverseLaplace.Boundary.Robin} 
when taking the limit $\eta \to \infty$ and it can also be derived with standard techniques, cf.\@ Ex.~1.1 in \cite{MelnikovMelnikov:2012}. 
However, \eqref{eq:InverseLaplace.Boundary.Robin} is divergent in the Neumann limit $\eta \to 0^+$.
Indeed, 
since any constant function $\psi(x) = \const$ 
is a solution to the field equation \eqref{eq:StaticKleinGordon} 
and the b.c.\@ \eqref{eq:HalfLine.RobinBoundaryCondition}, 
a unique Green's function for Neumann b.c.\@ does not exist, cf.\@ Ex.~1.2 in \cite{MelnikovMelnikov:2012}.

In the massive regime,
the term $D^{-1}_{\labelMink}(x, y)$  
can be found via contour integration along a semi-circle $C$ 
either in the upper or lower half of the complex $p$-plane 
for $x - y > 0$ (upper sign, positive orientation) and $x - y < 0$ (lower sign, negative orientation), respectively
(see  Fig.~\ref{fig:InverseHelmholtz.Minkowskian.Contour}),
as follows
\begin{eqnarray}
\label{eq:InverseHelmholtz.Minkowskian.Computation}
    D^{-1}_{\labelMink}(x, y)
  &=&
    \frac{1}{2 \pi}
    \left(
      \oint_{C} k(p) \id{p}
      - \lim_{R \to \infty} \int_{0}^{\pm \pi}
        k\bigl( R \e^{\i t} \bigr) \mult \i R \mult \e^{\i t}
      \id{t}
    \right)
  \eqend{,}
  \qquad
    k(p)
  \,\coloneq\, \frac{\e^{\i p (x - y)}}{p^2 + m^2}
  \eqend{,}
\nonumber\\
  &=& \pm \i \lim_{p \to \pm \i m}
    \frac{\e^{\i p (x - y)}}{p \pm \i m}
  \eqend{.}
\end{eqnarray}
where we used Jordan's lemma for the integral along the arc.
For $x = y$, the integral becomes
\begin{equation}
\label{eq:InverseHelmholtz.Minkowskian.Computation.Diagonal}
    D^{-1}_{\labelMink}(x, x)
  = \frac{1}{2 \pi}
    \int_{-\infty}^{\infty}
      \frac{\id{p}}{p^2 + m^2}
  =  \left. \frac{1}{2 \pi m} \arctan\left( \frac{p}{m} \right)
    \right\rvert_{-\infty}^{\infty}
  = \frac{1}{2 m}
  \eqend{.}
\end{equation}
Putting all results together, we have
\begin{equation}
\label{eq:InverseHelmholtz.Minkowskian}
    D^{-1}_{\labelMink}(x, y)
  = \frac{\e^{- m \abs{x - y}}}{2 m}
  \eqend{.}
\end{equation}

The term $D^{-1}_{\textlabel{bdy}}(x, y)$, defined in \eqref{eq:FractionalModifiedHelmholtz.HalfLine.BoundaryPart}, 
specified for either the Neumann b.c.\@ ($s_{\textlabel{bdy}} = 1$) or the Dirichlet b.c.\@ ($s_{\textlabel{bdy}} = -1$) is computed analogously to \eqref{eq:InverseHelmholtz.Minkowskian}, 
and the result reads
\begin{equation}
\label{eq:InverseHelmholtz.Boundary.Special}
    D^{-1}_{\textlabel{bdy}}(x, y)
  = \frac{s_{\textlabel{bdy}}}{2 \pi}
    \int_{-\infty}^{\infty}
      \frac{\e^{-\i p (x + y)}}{p^2 + m^2}
    \id{p}
  = s_{\textlabel{bdy}} \mult \frac{\e^{- m (x + y)}}{2 m}
  \eqend{.}
\end{equation}
For a generic Robin b.c.\@, we perform a similar integral, 
again with $x + y > 0$, and we close the contour 
$C$ in the lower half plane, as shown in Fig.~\ref{fig:InverseHelmholtz.Boundary.Contour}.
In particular, we introduce
\begin{equation}
    k(p)
  \coloneq \frac{1}{p^2 + m^2}
    \mult \frac{p + \i \eta}{p - \i \eta}
    \mult \e^{- \i p (x + y)}
  \eqend{,}
\end{equation}
to use in the contour integration:
\begin{eqnarray}
\label{eq:InverseHelmholtz.Boundary.Robin}
    D^{-1}_{\textlabel{bdy}}(x, y)
  &=& \frac{1}{2 \pi}
    \left(
      \ointclockwise_{C} k(p) \id{p}
      - \lim_{R \to \infty} \int_{0}^{-\pi}
        k\bigl( R \e^{\i t} \bigr) \mult \i R \mult \e^{\i t}
      \id{t}
    \right)
\nonumber\\
  &=& - \i 
  \lim_{p \to - \i m}
    \frac{1}{p - \i m}
    \mult \frac{p + \i \eta}{p - \i \eta}
    \mult \e^{- \i p (x + y)}
  \,=\, \frac{1}{2 m}
    \mult \frac{m - \eta}{m + \eta}
    \mult \e^{- m (x + y)}
  \eqend{,}
\end{eqnarray}
which can also be found through standard techniques
(cf.\@ Ex.~1.4 \cite{MelnikovMelnikov:2012}).
The special cases of Neumann and Dirichlet b.c.\@ in \eqref{eq:InverseHelmholtz.Boundary.Special} agree 
with the limits $\eta \to 0^+$ and $\eta \to \infty$ of \eqref{eq:InverseHelmholtz.Boundary.Robin}, respectively.

In the massless limit $m \to 0^+$, note that the full kernel $D^{-1}_{\labelMink}(x, y) + D^{-1}_{\textlabel{bdy}}(x, y)$, 
that is the sum of \eqref{eq:InverseHelmholtz.Minkowskian} 
with 
\eqref{eq:InverseHelmholtz.Boundary.Special} or \eqref{eq:InverseHelmholtz.Boundary.Robin}  
for the Dirichlet or the Robin b.c.\@ respectively,  
takes the indeterminate form $\infty - \infty$;
hence we bring it to the form $0 / 0$ first 
and then use L'H\^opitals' rule to recover the corresponding combination of \eqref{eq:InverseLaplace.Minkowskian} with \eqref{eq:InverseLaplace.Boundary.Robin} or \eqref{eq:InverseLaplace.Boundary.Special}.
In the case of the  Neumann b.c.\@, the analogous limit diverges because there is no unique Green's function due to the arbitrary constant term (zero mode), as mentioned above.

\section{Complex structure and two-point functions}
\label{appx:ComplexStructure.TwoPointFunctions}

The complex structure $I$ in \eqref{eq:BosonHilbertSpace.ComplexStructure} 
on the complex Hilbert space
is related to the two-point function, 
as shown in \cite{Froeb:2025}.
In this appendix, we first compute the operator kernels of $D^{-1/2}$ and $D^{+1/2}$, which are the upper right  
and lower left block of the complex structure respectively.
This computation is discussed in the massless and massive regime separately (see Sec.~\ref{appx:ComplexStructure.Massless} 
and Sec.~\ref{appx:ComplexStructure.Massive} respectively).
In Sec.~\ref{appx:ComplexStructure.TwoPointFunctions.Comparison},
the resulting expressions are compared with existing results
in the literature \cite{LiguoriMitchev:1998,MintchevPilo:2001}

\subsection{The complex structure in the massless regime}
\label{appx:ComplexStructure.Massless}

In the case of the massless bosons, 
the integral kernel of $D^{-1/2}_{\labelMink}$ is divergent everywhere; hence a regularisation parameter $\mu > 0$ must be introduced in order to have a finite result. In particular, we have 
\begin{eqnarray}
\label{eq:InverseRootLaplace.Minkowskian}
    D^{-\frac{1}{2}}_{\labelMink}(x, y)
  &=& \frac{1}{2 \pi}
    \int_{-\infty}^{\infty}
      \frac{\e^{\i p (x - y)}}{\abs{p}}
    \id{p}
  \,=\, \frac{1}{\pi}
    \int_{\mu}^{\infty}
      \frac{\cos\bigl( p \abs{x - y} \bigr)}{p}
    \id{p}
\nonumber\\
  &=& \frac{1}{\pi}
    \int_{\mu \abs{x - y}}^{\infty}
      \frac{\cos q}{q}
    \id{q}
  \,=\, - \frac{1}{\pi}
    \operatorname{Ci}\bigl( \mu \abs{x - y} \bigr)
\nonumber\\
  &=& - \frac{1}{\pi}
    \log\bigl( \mu \abs{x - y} \bigr)
    - \frac{\upgamma}{\pi}
    + \Ord\bigl( \mu^2 \bigr)
  \eqend{,}
\end{eqnarray}
where we have employed the cosine integral $\operatorname{Ci}$ 
(see \citeDLMFeq{6.2}{11}) 
and its series expansion for small values $\mu$, 
which has a logarithmic divergence at leading order 
and the Euler's constant $\upgamma$ occurs 
in the first subleading order.
As for $D^{-1/2}_{\textlabel{bdy}}$
(following from the definition \eqref{eq:FractionalModifiedHelmholtz.HalfLine.BoundaryPart} 
with $\nu = -1/2$)
in the case of Neumann and Dirichlet b.c.\@, 
the computation is analogous to the one reported above for $D^{-1/2}_{\labelMink}$ and gives
\begin{eqnarray}
\label{eq:InverseRootLaplace.Boundary.Special}
    D^{-\frac{1}{2}}_{\textlabel{bdy}}(x, y)
  &=& \frac{s_{\textlabel{bdy}}}{2 \pi}
    \int_{-\infty}^{\infty}
      \frac{\e^{-\i p (x + y)}}{\abs{p}} 
    \id{p}
  \,=\, - \frac{1}{\pi}
    \operatorname{Ci}\bigl( \mu (x + y) \bigr)
\nonumber\\
  &=& s_{\textlabel{bdy}} \left(
      - \frac{1}{\pi}
      \log\bigl( \mu (x + y) \bigr)
      - \frac{\upgamma}{\pi}
      + \Ord\bigl( \mu^2 \bigr)
    \right)
  \eqend{,}
\end{eqnarray}
where $s_{\textlabel{bdy}} = 1$ for the Neumann b.c.\@ and $s_{\textlabel{bdy}} = -1$ for the Dirichlet b.c.

While the regulator $\mu$ must be introduced to solve the integrals for the Neumann b.c.\@, this regularisation is not needed for the Dirichlet b.c. 
Indeed, in this case, 
the full kernel of the operator 
$D^{-1/2}(x, y)$ defined in \eqref{eq:FractionalModifiedHelmholtz.HalfLine.MomentumIntegral.Split}
is the Fourier sine transform given in~\citeDLMFtab{1.14}{3},
namely
\begin{eqnarray}
\label{eq:InverseRootLaplace.Dirichlet}
    D^{-\frac{1}{2}}(x, y)
  &=& \frac{1}{2 \pi}
    \int_{-\infty}^{\infty}
      \frac{1}{\abs{p}}
      \left(
        \e^{\i p (x - y)}
        - \e^{-\i p (x + y)}
      \right)
    \id{p}
\nonumber
\\
  &=& 
  \frac{2}{\pi}
    \int_{0}^{\infty}
      \frac{1}{\abs{p}}
      \sin(p x)
      \sin(p y)
    \id{p}
  \,=\, - \frac{1}{\pi}
    \log\left( \frac{\abs{x - y}}{x + y} \right)
  \eqend{.}
\end{eqnarray}

\begin{figure}[t!]
  \centering
  \begin{subfigure}[t]{0.45\textwidth}
    \centering
    \includegraphics{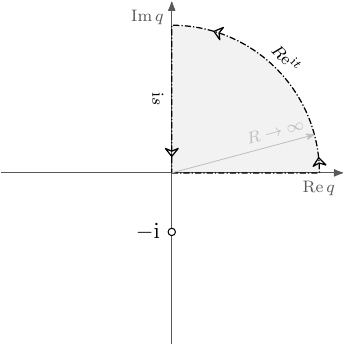}
    \caption{\label{fig:InverseRootLaplace.Contour} 
    Contour for an integral in \eqref{eq:InverseRootLaplace.Boundary.Robin.Computation}.}
  \end{subfigure}
  \hfill
  \begin{subfigure}[t]{0.45\textwidth}
    \centering
    \includegraphics{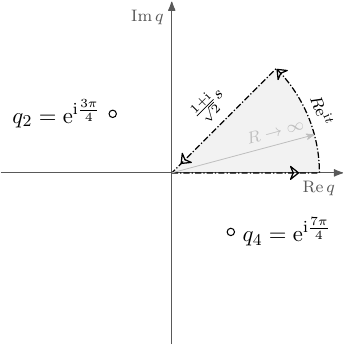}
    \caption{\label{fig:InverseFourthRootLaplace.Contour} 
    Contour for an integral in \eqref{eq:InverseFourthRootLaplace.Boundary.Robin.Computation}.
    For Robin b.c.\@, the two poles at $q_2$ and $q_4$ occur. 
    }
  \end{subfigure}
  \caption{\label{fig:InverseRootsLaplace.Contour} 
  Contours and poles for $D^{-\frac{1}{2}}_{\textlabel{bdy}}$ and $D^{-\frac{1}{4}}_{\textlabel{bdy}}$
  in the complex plane of the rescaled momentum, 
  in the left and right panel respectively.}
\end{figure}

Also for Robin b.c.\@ we can find a solution without introducing the regularization parameter $\mu$, as we have that
\begin{eqnarray}
\label{eq:InverseRootLaplace.Boundary.Robin.Computation}
    D^{-\frac{1}{2}}_{\textlabel{bdy}}(x, y)
  &=& 
  \frac{1}{2 \pi}
    \int_{-\infty}^{\infty}
      \frac{1}{\abs{p}}
      \mult \frac{p + \i \eta}{p - \i \eta}
      \mult \e^{- \i p (x + y)}
    \id{p}
\nonumber\\
  &=& - \frac{1}{2 \pi}
    \int_{-\infty}^{\infty}
      \frac{1}{\abs{p}}
      \mult \e^{- \i p (x + y)}
    \id{p}
    - \frac{1}{\pi}
    \int_{-\infty}^{0}
      \frac{\e^{- \i p (x + y)}}{p - \i \eta}
    \id{p}
    + \frac{1}{\pi}
    \int_{0}^{\infty}
      \frac{\e^{- \i p (x + y)}}{p - \i \eta}
    \id{p}
\nonumber\\
  &=& - \frac{1}{2 \pi}
    \int_{-\infty}^{\infty}
      \frac{1}{\abs{p}}
      \mult \e^{- \i p (x + y)}
    \id{p}
    + \frac{2}{\pi}
    \Re\int_{0}^{\infty}
      \frac{\e^{\i r q}}{q + \i}
    \id{q}
  \eqend{,}
  \qquad
    k(q)
  \,\coloneq\, \frac{\e^{\i r q}}{q + \i}
  \eqend{,}
\end{eqnarray}
where the first integral in the third line 
is identical to the expression found for the Dirichlet b.c.\@ in \eqref{eq:InverseRootLaplace.Boundary.Special}; hence it combines with $D^{-1/2}_{\labelMink}$ to yield the distribution \eqref{eq:InverseRootLaplace.Dirichlet}.
We evaluate the  second integral in the third line by first introducing $r \coloneq \eta (x + y)$ and $q \coloneq p/\eta$
and then employing a complex integration with integrand $k(q)$
over the contour shown in Fig.~\ref{fig:InverseRootLaplace.Contour} 
as follows
\begin{eqnarray}
    \label{eq:InverseRootLaplace.Boundary.Robin.ContourIntegral}
    \int_{0}^{\infty} k(q) \id{q}
  &=&
    \ointctrclockwise_C k(q) \id{q}
    - \lim_{R \to \infty} \int_{0}^{\frac{\pi}{2}}
      k(R \e^{\i t})
      \mult \i R \mult \e^{\i t}
    \id{t}
    - \i \lim_{R \to \infty} \int_{R}^{0}
      k(\i s)
    \id{s}
\nonumber
\\
  &=&
    \int_{0}^{\infty}
      \frac{\e^{-r s}}{s + 1}
    \id{s}
  \eqend{,}
\end{eqnarray}
where,  in the intermediate expression,
the first two terms vanish, 
while the remaining real integral can be written 
through the exponential integral function $\Ei$
(see \citeDLMFeq{6.2}{2} and \citeDLMFeq{6.2}{6}) as 
\begin{equation}
\label{eq:InverseRootLaplace.Boundary.Robin.ContourIntegral.ExponentialIntegral}
    \int_{0}^{\infty}
      \frac{\e^{-r s}}{s + 1}
    \id{s}
  = \e^{r}
    \int_{r}^{\infty}
      \frac{\e^{-t}}{t}
    \id{t}
  = \e^{r} \operatorname{E}_1(r)
  = - \e^{r} \Ei(-r)
  \eqend{.}
\end{equation}
Thus, $D^{-1/2}$ for Robin b.c.\@ is the tempered distribution given by 
\begin{equation}
\label{eq:InverseRootLaplace.Boundary.Robin}
    D^{-\frac{1}{2}}(x, y)
  = - \frac{1}{\pi}
    \log\left( \frac{\abs{x - y}}{x + y} \right)
    - \frac{2}{\pi} \mult \e^{\eta (x + y)} \Ei\bigl( -\eta (x + y) \bigr)
  \eqend{.}
\end{equation}
In the limiting case of Dirichlet b.c.\@, 
this expression 
becomes \eqref{eq:InverseRootLaplace.Boundary.Special}
because 
$\e^r \Ei(-r)$ vanishes when $r \to \infty$.
As for the limiting case of the Neumann b.c.\@, 
taking the zeroth order of the series expansion 
at $\eta = 0$ of \eqref{eq:InverseRootLaplace.Boundary.Robin.ContourIntegral.ExponentialIntegral}, yields \citeDLMFeq{6.2}{7} 
\begin{equation}
\label{eq:InverseRootLaplace.ExponentialIntegral.Series0}
    - \e^{r} \Ei(-r)
  = - \log(r)
    - \upgamma
    + \Ord\bigl( \eta \bigr)
  \eqend{,}
\end{equation}
which diverges logarithmically as $\eta \to 0^+$; 
hence we have to identify $\eta = \mu$ to recover 
\eqref{eq:InverseRootLaplace.Boundary.Special}.

For the other component $D^{+1/2}$ of the complex structure, we first compute
the term $D^{+1/2}_{\labelMink}$, which is given by the distributional Fourier transform 
(see Ch.~II, Sec.~2.3, Eq.~(14) in \cite{GelfandShilov:1964}) and reads
\begin{equation}
\label{eq:RootLaplace.Minkowskian}
    D^{+\frac{1}{2}}_{\labelMink}(x, y)
  = \frac{1}{2 \pi}
    \int_{-\infty}^{\infty}
      \abs{p} \e^{\i p (x - y)}
    \id{p}
  = - \frac{1}{\pi (x - y)^2}
  \eqend{.}
\end{equation}
The contribution $D^{+1/2}_{\textlabel{bdy}}$ for the Neumann b.c.\@ 
($s_{\textlabel{bdy}} = 1$) 
and Dirichlet b.c.\@ ($s_{\textlabel{bdy}} = -1$) 
is provided by a similar Fourier integral,
\begin{equation}
\label{eq:RootLaplace.Boundary.Special}
    D^{+\frac{1}{2}}_{\textlabel{bdy}}(x, y)
  = \frac{s_{\textlabel{bdy}}}{2 \pi}
    \int_{-\infty}^{\infty}
      \abs{p} \e^{-\i p (x + y)}
    \id{p}
  = - \frac{s_{\textlabel{bdy}}}{\pi (x + y)^2}
  \eqend{.}
\end{equation}
The tempered distribution $D^{+1/2}_{\textlabel{bdy}}(x, y)$ 
for the Robin b.c.\@ follows from a very similar computation to the one discussed above for
$D^{-1/2}_{\textlabel{bdy}}(x, y)$.
We set $r \coloneq \eta (x + y)$, $q \coloneq \frac{p}{\eta}$ and integrate over the contour in Fig.~\ref{fig:InverseRootLaplace.Contour},
\begin{eqnarray}
\label{eq:RootLaplace.Boundary.Robin}
    D^{+\frac{1}{2}}_{\textlabel{bdy}}(x, y)
  &=& \frac{1}{2 \pi}
    \int_{-\infty}^{\infty}
      \abs{p}
      \frac{p + \i \eta}{p - \i \eta}
      \e^{- \i p (x + y)\abs{p}}
    \id{p}
  \,=\, \frac{\eta^2}{2 \pi}
    \int_{-\infty}^{\infty}
      \abs{q}
      \frac{q + \i}{q - \i}
      \e^{- \i r q}
    \id{q}
\nonumber\\
  &=& \frac{\eta^2}{\pi}
    \Biggl(
      \int_{0}^{\infty}
        q \mult \e^{-\varepsilon q}
        \cos(r q)
      \id{q}
      - \i \int_{0}^{\infty}
        \frac{q}{q + \i}
        \mult \e^{\i r q}
      \id{q}
      + \i \int_{0}^{\infty}
        \frac{q}{q - \i}
        \mult \e^{- \i r q}
      \id{q}
    \Biggr)
\nonumber\\
  &=& \frac{\eta^2}{\pi}
    \Biggl(
      \int_{0}^{\infty}
        q \cos(r q)
      \id{q}
      + 2 \int_{0}^{\infty}
        \sin(r q)
      \id{q}
      - 2 \Re \int_{0}^{\infty}
        \frac{\e^{\i r q}}{q + \i}
      \id{q}
    \Biggr)
\nonumber\\
  &=& - \frac{1}{\pi (x + y)^2}
    + \frac{2 \eta}{\pi (x + y)}
    + \frac{2 \eta^2}{\pi} \mult \e^{\eta (x + y)} \Ei\bigl( -\eta (x + y) \bigr)
  \eqend{.}
\end{eqnarray}
The exponential integral $\Ei$ is the result of a complex contour integration analogous to the one occurring in $D^{-1/2}$.
The series expansion of \eqref{eq:RootLaplace.Boundary.Robin}
as $\eta \to 0^+$ and $\eta \to \infty$ 
is given respectively by
\begin{equation}
\label{eq:RootLaplace.Boundary.Robin.Series}
    D^{+\frac{1}{2}}_{\textlabel{bdy}}(x, y)
  = -\frac{1}{\pi (x + y)^2} + \Ord\bigl( \eta \bigr)
  \eqend{,}
  \qquad
    D^{+\frac{1}{2}}_{\textlabel{bdy}}(x, y)
  = \frac{1}{\pi (x + y)^2} + \Ord\bigl( \eta^{-1} \bigr)
  \eqend{,}
\end{equation}
which agrees with \eqref{eq:RootLaplace.Boundary.Special} in the limiting cases of Neumann b.c.\@ 
($s_{\textlabel{bdy}} = 1$) 
and Dirichlet b.c.\@ ($s_{\textlabel{bdy}} = -1$).

\subsection{The complex structure in the massive regime}
\label{appx:ComplexStructure.Massive}

In the massive regime, 
the term $D^{-1/2}_{\labelMink}$ 
in expression \eqref{eq:FractionalModifiedHelmholtz.HalfLine.MomentumIntegral.Split} of $D^{-1/2}$ reads
\begin{equation}
\label{eq:InverseRootHelmholtz.Minkowskian}
    D^{-\frac{1}{2}}_{\labelMink}(x, y)
  = \frac{1}{2 \pi}
    \int_{-\infty}^{\infty}
      \frac{\e^{\i p (x - y)}}{\sqrt{p^2 + m^2}} 
    \id{p}
  = \frac{1}{\pi}
    \int_{0}^{\infty}
      \frac{\cos\bigl( \abs{x - y} p \bigr)}{\sqrt{p^2 + m^2}}
    \id{p}
  = \frac{\BesselK0\bigl( m \abs{x - y} \bigr)}{\pi}
  \eqend{,}
\end{equation}
where the final integral is the representation~\citeDLMFeq{10.32}{6} of the modified Bessel function of the second kind.

As for the term $D^{-1/2}_{\textlabel{bdy}}$,
for Neumann and Dirichlet b.c.\@
it is computed similarly, finding
\begin{equation}
\label{eq:InverseRootHelmholtz.Boundary.Special}
    D^{-\frac{1}{2}}_{\textlabel{bdy}}(x, y)
  = \frac{s_{\textlabel{bdy}}}{2 \pi}
    \int_{-\infty}^{\infty}
      \frac{\e^{- \i p (x + y)}}{\sqrt{p^2 + m^2}}
    \id{p}
  = s_{\textlabel{bdy}} \frac{\BesselK0\bigl( m (x + y) \bigr)}{\pi}
  \eqend{.}
\end{equation}
When Robin b.c.\@ are imposed, we find 
\begin{eqnarray}
\label{eq:InverseRootHelmholtz.Boundary.Robin}
    D^{-\frac{1}{2}}_{\textlabel{bdy}}(x, y)
  &=& \frac{1}{2 \pi}
    \int_{-\infty}^{\infty}
      \frac{p + \i \eta}{p - \i \eta}
      \mult \frac{\e^{- \i p (x + y)}}{\sqrt{p^2 + m^2}}
    \id{p}
\nonumber\\
  &=& \frac{1}{\pi}
    \int_{0}^{\infty}
      \frac{\cos\bigl( p (x + y) \bigr)}{\sqrt{p^2 + m^2}}
    \id{p}
    - \frac{\i \eta}{\pi}
    \int_{-\infty}^{\infty}
      \frac{1}{p + \i \eta}
      \mult \frac{\e^{\i p (x + y)}}{\sqrt{p^2 + m^2}}
    \id{p}
\nonumber\\
  &=& 
  \frac{1}{\pi} \Big[
  \BesselK0\bigl( m (x + y) \bigr) 
  - 
  2 B_0(\eta, m; x + y)
  \Big]
  \eqend{,}
\end{eqnarray}
where we used the generalization of Basset's integral 
\begin{equation}
\label{eq:InverseRootHelmholtz.BoundaryIntegral}
    B_0(\eta, m; x + y)
  = \eta \mult \e^{\eta (x + y)}
    \int_{x + y}^{\infty}
      \e^{-\eta t}
      \mult \BesselK0(m t)
    \id{t}
  \eqend{,}
\end{equation}
which is the $\nu = 0$ case of the expression \eqref{eq:BoundaryIntegral.DifferentialEquation.Solution}, 
as discussed in Appendix~\ref{appx:GeneralizedBassetIntegral}.
In the special case $\eta = m$, it has the closed form
\begin{equation}
\label{eq:InverseRootHelmholtz.BoundaryIntegral1}
    B_0(m, m; x + y)
  = m \mult (x + y) 
    \Bigl[
      \BesselK1\bigl( m (x + y) \bigr)
      - \BesselK0\bigl( m (x + y) \bigr)
    \Bigr]
  \eqend{.}
\end{equation}
In Appendix~\ref{appx:GeneralizedBassetIntegral} we show that
\begin{equation}
\label{eq:InverseRootHelmholtz.BoundaryIntegral.SpecialLimits}
    \lim_{\eta \to 0^+} B_0(\eta, m; x + y)
  = 0
  \eqend{,}
  \qquad
    \lim_{\eta \to \infty} B_0(\eta, m; x + y)
  = \BesselK0\bigl( m (x + y) \bigr)
  \eqend{.}
\end{equation}
This implies that, 
in the limiting cases of the Neumann b.c.\@ and Dirichlet b.c.\@,
the result \eqref{eq:InverseRootHelmholtz.Boundary.Robin} becomes  \eqref{eq:InverseRootHelmholtz.Boundary.Special}.
As for the massless limit of \eqref{eq:InverseRootHelmholtz.Boundary.Robin}, 
we recover the result \eqref{eq:InverseRootLaplace.Boundary.Robin}
by using \eqref{eq:BoundaryIntegral.MasslessLimit.ZeroIndex}.

Finally, we compute the operator $D^{+1/2}$ in the massive regime.
For the tempered distribution $D^{+1/2}_{\labelMink}(x, y)$,
we find 
\begin{eqnarray}
\label{eq:RootHelmholtz.Minkowskian}
    D^{+\frac{1}{2}}_{\labelMink}(x, y)
  &=& \frac{1}{\pi}
    \int_{0}^{\infty}
      \sqrt{p^2 + m^2} \mult \cos\bigl( \abs{x - y} p \bigr)
    \id{p}
  \,=\, - \frac{m^2}{\pi (x - y)^2}
    \int_{0}^{\infty}
      \frac{\cos\bigl( \abs{x - y} p \bigr)}{(p^2 + m^2)^{\frac{3}{2}}}
    \id{p}
\nonumber\\
  &=& - \frac{m \mult \BesselK1\bigl( m \abs{x - y} \bigr)}{\pi \abs{x - y}}
  \eqend{,}
\end{eqnarray}
where integration by parts has been used twice, and then Basset's integral~\citeDLMFeq{10.32}{11}.
The contribution $D^{+1/2}_{\textlabel{bdy}}$ for the Neumann or Dirichlet b.c.\@ are computed analogously, 
\begin{equation}
\label{eq:RootHelmholtz.Boundary.Special}
    D^{+\frac{1}{2}}_{\textlabel{bdy}}(x, y)
  = \frac{s_{\textlabel{bdy}}}{2 \pi}
    \int_{-\infty}^{\infty}
      \sqrt{p^2 + m^2}
      \mult \e^{- \i p (x + y)}
    \id{p}
  = - s_{\textlabel{bdy}}
    \mult \frac{m \mult \BesselK1\bigl( m (x + y) \bigr)}{\pi (x + y)}
  \eqend{.}
\end{equation}
When the Robin b.c.\@ is imposed, we get
\begin{eqnarray}
\label{eq:RootHelmholtz.Boundary.Robin}
    D^{+\frac{1}{2}}_{\textlabel{bdy}}(x, y)
  &=& \frac{1}{2 \pi}
    \int_{-\infty}^{\infty}
      \frac{p + \i \eta}{p - \i \eta}
      \mult \sqrt{p^2 + m^2} \mult \e^{- \i p (x + y)}
    \id{p}
\nonumber\\
  &=& - \frac{m^2}{\pi (x + y)^2}
    \int_{0}^{\infty}
      \frac{\cos\bigl( p (x + y) \bigr)}{(p^2 + m^2)^{\frac{3}{2}}}
    \id{p}
    - \frac{\i \eta}{\pi}
    \int_{-\infty}^{\infty}
      \frac{1}{p + \i \eta}
      \mult \frac{\e^{\i p (x + y)}}{(p^2 + m^2)^{-\frac{1}{2}}}
    \id{p}
\nonumber\\
  &=& - \frac{
      m \Bigl[
        \BesselK1\bigl( m (x + y) \bigr)
        - 2 B_{-1}(\eta, m, x + y)
      \Bigr]
    }{\pi \mult (x + y)}
  \eqend{,}
\end{eqnarray}
where we use the generalised Basset integral, which is equivalent to
\begin{equation}
\label{eq:RootHelmholtz.BoundaryIntegral}
    B_{-1}(\eta, m; x + y)
  = \eta \mult (x + y) \mult \e^{\eta (x + y)}
    \int_{x + y}^{\infty}
      \frac{\e^{-\eta t}}{t}
      \mult \BesselK1(m t)
    \id{t}
  \eqend{,}
\end{equation}
as discussed in Appendix~\ref{appx:GeneralizedBassetIntegral}.
In the special case $\eta = m$, we have
\begin{equation}
\label{eq:RootHelmholtz.BoundaryIntegral1}
    B_{-1}(m, m; x + y)
  = m (x + y) \Bigl[ \BesselK1\bigl( m (x + y) \bigr)
    - \BesselK0\bigl( m (x + y) \bigr)
    \Bigr]
  \eqend{.}
\end{equation}
As consistency check of \eqref{eq:RootHelmholtz.Boundary.Robin}, 
we consider again the limiting regimes given by Neumann b.c.\@ and Dirichlet b.c.\@, finding respectively that
\begin{equation}
\label{eq:RootHelmholtz.BoundaryIntegral.SpecialLimits}
    \lim_{\eta \to 0^+} B_{-1}(\eta, m; x + y)
  = 0
  \eqend{,}
  \qquad
    \lim_{\eta \to \infty} B_{-1}(\eta, m; x + y)
  = \BesselK1\bigl( m (x + y) \bigr)
  \eqend{,}
\end{equation}
as discussed in Appendix~\ref{appx:GeneralizedBassetIntegral}.
The massless limit of \eqref{eq:RootHelmholtz.Boundary.Robin}
is obtained by using 
\eqref{eq:BoundaryIntegral.MasslessLimit.NegativeIndex}
and the result reads
\begin{eqnarray}
\label{eq:RootHelmholtz.Boundary.Robin.MasslessLimit}
    \lim_{m \to 0^+}
      D^{+\frac{1}{2}}_{\textlabel{bdy}}(x, y)
  &=& - \frac{m}{\pi (x + y)^2}
    + \frac{2 \eta^2}{\pi}
    \mult \e^{\eta (x + y)}
    \mult \upGamma\bigl( -1, \eta (x + y) \bigr)
\nonumber\\
  &=& - \frac{m}{\pi (x + y)^2}
    + \frac{2 \eta}{\pi (x + y)}
    - \frac{2 \eta^2}{\pi}
    \mult \e^{\eta (x + y)}
    \mult \upGamma\bigl( 0, \eta (x + y) \bigr)
  \eqend{,}
\end{eqnarray}
which is equivalent to the result \eqref{eq:RootLaplace.Boundary.Robin} for the massless field
because of the identity \citeDLMFeq{6.11}{1}.

\subsection{Comparison of the complex structure and the two-point functions}
\label{appx:ComplexStructure.TwoPointFunctions.Comparison}

It is instructive to compare our results 
for the complex structure in the massless regime
with the expressions of the two-point functions reported 
in \cite{LiguoriMitchev:1998} and \cite{MintchevPilo:2001}.

The two-point functions in \cite{LiguoriMitchev:1998} are written
in terms of the following tempered distributions 
\begin{equation}
\label{eq:ChiralBoson.2PointFunction.Components.Distributions}
    u(\xi)
  \coloneq - \frac{1}{\pi} \log \abs{\xi}
    - \frac{\i}{2} \sgn \xi
  \eqend{,}
  \qquad
    v_{\pm}(\xi)
  \coloneq \lim_{\varepsilon \to 0^+} \frac{2}{\pi} \e^{-\xi}
    \Ei(\xi \pm \i \varepsilon)
  \eqend{,}
\end{equation}
and in light-cone coordinates 
$t_y - x$, $t_y - y$ for right-movers (index $\mathrm{R}$) 
and $t_x + x$, $t_y + y$ for left-movers (index $\mathrm{L}$).
In particular, we have the right--right component
\begin{equation}
\label{eq:ChiralBoson.2PointFunction.RRComponent}
    \bigl\langle \varphi_{\textlabel{R}}(t_x - x) \mult \varphi_{\textlabel{R}}(t_y - y) \bigr\rangle
  = u\bigl( \mu (t_x - x - t_y + y) \bigr)
  \eqend{,}
\end{equation}
the right--left component
\begin{equation}
\label{eq:ChiralBoson.2PointFunction.RLComponent}
    \bigl\langle \varphi_{\textlabel{R}}(t_x - x) \mult \varphi_{\textlabel{L}}(t_y + y) \bigr\rangle
  = \begin{dcases}
      u\bigl( \mu (t_x - x - t_y - y) \bigr)
    & \quad \eta = 0
    \eqend{,}\\
      - u\bigl( \mu (t_x - x - t_y - y) \bigr)
    \\\qquad{}
      - v_-\bigl( \eta (t_x - x - t_y - y) \bigr)
    & \quad \eta \in (0, \infty)
    \eqend{,}\\
      - u\bigl( \mu (t_x - x - t_y - y) \bigr)
    & \quad \eta \to \infty
    \eqend{,}
    \end{dcases}
\end{equation}
the left--right component
\begin{equation}
\label{eq:ChiralBoson.2PointFunction.LRComponent}
    \bigl\langle \varphi_{\textlabel{L}}(t_x + x) \mult \varphi_{\textlabel{R}}(t_y - y) \bigr\rangle
  = \begin{dcases}
      u\bigl( \mu (t_x + x - t_y + y) \bigr)
    & \quad \eta = 0
    \eqend{,}\\
      - u\bigl( \mu (t_x + x - t_y + y) \bigr)
    \\\qquad{}
      - v_+\bigl( -\eta (t_x + x - t_y + y) \bigr)
    & \quad  \eta \in (0, \infty)
    \eqend{,}\\
      - u\bigl( \mu (t_x + x - t_y + y) \bigr)
    & \quad \eta \to \infty
    \eqend{,}
    \end{dcases}
\end{equation}
and the left--left component
\begin{equation}
\label{eq:ChiralBoson.2PointFunction.LLComponent}
    \bigl\langle \varphi_{\textlabel{L}}(t_x + x) \mult \varphi_{\textlabel{L}}(t_y + y) \bigr\rangle
  = u\bigl( \mu (t_x + x - t_y - y) \bigr)
  \eqend{.}
\end{equation}

Since our analysis is based on the field $\varphi$ and its time derivative $\dot{\varphi}$ on the $t=0$ Cauchy hypersurface,
we first transform this set of two-point functions 
into the one involving the field $\varphi(t, x)$ and its dual fields $\tilde{\varphi}(t, x)$ \cite{LiguoriMitchev:1998}, i.e.\@
\begin{equation}
\label{eq:ChiralBoson.2PointFunction.Transformed}
    \begin{pmatrix}
      \langle \varphi \varphi \rangle
      & \langle \varphi \tilde{\varphi} \rangle \\
      \langle \tilde{\varphi} \varphi \rangle
      & \langle \tilde{\varphi} \tilde{\varphi} \rangle
    \end{pmatrix}
  = \frac{1}{4}
    \begin{pmatrix}
      1 & 1 \\
      1 & -1
    \end{pmatrix}
    \begin{pmatrix}
      \langle \varphi_{\textlabel{R}} \varphi_{\textlabel{R}} \rangle
      & \langle \varphi_{\textlabel{R}} \varphi_{\textlabel{L}} \rangle \\
      \langle \varphi_{\textlabel{L}} \varphi_{\textlabel{R}} \rangle
      & \langle \varphi_{\textlabel{L}} \varphi_{\textlabel{L}} \rangle
    \end{pmatrix}
    \begin{pmatrix}
      1 & 1 \\
      1 & -1
    \end{pmatrix}
  \eqend{,}
\end{equation}
and then consider only the element 
\begin{eqnarray}
\label{eq:ChiralBoson.2PointFunction.Transformed.FieldFieldComponent}
    \langle \varphi(t_x, x) \mult \varphi(t_y, y) \rangle
  &=& \frac{1}{4}
    \Bigl(
      \bigl\langle \varphi_{\textlabel{R}}(t_x - x) \mult \varphi_{\textlabel{R}}(t_y - y) \bigr\rangle
      + \bigl\langle \varphi_{\textlabel{R}}(t_x - x) \mult \varphi_{\textlabel{L}}(t_y + y) \bigr\rangle
\nonumber\\
  && \quad{}
      + \bigl\langle \varphi_{\textlabel{L}}(t_x + x) \mult \varphi_{\textlabel{R}}(t_y - y) \bigr\rangle
      + \bigl\langle \varphi_{\textlabel{L}}(t_x + x) \mult \varphi_{\textlabel{L}}(t_y + y) \bigr\rangle
    \Bigr)
  \eqend{,}
  \quad
\end{eqnarray}
which provides the following two-point functions at equal times
\begin{equation}
\label{eq:ChiralBoson.2PointFunction.EqualTime}
    \langle \varphi(x) \mult \varphi(y) \rangle
  = \lim_{t_y \to t_x} \langle \varphi(t_x, x) \mult \varphi(t_y, y) \rangle
  \eqend{,}
  \quad
    \langle \dot{\varphi}(x) \mult \dot{\varphi}(y) \rangle
  = \lim_{t_y \to t_x}
    \ppderiv{\langle \varphi(t_x, x) \mult \varphi(t_y, y) \rangle}{t_x}{t_y}
  \eqend{.}
\end{equation}
From \eqref{eq:ChiralBoson.2PointFunction.RRComponent}--\eqref{eq:ChiralBoson.2PointFunction.LLComponent}
and \eqref{eq:ChiralBoson.2PointFunction.Transformed.FieldFieldComponent}, we compute \eqref{eq:ChiralBoson.2PointFunction.EqualTime},
which yields
\begin{equation}
\label{eq:ChiralBoson.2PointFunction.EqualTime.FieldFieldComponent}
    \langle \varphi(x) \mult \varphi(y) \rangle
=
\begin{dcases}
      - \frac{1}{2 \pi} \log\bigl( \mu^2 \abs{x - y} (x + y) \bigr)
      & \quad \eta = 0
    \eqend{,}\\
      - \frac{1}{2 \pi} \log\biggl( \frac{\abs{x - y}}{x + y} \biggr)
      - \frac{1}{\pi} \mult \e^{\eta (x + y)} \Ei\bigl( - \eta (x + y) \bigr)
      & \quad \eta \in (0, \infty)
    \eqend{,}\\
      - \frac{1}{2 \pi} \log\biggl( \frac{\abs{x - y}}{x + y} \biggr)
      & \quad \eta \to \infty
    \eqend{,}
    \end{dcases}
\end{equation}
and
\begin{equation}
\label{eq:ChiralBoson.2PointFunction.EqualTime.VelocityVelocityComponent}
    \langle \dot{\varphi}(x) \mult \dot{\varphi}(y) \rangle
=
 \begin{dcases}
      - \frac{1}{2 \pi (x - y)^2}
      - \frac{1}{2 \pi (x + y)^2}
      & \quad \eta = 0
    \eqend{,}\\
      - \frac{1}{2 \pi (x - y)^2}
      - \frac{1}{2 \pi (x + y)^2}
      + \frac{\eta}{\pi (x + y)}
    \\\qquad{}
      + \frac{\eta^2}{\pi} \mult \e^{\eta (x + y)} \Ei\bigl( - \eta (x + y) \bigr)
      & \quad \eta \in (0, \infty)
    \eqend{,}\\
      - \frac{1}{2 \pi (x - y)^2}
      + \frac{1}{2 \pi (x + y)^2}
      & \quad \eta \to \infty
    \eqend{.}
    \end{dcases}
\end{equation}
Since the imaginary parts drop out in the computation,
these expressions are real.
Instead, the following two-point function at equal times is purely imaginary
\begin{equation}
\label{eq:ChiralBoson.2PointFunction.EqualTime.FieldVelocityComponent}
    \langle \varphi(x) \mult \dot{\varphi}(y) \rangle
  = \lim_{t_y \to t_x}
    \pderiv{\langle \varphi(t_x, x) \mult \varphi(t_y, y) \rangle}{t_y}
  = \frac{\i}{2} \mult \updelta(x - y)
  = -\langle \dot{\varphi}(x) \varphi(y) \rangle
  \eqend{,}
\end{equation}
in agreement with the canonical commutation relation
between $\varphi$ and $\dot{\varphi}$.

In Sec.~\ref{appx:ComplexStructure.Massive} we found that
\begin{equation}
\label{eq:ComplexStructure.Blocks}
    D^{\pm\frac{1}{2}}(x, y)
  = \begin{dcases}
      u^{\pm}_m(x - y)
      + u^{\pm}_m(x + y)
    & 
    \quad \eta = 0
    \eqend{,}
    \\
      u^{\pm}_m(x - y)
      \pm u^{\pm}_m(x + y)
      + v^{\pm}_m(x + y)
    & 
    \quad \eta \in (0, \infty)
    \eqend{,}
    \\
      u^{\pm}_m(x - y)
      - u^{\pm}_m(x + y)
    & 
    \quad \eta \to \infty
    \eqend{,}
  \end{dcases}
\end{equation}
in terms of the tempered distributions 
\begin{eqnarray}
\label{eq:ComplexStructure.Blocks.Distributions.um}
    u^-_m(x) 
  &\coloneq& \frac{\BesselK0\bigl( m \abs{x} \bigr)}{\pi}
  \,=\, - \frac{\log\bigl( m \abs{x} \bigr) + \upgamma}{\pi}
    + \Ord(m^2)
  \eqend{,}
\\
\label{eq:ComplexStructure.Blocks.Distributions.up}
    u^+_m(x) 
  &\coloneq& - \frac{m \mult \BesselK1\bigl( m \abs{x} \bigr)}{\pi \abs{x}}
  \,=\, - \frac{1}{\pi x^2}
    + \Ord(m^2)
  \eqend{,}
\\
\label{eq:ComplexStructure.Blocks.Distributions.vm}
    v^-_m(x)
  &\coloneq& \frac{1}{\pi}
    \int_{-\infty}^{\infty}
      \frac{p}{\sqrt{p^2 + m^2}}
      \mult \frac{\e^{-\i p x}}{p - \i \eta}
    \id{p}
  \,=\, - \frac{2}{\pi} \mult \e^{\eta x} \mult \Ei(-\eta x)
    + \Ord(m^2)
  \eqend{,}
\\
\label{eq:ComplexStructure.Blocks.Distributions.vp}
    v^+_m(x)
  &\coloneq& \frac{\i \eta}{\pi}
    \int_{-\infty}^{\infty}
      \frac{\sqrt{p^2 + m^2}}{p - \i \eta}
      \mult \e^{-\i p x}
    \id{p}
  \,=\, \frac{2 \eta}{\pi x}
    + \frac{2 \eta^2}{\pi} \mult \e^{\eta x} \mult \Ei(-\eta x)
    + \Ord(m^2)
  \eqend{,}
\end{eqnarray}
where, for each distribution, 
we also report the first terms of its small mass expansion. 
Indeed, these terms provide the corresponding massless results 
obtained in Sec.~\ref{appx:ComplexStructure.Massless}
(when we identify the mass parameter $m$ with the regulator $\mu$ 
in the case of Neumann b.c.\@).
Comparing the results of the massless two-point function \eqref{eq:ChiralBoson.2PointFunction.EqualTime.FieldFieldComponent} and \eqref{eq:ChiralBoson.2PointFunction.EqualTime.VelocityVelocityComponent} with the small mass expansion of the 
complex structure \eqref{eq:ComplexStructure.Blocks}, 
we find
\begin{equation}
\label{eq:ComplexStructure.TwoPointFunctions.Indentification}
    D^{-\frac{1}{2}}(x, y)
  = 2 \langle \varphi(x) \mult \varphi(y) \rangle
  \eqend{,}
  \qquad
    D^{+\frac{1}{2}}(x, y)
  = 2 \langle \dot{\varphi}(x) \mult \dot{\varphi}(y) \rangle
  \eqend{,}
\end{equation}
which are in agreement with the identification given 
in Eq.~(2.39) of \cite{Froeb:2025}.

In the massive regime, the two-point function of $\varphi$ at equal times has been written in Eq.~(27) of \cite{MintchevPilo:2001}\footnote{In \cite{MintchevPilo:2001}, the expression in Eq.~(29) becomes $W_{m^2}(x) = (2 m)^{-1}$ in $1 + 1$ dimensions.} as follows
\begin{equation}
\label{eq:Brane.2PointFunction.FieldFieldComponent}
    \bigl\langle \varphi(x) \mult \varphi(y) \bigr\rangle
  = \frac{1}{4 \pi}
    \int_{0}^{\infty}
      \frac{\conj{\psi}_p(x) \mult \psi_p(y) }{\sqrt{m^2 + p^2}}
    \id{p}
  \eqend{,}
\end{equation}
in terms of the modes $\psi_p$, given in \eqref{eq:HalfLine.MomentumIntegral.Mode}.
The expression \eqref{eq:Brane.2PointFunction.FieldFieldComponent}
is identical to the definition of $D^{-1/2}$ from \eqref{eq:FractionalModifiedHelmholtz.HalfLine.MomentumIntegral.Definition}, up to the factor 2 occurring in \eqref{eq:ComplexStructure.TwoPointFunctions.Indentification}.

\section{The fourth root of the Green function}
\label{appx:InverseFourthRoots}

The numerical algorithm that we employ in this work 
appoximates the modular Hamiltonian \eqref{eq:BosonHilbertSpace.ModularHamiltonian}, 
which is composed of the two operators $M_{\pm}$ given in \eqref{eq:BosonHilbertSpace.ModularHamiltonian.Blocks}.
These operators are
written in terms of the operator $Z$ in \eqref{eq:BosonHilbertSpace.ModularHamiltonian.ArcothArg},
which involves the subspace projector $\Theta_A$ 
and the operators $D^{\pm1/4}$.
In this appendix, we discuss the computation of $D^{-1/4}$ 
in both the massless (Sec.~\ref{appx:InverseFourthRoots.Massless}) and massive regime (Sec.~\ref{appx:InverseFourthRoots.Massive}).

\subsection{Massless regime}
\label{appx:InverseFourthRoots.Massless}

In the massless regime, the tempered distribution $D^{-1/4}_{\labelMink}(x, y)$,
defined by \eqref{eq:FractionalModifiedHelmholtz.HalfLine.MinkowskianPart}
with $\nu = -1/4$, 
can be obtained as in 
Ch.~II, Sec.~2.3, Eq.~(12) of \cite{GelfandShilov:1964}, 
namely
\begin{equation}
\label{eq:InverseFourthRootLaplace.Minkowskian}
    D^{-\frac{1}{4}}_{\labelMink}(x, y)
  = \frac{1}{2 \pi}
    \int_{-\infty}^{\infty}
      \frac{\e^{\i p (x - y)}}{\sqrt{\abs{p}}}
    \id{p}
  = \frac{\upGamma\bigl( 1 - \frac{1}{2} \bigr) \sin\bigl( \frac{\pi}{4} \bigr)}{\pi}
    \abs{x - y}^{\frac{1}{2} - 1}
  = \frac{1}{\sqrt{2 \pi \abs{x - y}}}
  \eqend{.}
\end{equation}
Integrating this kernel against a test function yields a finite result via integration by parts.

The term $D^{-1/4}_{\textlabel{bdy}}(x, y)$, 
defined by \eqref{eq:FractionalModifiedHelmholtz.HalfLine.BoundaryPart}
with $\nu = -1/4$,
can be written
by introducing the variables $r \coloneq \sqrt{\eta (x + y)}$ 
and $q \coloneq \sqrt{p/\eta}$ as follows 
\begin{eqnarray}
\label{eq:InverseFourthRootLaplace.Boundary.Robin.Computation}
    D^{-\frac{1}{4}}_{\textlabel{bdy}}(x, y)
  &=& \frac{1}{2 \pi}
    \int_{-\infty}^{\infty}
      \frac{p + \i \eta}{p - \i \eta}
      \mult \frac{\e^{- \i p (x + y)}}{\sqrt{\abs{p}}}
    \id{p}
\nonumber\\
  &=& \frac{2 \sqrt{\eta}}{\pi}
    \int_{0}^{\infty}
      \frac{1}{2}
      \Bigl( \e^{\i r^2 q^2} + \e^{- \i r^2 q^2} \Bigr)
    \id{q}
    + \frac{4 \sqrt{\eta}}{\pi}
    \int_{0}^{\infty}
      \frac{1}{2 \i}
      \Biggl(
        \frac{\e^{\i r^2 q^2}}{q^2 + \i}
        - \frac{\e^{- \i r^2 q^2}}{q^2 - \i}
      \Biggr)
    \id{q}
    \nonumber
\\
  &=& \frac{2 \sqrt{\eta}}{\pi r}
    \mult \Re\biggl( \int_{0}^{\infty} \e^{\i q^2} \id{q} \biggr)
    + \frac{4 \sqrt{\eta}}{\pi}
    \mult \Im\biggl( \int_{0}^{\infty} k(q) \id{q} \biggr)
  \eqend{,}
  \qquad
    k(q)
  \,\coloneq\, \frac{\e^{\i r^2 q^2}}{q^2 + \i}
  \eqend{,}
  \qquad
\end{eqnarray}
where the function $k$ has poles at $q_2 \coloneq \e^{3 \pi \i / 4}$ and $q_4 \coloneq \e^{7 \pi \i / 4}$.

To compute the integrals in \eqref{eq:InverseFourthRootLaplace.Boundary.Robin.Computation}, we introduce the pizza slice contour $C$ depicted in Fig.~\ref{fig:InverseFourthRootLaplace.Contour} 
that follows the half line, continues in a $45^\circ$ arc, and returns along the ray of the unit vector 
$q_1 \coloneq \e^{\pi \i / 4}$ back towards the origin.
For the first integral in \eqref{eq:InverseFourthRootLaplace.Boundary.Robin.Computation}, 
we get
\begin{equation}
\label{eq:Integral.ExpIXSquared}
    \int_{0}^{\infty} \e^{\i q^2} \id{q}
  = \underbrace{\ointctrclockwise_C \e^{\i q^2} \id{q}}_{0}
    - \underbrace{\lim_{R \to \infty} \int_{0}^{\frac{\pi}{4}}
      \e^{\i R^2 \e^{2 \i t}}
      \i R \e^{\i t}
    \id{t}}_{0}
    - \int_{\infty}^{0}
      \e^{- s^2}
      \frac{1 + \i}{\sqrt{2}}
    \id{s}
  = \frac{\pi}{2} \frac{1 + \i}{\sqrt{2 \pi}}
  \eqend{,}
\end{equation}
where the contour integral does not enclose a pole and the arc-integral vanishes in the limit $R \to \infty$.
This also happens for the second integral term in \eqref{eq:InverseFourthRootLaplace.Boundary.Robin.Computation} and for this we get
\begin{equation}
\label{eq:Integral.ExpIRXSquaredOverXSquaredI}
    \int_{0}^{\infty} k(q) \id{q}
  = - \int_{\infty}^{0}
      k(q_1 s) \mult q_1
    \id{s}
  = \frac{1 - \i}{\sqrt{2}}\,
    \frac{\pi}{2} \mult \e^{r^2}
    \left(
      \frac{2}{\pi} \mult \e^{-r^2}
      \int_{0}^{\infty}
        \frac{\e^{-r^2 s^2}}{s^2 + 1}
      \id{s}
    \right)
  \eqend{,}
\end{equation}
where the integral expression within the parenthesis is the error function $\erfc (r)$ (see \citeDLMFeq{7.7}{1}), namely
\begin{equation}
\label{eq:ComplementaryErrorFunction}
    \erfc (r)
  = 1 - \frac{2}{\sqrt{\pi}} \int_{0}^{r} \e^{-t^2} \id{t}
  = \frac{2}{\sqrt{\pi}} \int_{r}^{\infty} \e^{-t^2} \id{t}
  \eqend{.}
\end{equation}
Thus, \eqref{eq:InverseFourthRootLaplace.Boundary.Robin.Computation}
becomes
\begin{equation}
\label{eq:InverseFourthRootLaplace.Boundary.Robin}
    D^{-\frac{1}{4}}_{\textlabel{bdy}}(x, y)
  = \frac{1}{\sqrt{2 \pi (x + y)}}
    - \sqrt{2 \eta} \mult \e^{\eta (x + y)}
    \mult \erfc\biggl( \sqrt{\eta (x + y)}\, \biggr)
  \eqend{.}
\end{equation}
In the special cases of Neumann b.c.\@ 
($s_{\textlabel{bdy}} = 1$) 
and Dirichlet b.c.\@ ($s_{\textlabel{bdy}} = -1$), 
the general computation \eqref{eq:InverseFourthRootLaplace.Boundary.Robin.Computation} reduces to the real part of the integral \eqref{eq:Integral.ExpIXSquared}, namely
\begin{equation}
\label{eq:InverseFourthRootLaplace.Boundary.Special}
    D^{-\frac{1}{4}}_{\textlabel{bdy}}(x, y)
  = \frac{s_{\textlabel{bdy}}}{2 \pi}
    \int_{-\infty}^{\infty}
      \frac{\e^{- \i p (x + y)}}{\sqrt{\abs{p}}}
    \id{p}
  = \frac{2 \mult s_{\textlabel{bdy}}}{\pi \sqrt{(x + y)}}
    \Re \int_{0}^{\infty} \e^{\i s^2} \id{s}
  = \frac{s_{\textlabel{bdy}}}{\sqrt{2 \pi (x + y)}}
  \eqend{,}
\end{equation}
which agrees with the limits $\eta \to 0^+$ and $\eta \to \infty$ of \eqref{eq:InverseFourthRootLaplace.Boundary.Robin}, respectively.

\subsection{Massive regime}
\label{appx:InverseFourthRoots.Massive}

In the massive regime, the kernel $D^{-1/4}_{\labelMink}(x, y)$ 
can be written in terms of the modified Bessel function of the second kind as 
\begin{eqnarray}
\label{eq:InverseFourthRootHelmholtz.Minkowskian}
    D^{-\frac{1}{4}}_{\labelMink}(x, y)
  &=& \frac{1}{2 \pi}
    \int_{-\infty}^{\infty}
      \frac{\e^{\i p (x - y)}}{\sqrt[4]{p^2 + m^2}}
    \id{p}
  \,=\, \frac{1}{\pi}
    \int_{0}^{\infty}
      \frac{\cos\bigl( \abs{x - y} p \bigr)}{\sqrt[4]{p^2 + m^2}}
    \id{p}
\nonumber\\
  &=& \frac{\sqrt[4]{2 m} \mult \BesselK{\frac{1}{4}}\bigl( m \abs{x - y} \bigr)
    }{\sqrt{\pi} \mult \upGamma(1/4) \sqrt[4]{\abs{x - y}}}
  \eqend{,}
\end{eqnarray}
which follows from Basset's integral~\citeDLMFeq{10.32}{11}.

As for the term $D^{-1/4}_{\textlabel{bdy}}(x, y)$, 
defined by \eqref{eq:FractionalModifiedHelmholtz.HalfLine.BoundaryPart}
with $\nu = -1/4$,
in the special cases of Neumann b.c.\@ 
($s_{\textlabel{bdy}} = 1$) 
and Dirichlet b.c.\@ ($s_{\textlabel{bdy}} = -1$),
it is computed analogously to $D^{-1/4}_{\labelMink}(x, y)$,
finding 
\begin{equation}
\label{eq:InverseFourthRootHelmholtz.Boundary.Special}
    D^{-\frac{1}{4}}_{\textlabel{bdy}}(x, y)
  = \frac{s_{\textlabel{bdy}}}{2 \pi}
    \int_{-\infty}^{\infty}
      \frac{\e^{- \i p (x + y)}}{\sqrt[4]{p^2 + m^2}}
    \id{p}
  = s_{\textlabel{bdy}}
    \frac{\sqrt[4]{2 m} \mult \BesselK{\frac{1}{4}}\bigl( m (x + y) \bigr)
    }{\sqrt{\pi} \mult \upGamma(1/4) \sqrt[4]{x + y}
    }
  \eqend{.}
\end{equation}
In the case of a generic Robin b.c.\@, 
the computation is similar to $D^{-1/2}$ in \eqref{eq:InverseRootHelmholtz.Boundary.Robin}. 
The integral has two contributions
\begin{eqnarray}
\label{eq:InverseFourthRootHelmholtz.Boundary.Robin}
    D^{-\frac{1}{4}}_{\textlabel{bdy}}(x, y)
  &=& \frac{1}{2 \pi}
    \int_{-\infty}^{\infty}
      \frac{p + \i \eta}{p - \i \eta}
      \mult \frac{\e^{- \i p (x + y)}}{\sqrt[4]{p^2 + m^2}}
    \id{p}
\nonumber\\
  &=& \frac{1}{\pi}
    \int_{0}^{\infty}
      \frac{\cos\bigl( p (x + y) \bigr)}{\sqrt[4]{p^2 + m^2}}
    \id{p}
    - \frac{\i \eta}{\pi}
    \int_{-\infty}^{\infty}
      \frac{1}{p + \i \eta}
      \mult \frac{\e^{\i p (x + y)}}{(p^2 + m^2)^{-\frac{1}{4} + \frac{1}{2}}}
    \id{p}
\nonumber\\
  &=& \sqrt[4]{\frac{2 m}{x + y}}
    \mult \frac{
      \BesselK{\frac{1}{4}}\bigl( m (x + y) \bigr)
      - 2 B_{-\frac{1}{4}}(\eta, m; x + y)
    }{\sqrt{\pi} \mult \upGamma(1/4)}
  \eqend{,}
\end{eqnarray}
where the integral denoted by  $B_{-1/4}(\eta, m; x + y)$
is analysed in Appendix~\ref{appx:GeneralizedBassetIntegral}.
In the special case of $\eta = m$, 
it has the explicit solution
\begin{equation}
\label{eq:InverseFourthRootHelmholtz.Boundary.Robin1}
    D^{-\frac{1}{4}}_{\textlabel{bdy}}(x, y)
  = \sqrt[4]{\frac{2 m}{x + y}}
    \mult \frac{
      (1 + 4 r) \mult \BesselK{\frac{1}{4}}(r)
      - 4 r \mult \BesselK{\frac{3}{4}}(r)
    }{\sqrt{\pi} \mult \upGamma(1/4)}
  \eqend{,}
    \qquad
    r \coloneq m (x + y)
  \eqend{.}
\end{equation}
Since we do not have a general solution of $B_{-1/4}(\eta, m; x + y)$ for all values of $\eta$ and $m$, 
in our numerical analysis for Robin b.c.\@ we only consider this special case where $\eta = m$.

\section{A generalization of Basset's integral}
\label{appx:GeneralizedBassetIntegral}

In this appendix, 
we discuss a generalization of Basset's integral
occurring as special cases in \eqref{eq:InverseFourthRootHelmholtz.RobinBoundary},  \eqref{eq:InverseRootHelmholtz.Boundary.Robin}, \eqref{eq:RootHelmholtz.Boundary.Robin} and \eqref{eq:InverseFourthRootHelmholtz.Boundary.Robin}.
The definition and differential equations are given in Sec.~\ref{appx:GeneralizedBassetIntegral.Definition},
while in Sec.~\ref{appx:GeneralizedBassetIntegral.Limits} 
we study special values and some relevant limits.

\subsection{Definition and differential equations}
\label{appx:GeneralizedBassetIntegral.Definition}

The generalized Basset integral $B_{\nu}(\eta, m; x)$,
where $\eta > 0$, $m > 0$ (corresponding to the boundary and mass parameters in our computations, respectively) 
and $x > 0$,
is defined as the following Fourier transform
\begin{eqnarray}
\label{eq:BoundaryIntegral.Definition}
    && B_{\nu}(\eta, m; x)
  \coloneq \frac{b_{\nu}(m, x)}{2}
    \int_{-\infty}^{\infty}
      \frac{\i \eta}{p + \i \eta}
      \frac{\e^{\i x p}}{(p^2 + m^2)^{\nu + \frac{1}{2}}}
    \id{p}
  \eqend{,}
  \quad
    b_{\nu}(m, x)
  \,\coloneq\, \frac{\upGamma\bigl( \nu + \frac{1}{2} \bigr)}{\sqrt{\pi}}
    \left( \frac{2 m}{x} \right)^{\nu}
  \eqend{,}
\nonumber\\
  &&= b_{\nu}(m, x)
    \int_{0}^{\infty}
      \frac{\eta}{(p^2 + m^2)^{\nu + \frac{1}{2}}}
      \mult \Im\Biggl( \frac{\e^{- \i x p}}{p - \i \eta} \Biggr)
    \id{p}
\nonumber\\
  &&= \frac{b_{\nu}(m, x)}{2 \i}
    \int_{0}^{\infty}
      \frac{\e^{- \i p x}}{(p^2 + m^2)^{\nu + \frac{1}{2}}}
      \mult \frac{\eta p + \i \eta^2}{p^2 + \eta^2}
    \id{p} 
    - \frac{b_{\nu}(m, x)}{2 \i}
    \int_{0}^{\infty}
      \frac{\e^{\i p x}}{(p^2 + m^2)^{\nu + \frac{1}{2}}}
      \mult \frac{\eta p - \i \eta^2}{p^2 + \eta^2}
    \id{p}
\nonumber\\
  &&= b_{\nu}(m, x)
    \int_{0}^{\infty}
      \frac{1}{(p^2 + m^2)^{\nu + \frac{1}{2}}}
      \mult \frac{\eta^2 \cos(x p) - \eta p \sin(x p)}{p^2 + \eta^2}
    \id{p}
\nonumber\\
  &&= b_{\nu}(m, x)
    \left( \pderiv{}{x} + \eta \right)
    \int_{0}^{\infty}
      \frac{1}{p^2 + \eta^2}
      \mult \frac{\cos(x p)}{(p^2 + m^2)^{\nu + \frac{1}{2}}}
    \id{p}
  \eqend{.}
\end{eqnarray}
Interchanging the integration and differentiation in the last step is possible (at least when $\nu > -1$) 
by using the dominated convergence theorem. 
The  integrand is continuously differentiable in $x, p$, and it is dominated by the function
\begin{equation}
\label{eq:GeneralizedBassetIntegral.Definition.DominatingFunction}
    f(p)
  \coloneq \frac{1}{(p^2 + m^2)^{\nu + \frac{1}{2}} (p^2 + \eta^2)}
  \geqslant 
  \abs{
      \frac{1}{(p^2 + m^2)^{\nu + \frac{1}{2}}}
      \mult \frac{\cos(x p)}{p^2 + \eta^2}
    }
  \eqend{,}
\end{equation}
which is integrable when $\nu > -1$.
For other values of $\nu$, $B_{\nu}(\eta, m; x)$ is only defined in the distributional sense.
Comparing the first line of \eqref{eq:BoundaryIntegral.Definition} with the standard form of Basset's integral \citeDLMFeq{10.32}{11}, we have an additional factor of $\i \eta / (p + \i \eta)$ in the integrand, which becomes one in the Dirichlet limit $\eta \to \infty$, recovering the standard integral of Basset.

Basset's original integral also appears in the inhomogeneous part of the following first-order differential equation
\begin{eqnarray}
\label{eq:BoundaryIntegral.DifferentialEquation.Derivation}
    \left( - \pderiv{}{x} + \eta \right)
    x^{\nu} B_{\nu}(\eta, m; x)
  &=& \frac{\i \eta}{2} \mult x^{\nu} \mult b_{\nu}(m, x)
    \int_{-\infty}^{\infty}
      \frac{-\i p + \eta}{p + \i \eta}
      \mult \frac{\e^{\i x p}}{(p^2 + m^2)^{\nu + \frac{1}{2}}}
    \id{p}
    \nonumber
\\
  &=& \eta \mult x^{\nu} \mult b_{\nu}(m, x)
    \int_{0}^{\infty}
      \frac{\cos(p\,x)}{(p^2 + m^2)^{\nu + \frac{1}{2}}}
    \id{p}
  \,=\, \eta \mult x^{\nu} \mult \BesselK{\nu}(m x)
  \eqend{,}
  \qquad
\end{eqnarray}
where we used again the dominated convergence theorem with the dominating function \eqref{eq:GeneralizedBassetIntegral.Definition.DominatingFunction}.
The solution of this differential equation yields an alternative integral representation for the function $B_{\nu}(\eta, m; x)$.
The homogeneous solution is $C \e^{\eta x}$ and we vary the constant $C$.
Since  the solution has to vanish at infinity, we have $\lim_{x \to \infty} C(x) \to 0$.
Then 
\begin{equation}
    - C'(x) \mult \e^{\eta x}
  = \eta \mult x^{\nu} \mult \BesselK{\nu}(m x)
  \quad \implies \quad
     C(x)
  = \eta \int_{x}^{\infty}
      \e^{-\eta t} \mult t^{\nu} \mult \BesselK{\nu}(m t)
    \id{t}
  \eqend{,}
\end{equation}
and
\begin{equation}
\label{eq:BoundaryIntegral.DifferentialEquation.Solution}
    B_{\nu}(\eta, m; x)
  = \frac{\eta \mult \e^{\eta x}}{x^{\nu}}
    \int_{x}^{\infty}
      \e^{-\eta t} \mult t^{\nu}
      \mult \BesselK{\nu}(m t)
    \id{t}
  = \frac{\eta \mult \e^{\eta x}}{m^{\nu + 1} \mult x^{\nu}}
    \int_{m x}^{\infty}
      \e^{-\frac{\eta}{m} s} \mult s^{\nu}
      \mult \BesselK{\nu}(s)
    \id{s}
  \eqend{,}
\end{equation}
which differs from a Laplace transform of $s^{\nu} \mult \BesselK{\nu}(s)$
by a non-vanishing lower limit.

By employing \eqref{eq:BoundaryIntegral.DifferentialEquation.Solution},
one finds the following recursive relations
\begin{eqnarray}
\label{eq:BoundaryIntegral.EtaDerivativeRecursion}
    \eta \pderiv{}{\eta} B_{\nu}(\eta, m; x)
  &=& - \nu B_{\nu}(\eta, m; x)
    + x \pderiv{}{x} B_{\nu}(\eta, m; x)
    + m x B_{\nu + 1}(\eta, m; x)
  \eqend{,}
\\
\label{eq:GeneralizedBassetIntegral.LambdaDerivativeRecursion}
    m \pderiv{}{m} B_{\nu}(\eta, m; x)
  &=& \nu B_{\nu}(\eta, m; x)
    - m x B_{\nu + 1}(\eta, m; x)
  \eqend{.}
\end{eqnarray}
From the sum of these two relations,
we notice that $B_{\nu}$ fulfils the following homogeneous, linear, partial differential equation
\begin{equation}
\label{eq:BoundaryIntegral.PartialDifferentialEquation}
    \left(
      \eta \pderiv{}{\eta}
      + m \pderiv{}{m}
      - x \pderiv{}{x}
    \right) B_{\nu}(\eta, m; x)
  = 0
  \eqend{.}
\end{equation}
Thus, $B_{\nu}$ is a homogeneous function of degree 
zero, meaning that it is invariant
under the rescaling $\eta \to \rho \eta$, $m \to \rho m$, $x \to \rho^{-1} x$.

\subsection{Special values and limiting regimes}
\label{appx:GeneralizedBassetIntegral.Limits}

When $m = \eta$ and $\nu \neq -\frac{1}{2}$,
the integral \eqref{eq:BoundaryIntegral.DifferentialEquation.Solution}
becomes (see \citeDLMFeq{10.43}{3}) 
\begin{equation}
\label{eq:BoundaryIntegral.ArgumentCoincidence}
    B_{\nu}(\eta, \eta; x)
  = \frac{\e^{\eta x}}{\eta^{\nu} x^{\nu}}
    \int_{\eta x}^{\infty}
      \e^{-s} \mult s^{\nu}
      \mult \BesselK{\nu}(s)
    \id{s}
  = \frac{\eta x}{2 \nu + 1}
    \Bigl(
      \BesselK{\nu + 1}(\eta x)
      - \BesselK{\nu}(\eta x)
    \Bigr)
  \eqend{.}
\end{equation}
Further special values occur when $\nu$ takes half odd integer values.
In  these cases 
the Bessel function in \eqref{eq:BoundaryIntegral.DifferentialEquation.Solution}
is given in terms of standard functions 
(exponentials and polynomials)
and the integral can be solved.
For example, for the cases $\nu = \pm \frac{1}{2}$, we use \citeDLMFeq{10.39}{2} and get
\begin{eqnarray}
\label{eq:BoundaryIntegral.HalfIndexCase}
    B_{\frac{1}{2}}(\eta, m; x)
  &=& \sqrt{\frac{\pi}{2 m x}}
    \mult \frac{\eta \mult \e^{\eta x}}{m}
    \int_{m x}^{\infty}
      \e^{-\left( \frac{\eta}{m} + 1 \right) s}
    \id{s}
  \,=\, \sqrt{\frac{\pi}{2 m x}}
    \mult \frac{\eta \mult \e^{-m x}}{\eta + m}
  \eqend{,}
\\
\label{eq:BoundaryIntegral.NegativeHalfIndexCase}
    B_{-\frac{1}{2}}(\eta, m; x)
  &=& \sqrt{\frac{\pi x}{2 m}}
    \mult \eta \mult \e^{\eta x}
    \int_{m x}^{\infty}
      \frac{1}{s}
      \mult \e^{-\left( \frac{\eta}{m} + 1 \right) s}
    \id{s}
  \,=\, \sqrt{\frac{\pi x}{2 m}}
    \mult \eta \mult \e^{\eta x}
    \mult \upGamma\bigl( 0, (\eta + m) x \bigr)
  \eqend{.}
  \qquad
\end{eqnarray}
Note that, setting $\nu = -\frac{1}{2}$ in \eqref{eq:BoundaryIntegral.Definition}, 
the coefficient function $b_{\nu}(m, x)$ is singular, 
but $B_{-1/2}$ has the explicit expression \eqref{eq:BoundaryIntegral.NegativeHalfIndexCase}, 
following from the integral representation \eqref{eq:BoundaryIntegral.DifferentialEquation.Solution}.

For all the limiting regimes that are discussed in the following, we use the dominated convergence theorem again, where $s^{\nu} \BesselK{\nu}(s)$ is the dominating integrable function in \eqref{eq:BoundaryIntegral.DifferentialEquation.Solution}.
The limiting regime given by the Neumann b.c.\@ ($\eta \to 0^+$) can be studied 
through \citeDLMFeq{10.43}{2} and \citeDLMFeq{10.43}{19}, 
finding 
\begin{equation}
\label{eq:BoundaryIntegral.EtaLimit0}
    \lim_{\eta \to 0^+}
      B_{\nu}(\eta, m; x)
  = \frac{\eta \mult \e^{\eta x}}{m^{\nu + 1} x^{\nu}}
    \int_{m x}^{\infty}
      \e^{-\frac{\eta}{m}}
      s^{\nu}
      \mult \BesselK{\nu}(s)
    \id{s}
  = 0 
  \eqend{.}
\end{equation}
As mentioned above, the limit corresponding to Dirichlet b.c.\@ is Basset's integral
\begin{equation}
  \label{eq:BoundaryIntegral.TimesEta.EtaLimitInfty}
    \lim_{\eta \to \infty}
      B_{\nu}(\eta, m; x)
  = b_{\nu}(m, x)
    \int_{0}^{\infty}
      \frac{\cos(x p)}{(p^2 + m^2)^{\nu + \frac{1}{2}}}
    \id{p}
  = \BesselK{\nu}(m x)
  \eqend{.}
\end{equation}

As for  small values of $m$, 
by employing \citeDLMFeq{10.30}{2},
we find three different cases, 
depending on the index $\nu$.
For $\nu < 0$, the 
asymptotic behaviour (denoted with the symbol $\sim$)
is given by 
\begin{equation}
\label{eq:BoundaryIntegral.MasslessLimit.NegativeIndex}
    B_{\nu}(\eta, m; x)
  \sim \frac{\upGamma(-\nu)}{2}
    \mult \eta \mult \e^{\eta x}
    \left( \frac{m}{2 x} \right)^{\nu}
    \int_{x}^{\infty}
      \e^{-\eta t} \mult t^{2 \nu}
    \id{t}
  = \frac{\upGamma(-\nu) \mult \e^{\eta x}}{2 \eta^{2 \nu}}
    \left( \frac{m}{2 x} \right)^{\nu}
    \upGamma(2 \nu + 1, \eta x)
  \eqend{,}
\end{equation}
and for $\nu > 0$, we get
\begin{equation}
\label{eq:BoundaryIntegral.MasslessLimit.PositiveIndex}
    B_{\nu}(\eta, m; x)
  \sim \frac{\upGamma(\nu)}{2}
    \mult \eta \mult \e^{\eta x}
    \left( \frac{2}{m x} \right)^{\nu}
    \int_{x}^{\infty}
      \e^{-\eta t}
    \id{t}
  = \frac{\upGamma(\nu)}{2}
    \left( \frac{m}{2 x} \right)^{\nu}
  \eqend{.}
\end{equation}
At $\nu = 0$, a logarithmic divergence occurs (see e.g.\@ \citeDLMFeq{10.30}{3}) as $m \to 0^+$
\begin{eqnarray}
\label{eq:BoundaryIntegral.MasslessLimit.ZeroIndex}
    B_0(\eta, m; x)
  &\sim& - \eta \mult \e^{\eta x}
    \int_{x}^{\infty}
      \e^{-\eta t}
      \biggl[ \upgamma + \log\biggl( \frac{m t}{2} \biggr) \biggr]
    \id{t}
\nonumber\\
  &\quad& =\, - \eta \mult \e^{\eta x}
    \left[
      \frac{\upgamma \mult \e^{-\eta x}}{\eta}
      + \frac{\e^{-\eta x}}{\eta}
      \log\biggl( \frac{m t}{2} \biggr)
      + \frac{1}{\eta}
      \int_{x}^{\infty}
        \frac{\e^{-\eta t}}{t}
      \id{t}
    \right]
\nonumber\\
  &\quad& =\, - \log\biggl( \frac{m t}{2} \biggr)
    - \upgamma
    - \e^{\eta x} \mult \upGamma(0, \eta x)
  \eqend{,}
\end{eqnarray}
where $\gamma$ is the Euler's constant.

In the limit $m \to \infty$, 
in \eqref{eq:BoundaryIntegral.DifferentialEquation.Solution}
we first employ the expansion of the modified Bessel function 
of the second kind 
from \citeDLMFeq{10.25}{3}, and then the expansion 
of the incomplete Gamma function from \citeDLMFeq{8.11}{2}, 
finding
\begin{eqnarray}
\label{eq:BoundaryIntegral.LargeLambda}
    B_{\nu}(\eta, m; x)
  &\sim& \sqrt{\frac{\pi}{2 m}}
    \frac{\eta \mult \e^{\eta x}}{x^{\nu}}
    \int_{x}^{\infty}
      \e^{-(\eta + m) t}
      \mult t^{\nu - \frac{1}{2}}
    \id{t}
  \,=\, \sqrt{\frac{\pi}{2 m}}
    \frac{
      \eta \mult \e^{\eta x}
      \mult \upGamma\bigl( \nu + \frac{1}{2}, (\eta + m) x \bigr)
    }{(\eta + m)^{\nu + \frac{1}{2}} x^{\nu}}
\nonumber\\
  &\sim& \sqrt{\frac{\pi}{2 m x}}
    \frac{\eta \mult \e^{-m x}}{\eta + m}
  \eqend{.}
\end{eqnarray}
These properties of 
$B_{\nu}(\eta, m; x)$ have been applied 
in Sec.~\ref{sec:HalfLine} and in the Appendices \ref{appx:ComplexStructure.TwoPointFunctions} and \ref{appx:InverseFourthRoots}.

\section{Operator discretization and smearing}
\label{appx:Discretization}

In this appendix, we show how to discretize the operator kernels with a set of box functions $e_i(x)$ with index $i \in \{0, 1, \dots, n - 1\}$ for a fixed discretization size (resolution) $n$ and a discretization grid such that $a_0 = 0$ and $b_i = b_i$, and $d_i \coloneq b_i - a_i > 0$ for all $i$,
providing further details to the discussion reported 
in Sec.~\ref{sec:HalfLine.Discretization}.
In particular, kernels of the types
$D^{-1/4}_{\labelMink}(x, y)$ and $D^{-1/4}_{\textlabel{bdy}}(x, y)$
are considered in 
Sec.~\ref{appx:Discretization.ConvolutionKernels}
and 
Sec.~\ref{appx:Discretization.AntiConvolutionKernels}
respectively.
In Sec.~\ref{appx:Discretization.TestFunctions}
we compute the smearing of the analytic references appearing
in Sec.~\ref{sec:Results.AdjacentInterval} and Sec.~\ref{sec:Results.SeparatedInterval} against the log-Gaussian functions \eqref{eq:LogGaussian}.

\subsection{Convolution operator discretization}
\label{appx:Discretization.ConvolutionKernels}

\begin{figure}
  \centering
  \begin{subfigure}[t]{0.45\textwidth}
    \centering
    \includegraphics{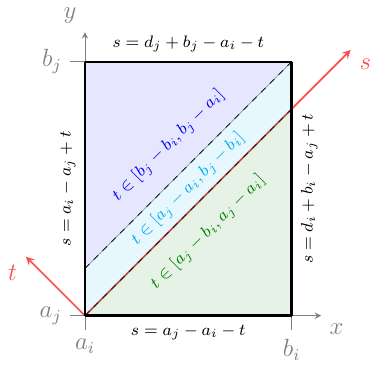}
    \caption{\label{fig:KernelDiscretization.Domain.Convolution} Kernel $D^{-1/4}_{\labelMink}(x, y)$,
    depending on $x - y$.}
  \end{subfigure}
  \hfill
  \begin{subfigure}[t]{0.45\textwidth}
    \centering
    \includegraphics{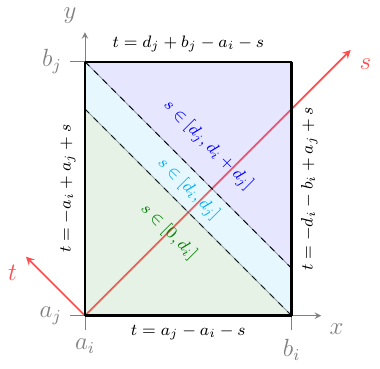}
    \caption{\label{fig:KernelDiscretization.Domain.AntiConvolution} Kernel $D^{-1/4}_{\textlabel{bdy}}(x, y)$, depending on $x + y$.}
  \end{subfigure}
  \caption{\label{fig:KernelDiscretization.Domain} Coordinates and subregions of the integration domain occurring in 
  $D^{-1/4}_{\labelMink}(x, y)$ (left) 
  and $D^{-1/4}_{\textlabel{bdy}}(x, y)$ (right), as discussed in 
  Sec.~\ref{appx:Discretization.ConvolutionKernels} 
  and in Sec.~\ref{appx:Discretization.AntiConvolutionKernels}, respectively.}
\end{figure}

To discretize the symmetric convolution kernel $D^{-1/4}_{\labelMink}(x, y) \eqcolon f^-(x - y)$ 
as computed in Appendix~\ref{appx:InverseFourthRoots}, we only need to compute the diagonal and the upper-triangular matrix elements, since the lower-triangular elements 
are identical to the corresponding upper-triangular elements.
The upper-triangular elements $D^{-1/4, (n, \Lambda)}_{\labelMink}$ have indices $i \leqslant j$ 
(hence $y - x \geqslant 0$) and are proportional to the integral
\begin{equation}
\label{eq:KernelDiscretization.Convolution}
    \mathcal{I}^-_{i j}
  = \int_{x = a_i}^{b_i} \int_{y = a_j}^{b_j}
      f^-\bigl( - (y - x) \bigr)
    \id{x} \id{y}
  \eqend{,}
\end{equation}
which is the first expression in \eqref{eq:KernelDiscretization.Integral.Split}.
To simplify this double-integral, we change coordinates $(x, y) \to (s, t)$ on the integral domain $[a_i, b_i] \times [a_j, b_j]$,
where $s = x - a_i + y - a_j$ and $t =y - x$; hence
\begin{equation}
\label{eq:KernelDiscretization.Convolution.Substitution}
    \d{s} \mult \d{t}
  = \left\lvert \pderiv{(s, t)}{(x, y)} \right\rvert
    \d{x} \mult \d{y}
  = \begin{vmatrix}
       1 & -1 \\
       1 &  1
    \end{vmatrix}
    \d{x} \mult \d{y}
  = 2 \mult \d{x} \mult \d{y}
  \eqend{.}
\end{equation}
The integral \eqref{eq:KernelDiscretization.Convolution} becomes
\begin{equation}
\label{eq:KernelDiscretization.Convolution.Substituted}
    \mathcal{I}^-_{i j}
  = \frac{1}{2} \iint_{\mathrm{rectangle}} f^-(-t)
    \id{s} \id{t}
  \eqend{,}
\end{equation}
where the rectangular domain has edge lengths $d_i \coloneq b_i - a_i$ and $d_j \coloneq b_j - a_j$.
The integrals in the two cases $d_i > d_j$ and $d_i \leqslant d_j$ reduce to the same formal expression, so we can consider just $d_i \leqslant d_j$ without loss of generality.
Since the integral kernel in \eqref{eq:KernelDiscretization.Convolution.Substituted} is a function of $t = y - x$, we split the integral domain into the three strips shown in Fig.~\ref{fig:KernelDiscretization.Domain.Convolution}.
The variables run over the ranges
\begin{equation}
\label{eq:KernelDiscretization.Convolution.Domain}
\begin{array}{ll}
    t \in [a_j - b_i, a_j - a_i]
    \eqend{,}
    &\qquad
    s \in [a_j - a_i - t, d_i + b_i - a_j + t]
    \eqend{,}
\\ 
    t \in [a_j - a_i, b_j - b_i]
    \eqend{,}
    &\qquad
    s \in [a_i - a_j + t, d_i + b_i - a_j + t]
    \eqend{,}
\\
    t \in [b_j - b_i, b_j - a_i]
    \eqend{,}
    &\qquad
    s \in [a_i - a_j + t, d_j + b_j - a_i - t]
    \eqend{,}
\end{array}
\end{equation}
and the integral over $s$
is an elementary integration for each strip.
Along the diagonal ($i = j$), the integral domain is always a square $d_i = d_j$ and the central strip vanishes. 
Since $f(t) = f(-t)$ in general, we get the diagonal elements
\begin{eqnarray}
\label{eq:KernelDiscretization.Convolution.DiagonalIntegral}
    \mathcal{I}^-_{i i}
  &=& \int_{-d_i}^{0}
      (d_i + t) \mult f^-(-t)
    \id{t}
    + \int_{0}^{d_i}
      (d_i - t) \mult f^-(-t)
    \id{t}
\nonumber\\
  &=& 2 d_i \int_{0}^{d_i}
      f^-(-t)
    \id{t}
    - 2 \int_{0}^{d_i}
      t \mult f^-(-t)
    \id{t}
      \eqend{.}
\end{eqnarray}
For the integral 
\eqref{eq:KernelDiscretization.Convolution.Substituted} above the diagonal ($i < j$), we find
\begin{eqnarray}
\label{eq:KernelDiscretization.Convolution.Integral}
    \mathcal{I}^-_{i j}
  &=&
  - \int_{a_j - b_i}^{a_j - a_i}
      (a_j - b_i - t) \mult f^-(-t)
    \id{t}
    + d_i \int_{a_j - a_i}^{b_j - b_i}
      f^-(-t)
    \id{t}
\nonumber\\
  &&{}
    + \int_{b_j - b_i}^{b_j - a_i}
      (b_j - a_i - t) \mult f^-(-t)
    \id{t}
  \eqend{,}
\end{eqnarray}
while below the diagonal we employ
the symmetry $\mathcal{I}^-_{j i} = \mathcal{I}^-_{i j}$. 
The solutions of these integrals for the operator $D^{-1/4}_{\labelMink}$ in the massless and massive regimes 
are reported in Sec.~\ref{sec:HalfLine.Discretization}.

\subsection{Discretization of the boundary contribution}
\label{appx:Discretization.AntiConvolutionKernels}

Now we study the term $D^{-1/4}_{\textlabel{bdy}}(x, y) \eqcolon f^+(x + y)$ 
as computed in Appendix~\ref{appx:InverseFourthRoots},
which is also symmetric in $x$ and $y$, 
by employing the same coordinate substitution \eqref{eq:KernelDiscretization.Convolution.Substitution},
finding 
\begin{equation}
    \mathcal{I}^+_{i j}
  = \frac{1}{2} \iint_{\mathrm{rectangle}} f^+(s + a_i + a_j)
    \id{s} \id{t}
  \eqend{,}
\end{equation}
where the integral kernel now depends only on the sum variable $s$.
Thus, it is convenient to 
swap the order of integration first and then 
split the integral domain into the three strips shown in Fig.~\ref{fig:KernelDiscretization.Domain.AntiConvolution},
which are given by 
\begin{equation}
\label{eq:KernelDiscretization.AntiConvolution.Domain}
\begin{array}{ll}
    s \in [0, d_i]
    \eqend{,}
    &\qquad
    t \in [a_j - a_i - s, -a_i + a_j + s]
    \eqend{,}
\\
    s \in [d_i, d_j]
    \eqend{,}
    &\qquad
    t \in [-d_i - b_i + a_j + s, -a_i + a_j + s]
    \eqend{,}
\\
    s \in [d_j, d_i + d_j]
    \eqend{,}
    &\qquad
    t \in [-d_i - b_i + a_j + s, d_j + b_j - a_i - s]
    \eqend{.}
\end{array}
\end{equation}
The inner integral over $t$ is again elementary.
As for the diagonal, where $i = j$,
we have that $f^+$ is always positive and therefore 
a special treatment is not needed. 
Thus, for any index pair, the integral $\mathcal{I}^+_{i j}$ can be simplified as follows
\begin{eqnarray}
\label{eq:KernelDiscretization.AntiConvolution.Integral.Simplification}
    \mathcal{I}^+_{i j}
  &=& \int_{0}^{d_i}
      s \mult f^+(s + a_i + a_j)
    \id{s}
    + d_i \int_{d_i}^{d_j}
      f^+(s + a_i + a_j)
    \id{s}
  \nonumber\\
  &&{} + \int_{d_j}^{d_i + d_j}
      (d_j + d_i - s) \mult f^+(s + a_i + a_j)
    \id{s}
  \eqend{,}
\end{eqnarray}
and then we perform another substitution $t = s + a_i + a_j$ (reusing the variable $t$), finding
\begin{eqnarray}
\label{eq:KernelDiscretization.AntiConvolution.Integral}
    \mathcal{I}^+_{i j}
  &=& \int_{a_i + a_j}^{b_i + a_j}
      (t - a_i - a_j) \mult  f^+(t)
    \id{t}
    + d_i \int_{b_i + a_j}^{b_j + a_i}
      f^+(t)
    \id{t}
    + \int_{b_j + a_i}^{b_i + b_j}
      (b_j + b_i - t) \mult  f^+(t)
    \id{t}
\nonumber\\
  &=& (a_i + a_j) \int_{b_i + a_j}^{a_i + a_j}
      f^+(t)
    \id{t}
    + d_i \int_{b_i + a_j}^{b_j + a_i}
      f^+(t)
    \id{t}
    + (b_j + b_i) \int_{b_j + a_i}^{b_i + b_j}
      f^+(t)
    \id{t}
\nonumber\\
  &&{} + \int_{a_i + a_j}^{b_i + a_j}
      t\mult  f^+(t)
    \id{t}
    + \int_{b_i + b_j}^{b_j + a_i}
      t \mult  f^+(t)
    \id{t}
  \eqend{.}
\end{eqnarray}
Note that this result is formally very similar to the expression for $\mathcal{I}^-_{i j}$ in Sec.~\ref{appx:Discretization.ConvolutionKernels}.
Hence, we combined the two results in \eqref{eq:KernelDiscretization.ElementIntegrals}.

\subsection{Smearing of analytic references with test functions}
\label{appx:Discretization.TestFunctions}

To compare the numerical results for $M_-^{(n, \Lambda)}$ with the analytic references considered in Sec.~\ref{sec:Results.AdjacentInterval} and \ref{sec:Results.SeparatedInterval},
we need to smear both the numerical data and the analytic references against test functions, for which we use log-Gaussian functions \eqref{eq:LogGaussian} in our analysis.
The smearing of the numerical results is given by \eqref{eq:HalfLine.ModularHamiltonian.SmearedBlock.CompactNotation} and we label it as $(h_{x_i}, M_- h_{x_j})^{(n, \Lambda)}$ in all plots, where $n$ is the number of box functions in the numerical approximation (mostly $n = 256$) and $\Lambda$ is the cutoff parameter of the discretization (mostly $\Lambda = 16 \ell$).

The smearing of any analytic reference is an integration against the log-Gaussian functions \eqref{eq:LogGaussian}.
These functions are approximately Gaussian near the peak if they are centered at $x_i \gg \sigma$, i.e.\@ far away from the boundary at $x = 0$. 
Around the peak we have $x \approx x_i$ and the parameter expansion
\begin{equation}
    \beta_i
  \coloneq \log \alpha_i
  = \frac{\sigma^2}{2 x_i^2}
    + \Ord\left( \frac{\sigma^3}{x_i^3} \right)
  \eqend{,}
\end{equation}
which follows from the definition of $\alpha_i$ in \eqref{eq:LogGaussian}.
Hence, the approximation of a log-Gaussian function \eqref{eq:LogGaussian} is given by
\begin{equation}
    h_{x_i}(x)
  \sim \sqrt[4]{\frac{x_i^2}{\pi \sigma^2 x^2}}
    \mult \exp\Biggl(
      - \frac{x_i^2}{2 \sigma^2}
      \biggl[ \frac{\sigma^2}{2 x_i^2} + \frac{x - x_i}{x_i} \biggr]^2
    \Biggr)
  \sim \frac{1}{\sqrt[4]{\pi \sigma^2}}
    \mult \exp\Biggl(
      -\frac{(x - x_i)^2}{2 \sigma^2}
    \Biggr)
  \eqend{.}
\end{equation}

The references that we consider in the
numerical analysis of Sec.~\ref{sec:Results.AdjacentInterval} and Sec.~\ref{sec:Results.SeparatedInterval}
are linear combinations of monomial operators 
(for some integers $n$) with kernel
\begin{equation}
\label{eq:MonomialReference.Kernel}
    X^n(x, y)
  \coloneq x^n \mult \updelta(x - y)
  \eqend{,}
\end{equation}
and the double pole operator with kernel
\begin{equation}
\label{eq:DoublePoleReference.Kernel}
    Y_{\textlabel{sc}}(x, y)
  \coloneq \frac{1}{x_{\textlabel{sc}}^2 + x^2}
    \mult \updelta(x - y)
  \eqend{.}
\end{equation}
In the following, we compute their smearing against the log-Gaussian functions by using the substitutions
\begin{equation}
\label{eq:ReferenceSmearing.Substitutions}
    t
  \coloneq \log x
  \eqend{,}
  \quad
    \tau_i
  \coloneq \beta_i - \log x_i
  \eqend{,}
  \quad
    \gamma_{i j}
  \coloneq \sqrt{\frac{4 \beta_i \beta_j}{\beta_i + \beta_j}}
  \eqend{,}
  \quad
    \tau_{i j}
  \coloneq \frac{\beta_j \tau_i + 4 \beta_i \beta_j + \beta_i \tau_j}{\beta_i + \beta_j}
  \eqend{.}
\end{equation}
As for the monomial operator $X^n$, 
the smearing with \eqref{eq:LogGaussian} 
at positions $x_i$ and $x_j$ is
\begin{eqnarray}
\label{eq:MonomialReference.Smearing}
    \innerProd[\labelReal]{h_{x_i}}{X^n h_{x_j}}
  &=& 
  \frac{1}{\sqrt[4]{4 \pi^2 \beta_i \beta_j}}
    \int_{0}^{\infty}
      \frac{x^n}{x}
      \mult \exp\bigggl(
        - \frac{\log^2\left( \alpha_i \frac{x}{x_i} \right)}{4 \beta_i}
        - \frac{\log^2\left( \alpha_j \frac{x}{x_j} \right)}{4 \beta_j}
      \bigggr)
    \id{x}
\nonumber\\
  &=& \frac{1}{\sqrt[4]{4 \pi^2 \beta_i \beta_j}}
    \int_{-\infty}^{\infty}
      \e^{n t}
      \mult \exp\Biggl(
        - \frac{t^2 + 2 \tau_i t + \tau_i^2}{4 \beta_i}
        - \frac{t^2 + 2 \tau_j t + \tau_j^2}{4 \beta_j}
      \Biggr)
    \id{t}
\nonumber\\
  &=& \sqrt{\frac{\sqrt{4 \beta_i \beta_j}}{\beta_i + \beta_j}}
    \mult \exp\Biggl(
      - \frac{
        (\tau_i - \tau_j)^2
      }{4 (\beta_i + \beta_j)}
    \Biggr)
    \mult \exp\Biggl(
      - n \frac{
        \beta_j \tau_i - n \beta_i \beta_j + \beta_i \tau_j
      }{\beta_i + \beta_j}
    \Biggr)
  \eqend{.}
  \qquad
\end{eqnarray}
In the diagonal case, namely for $i = j$, 
this simplifies to
\begin{equation}
\label{eq:MonomialReference.Smearing.Diagonal}
    \innerProd[\labelReal]{h_{x_i}}{X^n h_{x_i}}
  =
    \exp\Biggl(
      n \mult \frac{n \beta_i - 2 \tau_i}{2}
    \Biggr)
  =
  \alpha_i^{\frac{n (n - 2)}{2}} x_i^n
  \eqend{.}
\end{equation}

We carry out the same analysis for the double pole kernel \eqref{eq:DoublePoleReference.Kernel}  as follows
\begin{eqnarray}
\label{eq:DoublePoleReference.Smearing}
    \innerProd[\labelReal]{h_{x_i}}{Y_{\textlabel{sc}} \mult h_{x_j}}
  &=& \frac{1}{\sqrt[4]{4 \pi^2 \beta_i \beta_j}}
    \int_{0}^{\infty}
      \frac{1}{x}
      \mult \frac{1}{x_{\textlabel{sc}}^2 + x^2}
      \mult \exp\bigggl(
        - \frac{\log^2\left( \alpha_i \frac{x}{x_i} \right)}{4 \beta_i}
        - \frac{\log^2\left( \alpha_j \frac{x}{x_j} \right)}{4 \beta_j}
      \bigggr)\
    \id{x}
\nonumber\\
  &=& \frac{1}{\sqrt[4]{4 \pi^2 \beta_i \beta_j}}
    \int_{-\infty}^{\infty}
      \frac{\e^{-2 t}}{1 + x_{\textlabel{sc}}^2 \e^{-2 t}}
      \mult \exp\Biggl(
        - \frac{t^2 + 2 \tau_i t + \tau_i^2}{4 \beta_i}
        - \frac{t^2 + 2 \tau_j t + \tau_j^2}{4 \beta_j}
      \Biggr)
    \id{t}
\nonumber\\
  &=& \frac{\innerProd[\labelReal]{h_{x_i}}{X^{-2} h_{x_j}}}{\sqrt{\pi} \gamma_{i j}}
    \int_{-\infty}^{\infty}
      \frac{1}{1 + x_{\textlabel{sc}}^2 \e^{-2 t}}
      \mult \exp\Biggl(
        - \frac{\left(t + \tau_{i j} \right)^2
        }{\gamma_{i j}^2}
      \Biggr)
    \id{t}
  \eqend{,}
\end{eqnarray}
where the remaining integral is evaluated numerically.

In Sec.~\ref{sec:Results.AdjacentInterval} we have studied the modular Hamiltonian of the interval $A = (0, \ell)$ adjacent to the boundary, for which we have also considered 
two references:
a linear one corresponding to a left wedge on the line (LW)
and a quadratic one corresponding to BCFT,
given in \eqref{eq:LeftWedge.LinearReference} and \eqref{eq:LeftWedge.BCFTReference} respectively.
Their smearings are
\begin{eqnarray}
\label{eq:LeftWedge.LinearReference.Smearing}
    \innerProd[\labelReal]{h_{x_i}}{M_-^{\textlabel{LW}} h_{x_j}}
  &=& 2 \pi \biggl(
      \ell \innerProd[\labelReal]{h_{x_i}}{X^0 h_{x_j}}
      - \innerProd[\labelReal]{h_{x_i}}{X^1 h_{x_j}}
    \biggr)
  \eqend{,}
\\
\label{eq:LeftWedge.BCFTReference.Smearing}
    \innerProd[\labelReal]{h_{x_i}}{M_-^{\textlabel{BCFT}} h_{x_j}}
  &=& \frac{\pi}{\ell} \biggl(
      \ell^2 \innerProd[\labelReal]{h_{x_i}}{X^0 h_{x_j}}
      - \innerProd[\labelReal]{h_{x_i}}{X^2 h_{x_j}}
    \biggr)
  \eqend{.}
\end{eqnarray}
In the figures of Sec.~\ref{sec:Results.AdjacentInterval}, these references are shown with a dotted black line (LW) and a solid magenta line (BCFT).
We report some results along the diagonal, 
where $x_j = x_i$, in the top left panel of Fig.~\ref{fig:LeftWedge.Mm.Boundaries.Diagonals} 
and in the left panels of Fig.~\ref{fig:LeftWedge.Mm.Massive.Diagonals};
and similarly, along the anti-diagonal, 
where $x_j = x_{23 - i}$, 
in the bottom left panel of Fig.~\ref{fig:LeftWedge.Mm.Boundaries.Diagonals}.

In Sec.~\ref{sec:Results.SeparatedInterval}, 
we have explored
the modular Hamiltonian of the interval separated from the boundary $A = (d, d + \ell)$, 
introducing three references: 
the inner linear reference \eqref{eq:DoubleCone.InnerLinearReference}
corresponding to a right wedge on the line (RW), 
the outer linear reference \eqref{eq:DoubleCone.OuterLinearReference} 
corresponding to a left wedge on the line (LW), 
and the reference \eqref{eq:DoubleCone.LocalReference}
corresponding to the local term of the modular Hamiltonian for the massless Dirac found in \cite{MintchevTonni:2021a}.
The smearings of the linear references are
\begin{eqnarray}
\label{eq:DoubleCone.InnerLinearReference.Smearing}
    \innerProd[\labelReal]{h_{x_i}}{M_-^{\textlabel{RW}} h_{x_j}}
  &=& 2 \pi \biggl(
      \innerProd[\labelReal]{h_{x_i}}{X^1 h_{x_j}}
      - d \innerProd[\labelReal]{h_{x_i}}{X^0 h_{x_j}}
    \biggr)
  \eqend{,}
\\
\label{eq:DoubleCone.OuterLinearReference.Smearing}
    \innerProd[\labelReal]{h_{x_i}}{M_-^{\textlabel{LW}} h_{x_j}}
  &=& 2 \pi \biggl(
      (d + \ell) \innerProd[\labelReal]{h_{x_i}}{X^0 h_{x_j}}
      - \innerProd[\labelReal]{h_{x_i}}{X^1 h_{x_j}}
    \biggr)
  \eqend{.}
\end{eqnarray}
Similarly, for the smearing of the reference \eqref{eq:DoubleCone.LocalReference} we have
\begin{eqnarray}
\label{eq:DoubleCone.LocalReference.Smearing}
    \innerProd[\labelReal]{h_{x_i}}{M_-^{\textlabel{F,loc}} h_{x_j}}
  &=& \frac{\pi}{\ell} \biggl(
      \big(3 d^2 + 3 d \ell + \ell^2\big) \innerProd[\labelReal]{h_{x_i}}{X^0 h_{x_j}}
      - \innerProd[\labelReal]{h_{x_i}}{X^2 h_{x_j}}
  \nonumber\\
  && \qquad{}
      - d (d + \ell) (2 d + \ell)^2 \innerProd[\labelReal]{h_{x_i}}{Y_{\textlabel{sc}} h_{x_j}}
    \biggr)
  \eqend{.}
\end{eqnarray}
In the figures of Sec.~\ref{sec:Results.SeparatedInterval},
we use dotted black lines for the linear references 
and a dash-dotted magenta line for the fermionic local reference.
We report results along the diagonal, where $x_j = x_i$, in Fig.~\ref{fig:DoubleCone.Mm.Boundaries.Diagonal}, \ref{fig:DoubleCone.Mm.Massive.Diagonal} and \ref{fig:DoubleCone.Mm.Massive.Robin.Diagonal};
and similarly, we show some results along the curve \eqref{eq:ConjugateCurve}, 
where $x_j = x_{\textlabel{c}}(x_i)$, 
 in the left panels of Fig.~\ref{fig:DoubleCone.Mm.Boundaries.Conjugate}, 
\ref{fig:DoubleCone.Mm.Massive.Robin.Conjugate}
and in all the panels of Fig.~\ref{fig:DoubleCone.Mm.Massive.Conjugate}.

\section{On the second component of the modular Hamiltonian}
\label{appx:ModularHamiltonian.BCFT.Mp}

In this appendix, we compute the second component $M_+^{\textlabel{BCFT}}$ of the modular Hamiltonian \eqref{eq:BosonHilbertSpace.ModularHamiltonian}
of an adjacent interval
in the massless regime and a Dirichlet b.c.\@, 
corresponding to the BCFT result for $M_-^{\textlabel{BCFT}}$ given in \eqref{eq:LeftWedge.BCFTReference}.

By employing the relation \eqref{eq:BosonHilbertSpace.ModularHamiltonian.Blocks.Relation} 
and the operator $D^{+1/2}$ given as sum of  \eqref{eq:FractionalModifiedHelmholtz.HalfLine.MinkowskianPart} and \eqref{eq:FractionalModifiedHelmholtz.HalfLine.BoundaryPart} for $\nu = 1/2$,
the distributional kernel $M_+^{\textlabel{BCFT}}$ can be written 
through its momentum space representation as follows
\begin{eqnarray}
\label{eq:LeftWedge.ModularHamiltonian.Mp.Computation}
    M_+^{\textlabel{BCFT}}(x, y)
  &=& \bigl( D^{+\frac{1}{2}} M_- D^{+\frac{1}{2}} \bigr)(x, y)
\nonumber\\
  &=& \frac{1}{\pi}
    \int_{0}^{\infty}
      \left( \frac{1}{(x + u)^2} - \frac{1}{(x - u)^2} \right)
      \frac{\ell^2 - u^2}{\ell}
      \left( \frac{1}{(u + y)^2} - \frac{1}{(u - y)^2} \right)
    \id{u}
\nonumber\\
  &=& \frac{1}{4 \pi}
    \iint_{-\infty}^{\infty}
      \abs{p \mult q}
      \left( \e^{\i p x} - \e^{- \i p x} \right)
      \e^{- \i q y}
      \int_{0}^{\infty}
        \frac{\ell^2 - u^2}{\ell}
        \left( \e^{\i (q - p) u} - \e^{- \i (q + p) u} \right)
      \id{u}
    \id{p}
    \id{q}
\nonumber\\
  &=& - \frac{\ell}{2 \pi}
    \lim_{\varepsilon \to 0^+}
    \iint_{-\infty}^{\infty}
      \abs{p} \sin(p x)
      \left(
        \frac{\abs{q} \e^{- \i q y}}{q - p + \i \varepsilon}
        - \frac{\abs{q} \e^{- \i q y}}{q + p + \i \varepsilon}
      \right)
    \id{q}
    \id{p}
\nonumber\\
  &&{}
    + \frac{1}{\pi \ell}
    \lim_{\varepsilon \to 0^+}
    \iint_{-\infty}^{\infty}
      \abs{p} \sin(p x)
      \left(
        \frac{\abs{q} \e^{- \i q y}}{(q - p + \i \varepsilon)^3}
        - \frac{\abs{q} \e^{- \i q y}}{(q + p + \i \varepsilon)^3}
      \right)
    \id{q}
    \id{p}
  \eqend{,}
\end{eqnarray}
where we used the distributional relations for real numbers $r$ given by
\begin{equation}
\label{eq:HalfSpace.TemperedDistribution.DiracDelta}
    \int_{0}^{\infty}
      \e^{\i r u}
    \id{u}
  = \pi \updelta(r)
    + \pvalue \frac{\i}{r}
  = \lim_{\varepsilon \to 0^+} \frac{\i}{r + \i \varepsilon}
  \eqend{,}
\end{equation}
and
\begin{equation}
\label{eq:HalfSpace.TemperedDistribution.DiracDeltaSecondDeriv}
    \int_{0}^{\infty}
      u^2 \e^{\i r u}
    \id{u}
  = - \pi \updelta''(r)
    - \pvalue \frac{2 \i}{r^3}
  = \lim_{\varepsilon \to 0^+} \frac{- 2 \i}{(r + \i \varepsilon)^3}
  \eqend{.}
  \end{equation}
As for the remaining integrals in \eqref{eq:LeftWedge.ModularHamiltonian.Mp.Computation},
we first integrate over $q$ by closing a semi-circle contour $C$ in the lower half plane where $\e^{- \i q y}$ yields the decaying exponential $\e^{- R y}$ along the arc with radius $R$ (which is similar to Fig.~\ref{fig:InverseHelmholtz.Boundary.Contour} because $y > 0$) 
and therefore it does not contribute in the limit $R \to \infty$,
hence
\begin{eqnarray}
\label{eq:LeftWedge.ModularHamiltonian.Mp.ContourIntegration}
    M_+^{\textlabel{BCFT}}(x, y)
  &=& \frac{\ell}{2 \pi}
    \lim_{\varepsilon \to 0^+}
    \int_{-\infty}^{\infty}
      \abs{p} \sin(x p)
      \ointctrclockwise_{C}
      \Biggl(
        \frac{\abs{q} \e^{- \i y q}}{q - q_+}
        - \frac{\abs{q} \e^{- \i y q}}{q - q_-}
      \Biggr)
      \id{q}
    \id{p}
\nonumber\\
  &&{}
    + \frac{1}{\pi \ell}
    \lim_{\varepsilon \to 0^+}
    \int_{-\infty}^{\infty}
      \abs{p} \sin(x p)
      \ointctrclockwise_{C}
      \Biggl(
        \frac{\abs{q} \e^{- \i y q}}{(q - q_+)^3}
        - \frac{\abs{q} \e^{- \i y q}}{(q - q_-)^3}
      \Biggr)
      \id{q}
    \id{p}\eqend{.}
\end{eqnarray}
Since the contour encloses the two poles at $q_{\pm} = \pm p - \i \varepsilon$, we get
\begin{eqnarray}
\label{eq:LeftWedge.ModularHamiltonian.Mp.ContourIntegrationResult}
    M_+^{\textlabel{BCFT}}(x, y)
  &=& \frac{\ell}{2 \pi}
    \lim_{\varepsilon \to 0^+}
    \int_{-\infty}^{\infty}
      \abs{p} \sin(x p)
      \bigl[
        2 \pi \i \abs{q_+} \e^{- \i y q_+}
        - 2 \pi \i \abs{q_-} \e^{- \i y q_-}
      \bigr]
    \id{p}
\nonumber\\
  &&{}
    + \frac{1}{\pi \ell}
    \lim_{\varepsilon \to 0^+}
    \int_{-\infty}^{\infty}
      \abs{p} \sin(x p)
      \left.
        \frac{2 \pi \i}{2!}
        \deriv[2]{}{q}
        \Bigl( \abs{q} \e^{- \i y q} \Bigr)
      \right\rvert^{q = q_+}_{q = q_-}
    \id{p}
\nonumber\\
  &=& 2 \ell
    \int_{-\infty}^{\infty}
      p^2 \sin(x p) \sin(y p)
    \id{p}
\nonumber\\
  &&{}
    + \frac{\i}{\ell}
    \int_{-\infty}^{\infty}
      \abs{p} \sin(x p)
      \bigl[
        - 4 \i y \sgn{p} \cos(y p)
        + 2 \i y^2 \abs{p} \sin(y p)
      \bigr]
    \id{p}
  \eqend{.}
\end{eqnarray}
By using that $\abs{p} \sgn(p) = p$ and $\abs{p}^2 = p^2$, 
we recollect the various terms first 
and then apply the prosthaphaeresis formulas, 
finding a form for the integrands that allows 
to evaluate the integrals in terms of derivatives of Dirac deltas.
\begin{eqnarray}
\label{eq:LeftWedge.ModularHamiltonian.Mp}
    M_+^{\textlabel{BCFT}}(x, y)
  &=& \frac{\ell^2 - y^2}{\ell}
    \int_{0}^{\infty}
      p^2 \Bigl[ \cos\bigl( (x - y) p \bigr) - \cos\bigl( (x + y) p \bigr) \Bigr]
    \id{p}
\nonumber\\
  &&{}
    + \frac{2 y}{\ell}
    \int_{0}^{\infty}
      p \Bigl[ \sin\bigl( (x - y) p \bigr) + \sin\bigl( (x + y) p \bigr) \Bigr]
    \id{p}
\nonumber\\
  &=& - 2 \pi
    \biggl(
      \frac{\ell^2 - y^2}{2 \ell}
      \bigl[ \updelta''(x - y) - \updelta''(x + y) \bigr]
      - \frac{y}{\ell}
      \bigl[ \updelta'(x - y) + \updelta'(x + y) \bigr]
    \biggr)
  \eqend{,}
  \qquad
\end{eqnarray}
where the Dirac deltas at $x + y$ 
do not contribute to the operator because $x > 0$ and $y > 0$.
We remark that \eqref{eq:LeftWedge.ModularHamiltonian.Mp}  corresponds to the Legendre operator in Eq.~(3) in \cite{LongoMorsella:2023}
obtained for the massless field on the double cone,
in the special case of  
two-dimensional Minkowski spacetime.

\section{Additional results for the interval separated from the boundary}
\label{appx:AdditionalNumericalResults}

In this appendix we provide additional plots 
supporting the analysis of 
Sec.~\ref{sec:Results.SeparatedInterval.Massive}
for the modular Hamiltonian of the interval separated from the boundary for the massive scalar.

\begin{figure}
  \centering
  \includegraphics{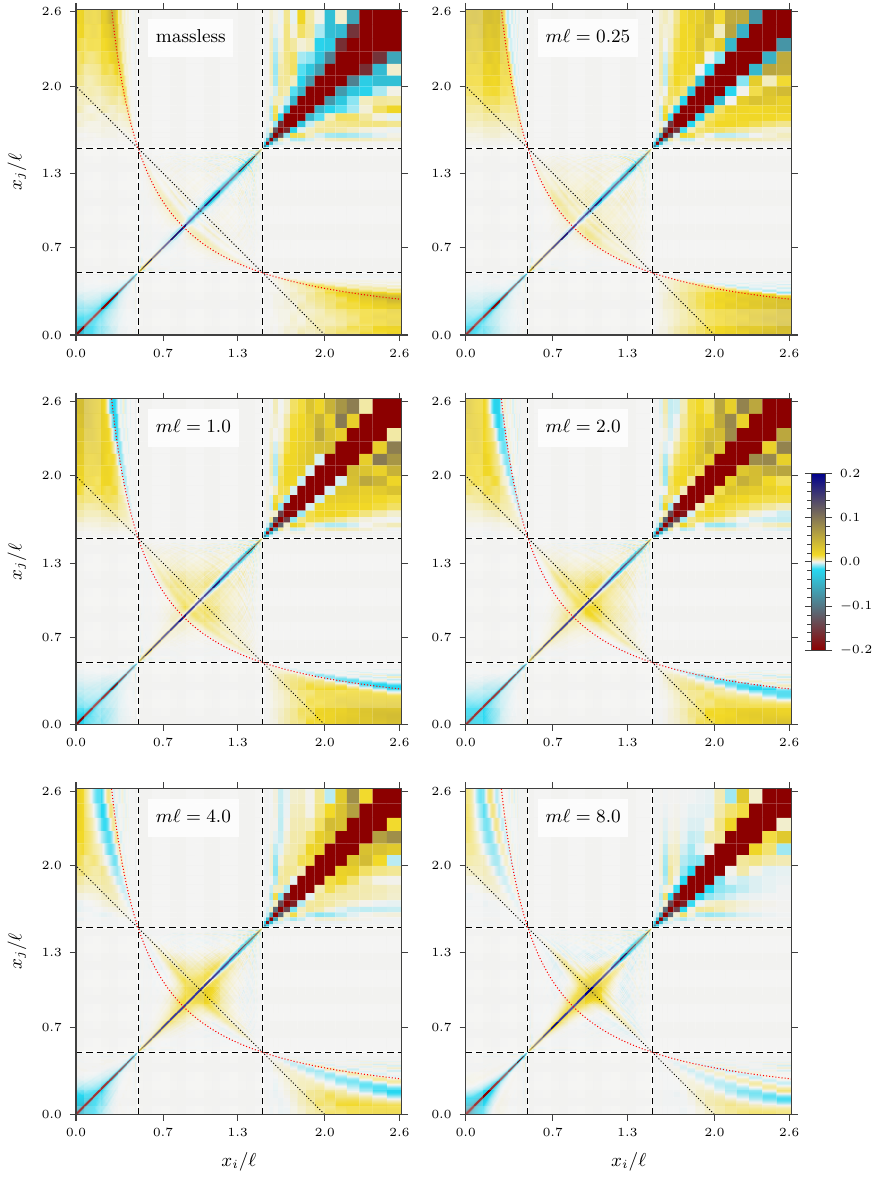}
  \caption{\label{fig:DoubleCone.Mm.Neumann.Matrices} 
    Interval at distance $d=\ell/2$ from the boundary, where Neumann b.c.\@ are imposed:
    Massive regime.
    Output of the numerical algorithm for 
    $M_-^{(256, 16\ell)} / \ell$ before the smearing \eqref{eq:HalfLine.ModularHamiltonian.SmearedBlock.CompactNotation} 
    through the log-Gaussian test functions \eqref{eq:LogGaussian.Discretization},
    for various values of $m\ell$.
    The corresponding results for Dirichlet b.c.\@
    are shown in Fig.~\ref{fig:DoubleCone.Mm.Dirichlet.Matrices}.}
\end{figure}

\begin{figure}
  \centering
  \includegraphics{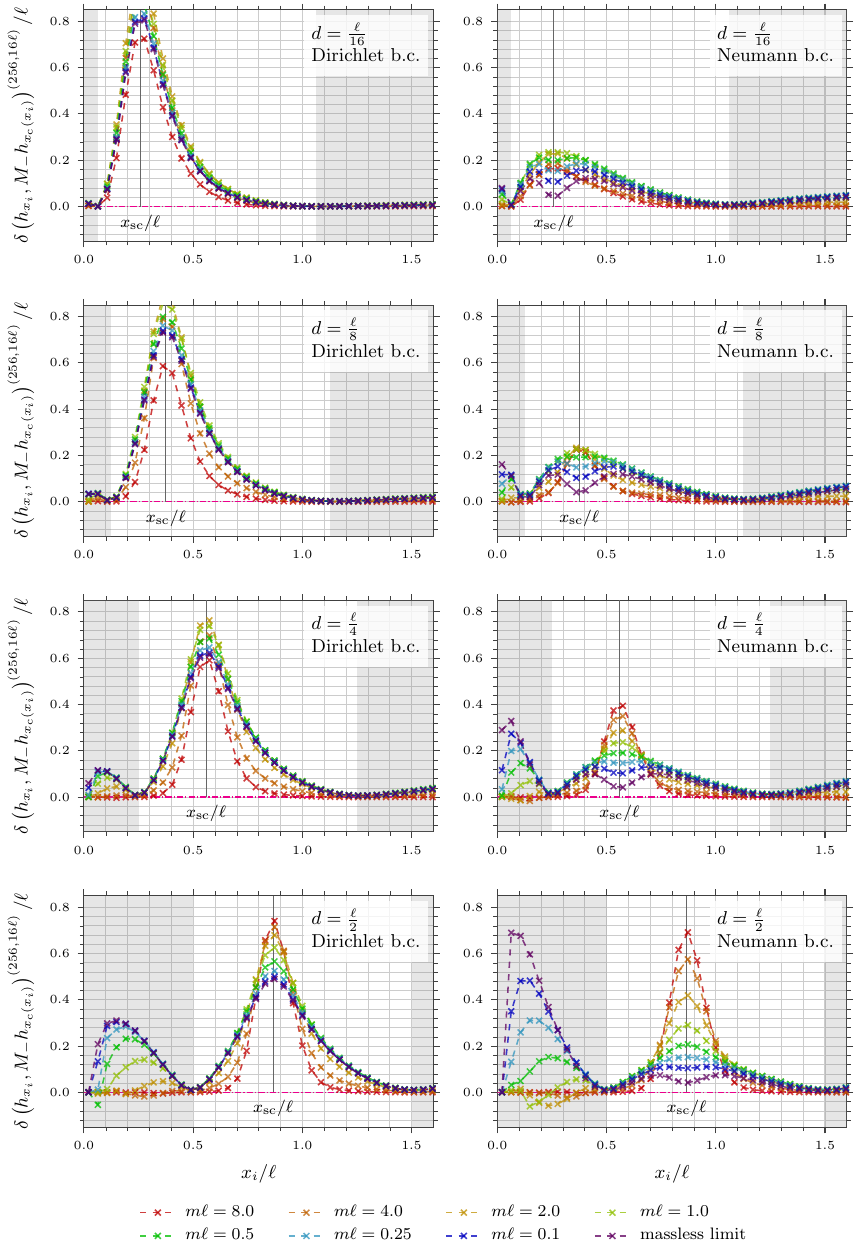}
  \caption{\label{fig:DoubleCone.errMm.Massive.Conjugate}
    Interval separated from the boundary, massive regime:
    Difference of the massive modular Hamiltonian \eqref{eq:DoubleCone.Mm.Error} along the red crosses in Fig.~\ref{fig:DoubleCone.Mm.Massless.SmearedMatrices}, 
    in the same setups of Fig.~\ref{fig:DoubleCone.Mm.Massive.Conjugate}.}
\end{figure}

In Fig.~\ref{fig:DoubleCone.Mm.Neumann.Matrices},
we report $M_-^{(256, 16\ell)} / \ell$ in the case of Neumann b.c.\@ before the smearing against log-Gaussian functions.
The analogous results for Dirichlet b.c.\@
are shown in Fig.~\ref{fig:DoubleCone.Mm.Dirichlet.Matrices}.
In both cases, the plot range has been reduced in order to highlight the non-local contributions away from the main diagonal.
Although there are minor differences in the non-local contributions, the main features are the same ones highlighted in the discussion of Fig.~\ref{fig:DoubleCone.Mm.Dirichlet.Matrices}.
In particular, we remark again the behaviour of the non-local contributions in $A$ as the mass varies from zero to 
a value such that $m \ell = 8.0$.
While in the top left panel of Fig.~\ref{fig:DoubleCone.Mm.Neumann.Matrices} they are mostly along the red dotted line, given by \eqref{eq:ConjugateCurve},
in the bottom right panel, i.e.\@ for the largest value of $m \ell$ that we have considered here, they are along the anti-diagonal (black dotted line).

In Fig.~\ref{fig:DoubleCone.errMm.Massive.Conjugate}, 
which supports Fig.~\ref{fig:DoubleCone.Mm.Massive.Conjugate},
we provide a more detailed comparison of the results for the modular Hamiltonian for the massive scalar when different b.c.\@ are imposed. 
Here we consider the difference \eqref{eq:DoubleCone.Mm.Error}
between the numeric results and the analytic reference \eqref{eq:DoubleCone.LocalReference}, 
extracted along the conjugate curve,
for either Dirichlet b.c.\@ (left panels) or 
Neumann b.c.\@ (right panels)
and for four different values of the distance $d$, 
corresponding to the four panel rows in the figure.
These plots are analogous to the right panels in
Fig.~\ref{fig:DoubleCone.Mm.Boundaries.Conjugate} in the massless regime.
The plots in Fig.~\ref{fig:DoubleCone.errMm.Massive.Conjugate} 
highlight the non-local nature of the modular Hamiltonian.
We  stress that the plots along the conjugate curve in Fig.~\ref{fig:DoubleCone.Mm.Massive.Conjugate} 
and the corresponding difference in Fig.~\ref{fig:DoubleCone.errMm.Massive.Conjugate} 
provide only a guiding idea for the non-local contribution 
to the modular Hamiltonian,
which could be helpful in finding 
the underlying analytic expressions.

\begin{figure}
  \centering
  \includegraphics{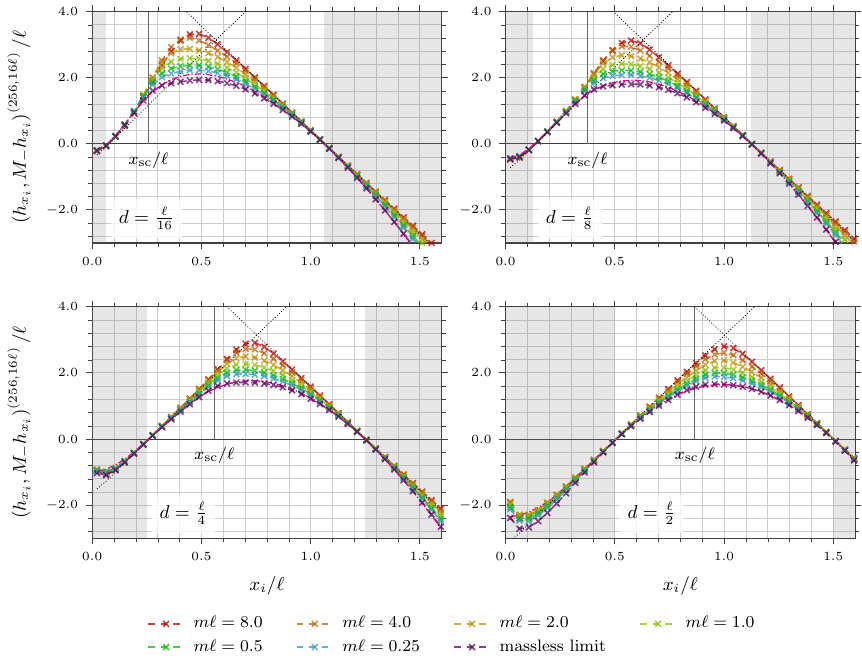}
  \caption{\label{fig:DoubleCone.Mm.Massive.Robin.Diagonal}
    Interval separated from the boundary for various $d/\ell$: 
    Massive regime for Robin b.c.\@ with $m = \eta$. 
    Diagonal (white crosses in Fig.~\ref{fig:DoubleCone.Mm.Massless.SmearedMatrices}) 
    of the modular Hamiltonian for different values of $m \ell$.}
\end{figure}

\begin{figure}
  \centering
  \includegraphics{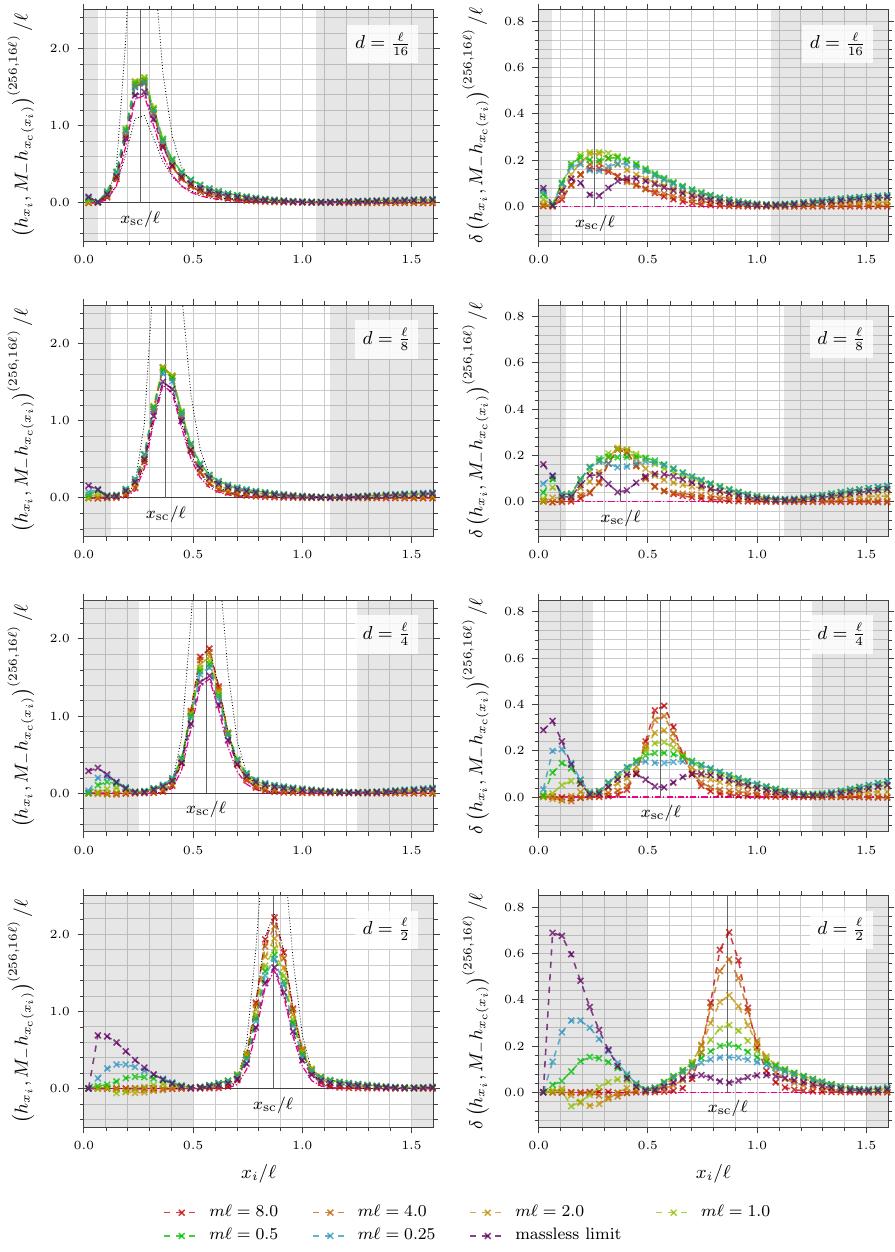}
  \caption{\label{fig:DoubleCone.Mm.Massive.Robin.Conjugate} 
    Interval separated from the boundary and massive regime.
    Numerical results for Robin b.c.\@ with $m = \eta$ along the red crosses in Fig.~\ref{fig:DoubleCone.Mm.Massless.SmearedMatrices}, 
    in the same setups of Fig.~\ref{fig:DoubleCone.Mm.Massive.Conjugate} and Fig.~\ref{fig:DoubleCone.errMm.Massive.Conjugate}.}
\end{figure}

Finally, 
in Fig.~\ref{fig:DoubleCone.Mm.Massive.Robin.Diagonal}
and in the left panels of Fig.~\ref{fig:DoubleCone.Mm.Massive.Robin.Conjugate}
we repeat the analysis performed for Fig.~\ref{fig:DoubleCone.Mm.Massive.Diagonal}
and Fig.~\ref{fig:DoubleCone.Mm.Massive.Conjugate} respectively
(in the case of Dirichlet b.c.\@ and Neumann b.c.\@)
also for the special case of Robin b.c.\@ with $\eta = m$.
Similarly, we include
the difference \eqref{eq:DoubleCone.Mm.Error} 
in the right panels of Fig.~\ref{fig:DoubleCone.Mm.Massive.Robin.Conjugate}.
Since $\eta = m$, the results for small masses 
are similar to corresponding ones for Neumann b.c.\@
in the right panels of Fig.~\ref{fig:DoubleCone.Mm.Massive.Diagonal}, Fig.~\ref{fig:DoubleCone.Mm.Massive.Conjugate} and  Fig.~\ref{fig:DoubleCone.errMm.Massive.Conjugate}.
For large separation distance $d$, 
some results for the interval subspace 
(see the unshaded region e.g.\@ in the bottom right panel of 
Fig.~\ref{fig:DoubleCone.Mm.Massive.Robin.Diagonal})
are comparable to the ones for the modular Hamiltonian 
of an interval in the line for the massive scalar 
in the ground state
\cite{BostelmannCadamuroMinz:2023}.

\newpage
\bibliographystyle{nb}
\bibliography{references}

\end{document}